%% file: main.tex
\pdfoutput=1

\documentclass[12pt,a4paper]{article}

\usepackage{ifthen} 
\usepackage{rotating}
\usepackage{dashrule}
\usepackage{array} 
\usepackage{dcolumn}
\usepackage{comment}
\usepackage{xcolor}

\usepackage{graphpap} 
\usepackage{makecell} 
\usepackage{rotating} 
\usepackage{tikz}
\usepackage{xcolor}
\usepackage{longtable} 

\usepackage{lscape}
\usepackage{multirow} 
\usepackage{mathrsfs}
\usepackage{mathtools} 
\usetikzlibrary{patterns}
\usepackage{nicefrac} 
\usepackage{transparent}
\usepackage{afterpage}

\newboolean{pdflatex}
\setboolean{pdflatex}{true} 

\newboolean{articletitles}
\setboolean{articletitles}{true} 

\newboolean{uprightparticles}
\setboolean{uprightparticles}{True} 

\def\paperauthors{LHCb collaboration} 
\def\paperasciititle{Study of Bc+ decays to charmonia and three light hadrons} 
\def\papertitle{Study of \Bc decays to charmonia and three light hadrons} 

\def\paperkeywords{{High Energy Physics}, {LHCb}} 
\def\papercopyright{\the\year\ CERN for the benefit of the LHCb collaboration} 
\def\paperlicence{CC BY 4.0 licence}
\def\paperlicenceurl{https://creativecommons.org/licenses/by/4.0/}

\usepackage{wasysym}

\DeclareMathOperator*{\bigplus}{\scalerel*{+}{\sum}}
\usepackage{scalerel}

\makeatletter
\g@addto@macro\bfseries{\boldmath}
\makeatother

\input{preamble}
\usepackage{longtable} 

\newcolumntype{d}[1]{D{,}{\,\pm\,}{#1} }
\newcolumntype{f}[1]{D{,}{.}{#1} }

\begin{document}

\renewcommand{\thefootnote}{\fnsymbol{footnote}}
\setcounter{footnote}{1}

\input{title-LHCb-PAPER}


\renewcommand{\thefootnote}{\arabic{footnote}}
\setcounter{footnote}{0}



\pagestyle{plain} 
\setcounter{page}{1}
\pagenumbering{arabic}


%

\input{introduction}
\input{detector}
\input{selection}
\input{sig_eff}
\input{systematics}

\input{results}
\input{acknowledgements}

\clearpage 
\appendix

\input{covariances}





\usetikzlibrary{patterns}

\clearpage
\addcontentsline{toc}{section}{References}
\bibliographystyle{LHCb}
\bibliography{main,standard,LHCb-PAPER,LHCb-CONF,LHCb-DP,LHCb-TDR}
 
\newpage

\input{Authorship_LHCb-PAPER-2021-034}

\end{document}

%% file: preamble.tex
\usepackage[top=1in, bottom=1.25in, left=1in, right=1in]{geometry}

\usepackage{microtype}
\usepackage{lineno}  
\usepackage{xspace} 
\usepackage{caption} 

\usepackage{graphicx}  
\usepackage{color}
\usepackage{colortbl}
\graphicspath{{./figs/}} 
\DeclareGraphicsExtensions{.pdf,.PDF,png,.PNG}

\usepackage{amsmath} 
\usepackage{amssymb}
\usepackage{amsfonts}
\usepackage{upgreek} 

\newcommand*\patchAmsMathEnvironmentForLineno[1]{%
\expandafter\let\csname old#1\expandafter\endcsname\csname #1\endcsname
\expandafter\let\csname oldend#1\expandafter\endcsname\csname
end#1\endcsname
 \renewenvironment{#1}%
   {\linenomath\csname old#1\endcsname}%
   {\csname oldend#1\endcsname\endlinenomath}%
}
\newcommand*\patchBothAmsMathEnvironmentsForLineno[1]{%
  \patchAmsMathEnvironmentForLineno{#1}%
  \patchAmsMathEnvironmentForLineno{#1*}%
}
\AtBeginDocument{%
\patchBothAmsMathEnvironmentsForLineno{equation}%
\patchBothAmsMathEnvironmentsForLineno{align}%
\patchBothAmsMathEnvironmentsForLineno{flalign}%
\patchBothAmsMathEnvironmentsForLineno{alignat}%
\patchBothAmsMathEnvironmentsForLineno{gather}%
\patchBothAmsMathEnvironmentsForLineno{multline}%
\patchBothAmsMathEnvironmentsForLineno{eqnarray}%
}

\usepackage{hyperxmp}

\usepackage[pdftex,
            pdfauthor={\paperauthors},
            pdftitle={\paperasciititle},
            pdfkeywords={\paperkeywords},
            pdfcopyright={Copyright (C) \papercopyright},
            pdflicenseurl={\paperlicenceurl}]{hyperref}

\usepackage[colorinlistoftodos,textsize=scriptsize]{todonotes}

\usepackage[all]{hypcap} 

\input{lhcb-symbols-def} 

\usepackage{cite} 
\usepackage{mciteplus}

%% file: lhcb-symbols-def.tex
\setboolean{uprightparticles}{true} 
\usepackage{xspace} 
\usepackage{upgreek}

\def\lhcb   {\mbox{LHCb}\xspace}

\def\cdf    {\mbox{CDF}\xspace}

\def\MagUp {\mbox{\em Mag\kern -0.05em Up}\xspace}

\ifthenelse{\boolean{uprightparticles}}%
{

 \def\Pmu         {\ensuremath{\upmu}\xspace}

 \def\Ppi         {\ensuremath{\uppi}\xspace}                 
                  
 \def\Prho        {\ensuremath{\uprho}\xspace}

 \def\Pphi        {\ensuremath{\upphi}\xspace}

 \def\Ppsi        {\ensuremath{\uppsi}\xspace}

 \def\PDelta      {\ensuremath{\Delta}\xspace}                 
 \def\PXi         {\ensuremath{\Xi}\xspace}                 
 \def\PLambda     {\ensuremath{\Lambda}\xspace}                 
 \def\PSigma      {\ensuremath{\Sigma}\xspace}                 
 \def\POmega      {\ensuremath{\Omega}\xspace}                 
 \def\PUpsilon    {\ensuremath{\Upsilon}\xspace}

 \def\PB      {\ensuremath{\mathrm{B}}\xspace}                 
                  
 \def\PD      {\ensuremath{\mathrm{D}}\xspace}

 \def\PJ      {\ensuremath{\mathrm{J}}\xspace}                 
 \def\PK      {\ensuremath{\mathrm{K}}\xspace}

 \def\PR      {\ensuremath{\mathrm{R}}\xspace}

 \def\PW      {\ensuremath{\mathrm{W}}\xspace}                 
 \def\PX      {\ensuremath{\mathrm{X}}\xspace}                 
 \def\PY      {\ensuremath{\mathrm{Y}}\xspace}                 
                  
 \def\Pa      {\ensuremath{\mathrm{a}}\xspace}                 
 \def\Pb      {\ensuremath{\mathrm{b}}\xspace}                 
 \def\Pc      {\ensuremath{\mathrm{c}}\xspace}

 \def\Pf      {\ensuremath{\mathrm{f}}\xspace}                 
                  
 \def\Ph      {\ensuremath{\mathrm{h}}\xspace}                 
 \def\Pi      {\ensuremath{\mathrm{i}}\xspace}

 \def\Pp      {\ensuremath{\mathrm{p}}\xspace}

 \def\Ps      {\ensuremath{\mathrm{s}}\xspace}

 \def\thebaroffset{0.0em}
}
{

 \def\Pmu         {\ensuremath{\mu}\xspace}

 \def\Ppi         {\ensuremath{\pi}\xspace}                 
                  
 \def\Prho        {\ensuremath{\rho}\xspace}

 \def\Pphi        {\ensuremath{\phi}\xspace}

 \def\Ppsi        {\ensuremath{\psi}\xspace}                 
                  
 \mathchardef\PDelta="7101
 \mathchardef\PXi="7104
 \mathchardef\PLambda="7103
 \mathchardef\PSigma="7106
 \mathchardef\POmega="710A
 \mathchardef\PUpsilon="7107
                  
 \def\PB      {\ensuremath{B}\xspace}                 
                  
 \def\PD      {\ensuremath{D}\xspace}

 \def\PJ      {\ensuremath{J}\xspace}                 
 \def\PK      {\ensuremath{K}\xspace}

 \def\PR      {\ensuremath{R}\xspace}

 \def\PW      {\ensuremath{W}\xspace}                 
 \def\PX      {\ensuremath{X}\xspace}                 
 \def\PY      {\ensuremath{Y}\xspace}                 
                  
 \def\Pa      {\ensuremath{a}\xspace}                 
 \def\Pb      {\ensuremath{b}\xspace}                 
 \def\Pc      {\ensuremath{c}\xspace}

 \def\Pf      {\ensuremath{f}\xspace}                 
                  
 \def\Ph      {\ensuremath{h}\xspace}                 
 \def\Pi      {\ensuremath{i}\xspace}

 \def\Pp      {\ensuremath{p}\xspace}

 \def\Ps      {\ensuremath{s}\xspace}

 \def\thebaroffset{0.18em}
}
\newcommand{\offsetoverline}[2][\thebaroffset]{\kern #1\overline{\kern -#1 #2}}%

\makeatletter
\ifcase \@ptsize \relax
  \newcommand{\miniscule}{\@setfontsize\miniscule{4}{5}}
\or
  \newcommand{\miniscule}{\@setfontsize\miniscule{5}{6}}
\or
  \newcommand{\miniscule}{\@setfontsize\miniscule{5}{6}}
\fi
\makeatother

\DeclareRobustCommand{\optbar}[1]{\shortstack{{\miniscule (\rule[.5ex]{1.25em}{.18mm})}
  \\ [-.7ex] $#1$}}

\def\mumu       {{\ensuremath{\Pmu^+\Pmu^-}}\xspace}

\def\Wp     {{\ensuremath{\PW^+}}\xspace}

\def\squark    {{\ensuremath{\Ps}}\xspace}

\def\cquark    {{\ensuremath{\Pc}}\xspace}

\def\bquark    {{\ensuremath{\Pb}}\xspace}

\def\pion   {{\ensuremath{\Ppi}}\xspace}
\def\piz    {{\ensuremath{\pion^0}}\xspace}
\def\pip    {{\ensuremath{\pion^+}}\xspace}
\def\pim    {{\ensuremath{\pion^-}}\xspace}

\def\rhomeson {{\ensuremath{\Prho}}\xspace}
\def\rhoz     {{\ensuremath{\rhomeson^0}}\xspace}

\def\kaon    {{\ensuremath{\PK}}\xspace}
\def\Kbar    {{\ensuremath{\offsetoverline{\PK}}}\xspace}

\def\KorKbar {\kern \thebaroffset\optbar{\kern -\thebaroffset \PK}{}\xspace}

\def\Kp      {{\ensuremath{\kaon^+}}\xspace}
\def\Km      {{\ensuremath{\kaon^-}}\xspace}

\def\KS      {{\ensuremath{\kaon^0_{\mathrm{S}}}}\xspace}

\def\Kstarz  {{\ensuremath{\kaon^{*0}}}\xspace}
\def\Kstarzb {{\ensuremath{\Kbar{}^{*0}}}\xspace}

\def\aone     {{\ensuremath{\Pa_1(1260)^+}}\xspace}

\newcommand{\phiz}{\ensuremath{\Pphi}\xspace}

\def\Dbar    {{\ensuremath{\offsetoverline{\PD}}}\xspace}
\def\D       {{\ensuremath{\PD}}\xspace}

\def\DorDbar {\kern \thebaroffset\optbar{\kern -\thebaroffset \PD}\xspace}
\def\Dz      {{\ensuremath{\D^0}}\xspace}
\def\Dzb     {{\ensuremath{\Dbar{}^0}}\xspace}

\def\Dm      {{\ensuremath{\D^-}}\xspace}

\def\Dstarp  {{\ensuremath{\D^{*+}}}\xspace}

\def\theDstarm{{\ensuremath{\D^{*-}}}\xspace}

\def\Ds      {{\ensuremath{\D^+_\squark}}\xspace}
\def\Dsp     {{\ensuremath{\D^+_\squark}}\xspace}
\def\Dsm     {{\ensuremath{\D^-_\squark}}\xspace}

\def\B       {{\ensuremath{\PB}}\xspace}

\def\BorBbar {\kern \thebaroffset\optbar{\kern -\thebaroffset \PB}\xspace}
\def\Bz      {{\ensuremath{\B^0}}\xspace}

\def\Bd      {{\ensuremath{\B^0}}\xspace}

\def\BdorBdbar {\kern \thebaroffset\optbar{\kern -\thebaroffset \Bd}\xspace}
\def\Bu      {{\ensuremath{\B^+}}\xspace}

\def\Bp      {{\ensuremath{\Bu}}\xspace}

\def\Bs      {{\ensuremath{\B^0_\squark}}\xspace}

\def\BsorBsbar {\kern \thebaroffset\optbar{\kern -\thebaroffset \Bs}\xspace}
\def\Bc      {{\ensuremath{\B_\cquark^+}}\xspace}

\def\jpsi     {{\ensuremath{{\PJ\mskip -3mu/\mskip -2mu\Ppsi\mskip 2mu}}}\xspace}
\def\psitwos  {{\ensuremath{\Ppsi{\rm{(2S)}}}}\xspace}

\def\Y#1S{\ensuremath{\PUpsilon{(#1S)}}\xspace}

\def\proton      {{\ensuremath{\Pp}}\xspace}

\def\LorLbar     {\kern \thebaroffset\optbar{\kern -\thebaroffset \PLambda}\xspace}

\def\BF         {{\ensuremath{\mathcal{B}}}\xspace}
\def\BR         {\BF}

\newcommand{\decay}[2]{\ensuremath{#1\!\to #2}\xspace} 

\def\to                 {\ensuremath{\rightarrow}\xspace}

\def\eps   {{\ensuremath{\varepsilon}}\xspace}

\def\BcTojpsitripi    {\decay{\Bc}{\jpsi\pip\pim\pip}}
\def\BcTopsitwostripi {\decay{\Bc}{\psitwos\pip\pim\pip}}
\def\BcTojpsikkpi    {\decay{\Bc}{\jpsi\Kp\Km\pip}}
\def\BcTopsitwoskkpi {\decay{\Bc}{\psitwos\Kp\Km\pip}}

\def\BcTojpsikpipi  {\decay{\Bc}{\jpsi\Kp\pim\pip}}
\def\BcTojpsikkk  {\decay{\Bc}{\jpsi\Kp\Km\Kp}}
\def\BcTopsitwospi  {\decay{\Bc}{\psitwos\pip}}

\def\Tojpsitripi    {\jpsi\pip\pim\pip}
\def\Topsitwostripi {\psitwos\pip\pim\pip}
\def\Tojpsikkpi   {\jpsi\Kp\Km\pip}
\def\Topsitwoskkpi {\psitwos\Kp\Km\pip}

\def\Tojpsikpipi {\jpsi\Kp\pim\pip}
\def\Tojpsikkk  {\jpsi\Kp\Km\Kp}
\def\Topsitwospi  {\psitwos\pip}

\def\psitwosTojpsipipi {\mbox{\decay{\psitwos}{\jpsi\pip\pim}}}

\def\JpsiPiPi{\jpsi\pip\pim}

\def\AT#1     {\ensuremath{A_{\mathrm{T}}^{#1}}\xspace}           

\def\C#1      {\ensuremath{\mathcal{C}_{#1}}\xspace}                       
\def\Cp#1     {\ensuremath{\mathcal{C}_{#1}^{'}}\xspace}                    
\def\Ceff#1   {\ensuremath{\mathcal{C}_{#1}^{\mathrm{(eff)}}}\xspace}        
\def\Cpeff#1  {\ensuremath{\mathcal{C}_{#1}^{'\mathrm{(eff)}}}\xspace}       
\def\Ope#1    {\ensuremath{\mathcal{O}_{#1}}\xspace}                       
\def\Opep#1   {\ensuremath{\mathcal{O}_{#1}^{'}}\xspace}                    

\newcommand{\nospaceunit}[1]{\ensuremath{\text{#1}}}       
\newcommand{\aunit}[1]{\ensuremath{\text{\,#1}}}       

\newcommand{\tev}{\aunit{Te\kern -0.1em V}\xspace}
\newcommand{\gev}{\aunit{Ge\kern -0.1em V}\xspace}
\newcommand{\mev}{\aunit{Me\kern -0.1em V}\xspace}
\newcommand{\kev}{\aunit{ke\kern -0.1em V}\xspace}
\newcommand{\ev}{\aunit{e\kern -0.1em V}\xspace}
\newcommand{\mevc}{\ensuremath{\aunit{Me\kern -0.1em V\!/}c}\xspace}
\newcommand{\gevc}{\ensuremath{\aunit{Ge\kern -0.1em V\!/}c}\xspace}
\newcommand{\mevcc}{\ensuremath{\aunit{Me\kern -0.1em V\!/}c^2}\xspace}
\newcommand{\kevcc}{\ensuremath{\aunit{ke\kern -0.1em V\!/}c^2}\xspace}
\newcommand{\gevcc}{\ensuremath{\aunit{Ge\kern -0.1em V\!/}c^2}\xspace}

\def\mum  {\ensuremath{\,\upmu\nospaceunit{m}}\xspace}

\def\fb   {\ensuremath{\aunit{fb}}\xspace}
\def\invfb   {\ensuremath{\fb^{-1}}\xspace}

\newcommand{\chisq}{\ensuremath{\chi^2}\xspace}

\newcommand{\chisqip}{\ensuremath{\chi^2_{\text{IP}}}\xspace}

\def\gsim{{~\raise.15em\hbox{$>$}\kern-.85em
          \lower.35em\hbox{$\sim$}~}\xspace}
\def\lsim{{~\raise.15em\hbox{$<$}\kern-.85em
          \lower.35em\hbox{$\sim$}~}\xspace}

\def\sPlot{\mbox{\em sPlot}\xspace}

\def\pt         {\ensuremath{p_{\mathrm{T}}}\xspace}

\def\evtgen     {\mbox{\textsc{EvtGen}}\xspace}

\def\geant      {\mbox{\textsc{Geant4}}\xspace}

\def\photos     {\mbox{\textsc{Photos}}\xspace}

\def\pythia     {\mbox{\textsc{Pythia}}\xspace}

\def\tell1  {TELL1\xspace}
\def\ukl1   {UKL1\xspace}



%% file: title-LHCb-PAPER.tex

\begin{titlepage}
\pagenumbering{roman}

\vspace*{-1.5cm}
\centerline{\large EUROPEAN ORGANIZATION FOR NUCLEAR RESEARCH (CERN)}
\vspace*{1.5cm}
\noindent
\begin{tabular*}{\linewidth}{lc@{\extracolsep{\fill}}r@{\extracolsep{0pt}}}
\ifthenelse{\boolean{pdflatex}}
{\vspace*{-1.5cm}\mbox{\!\!\!\includegraphics[width=.14\textwidth]{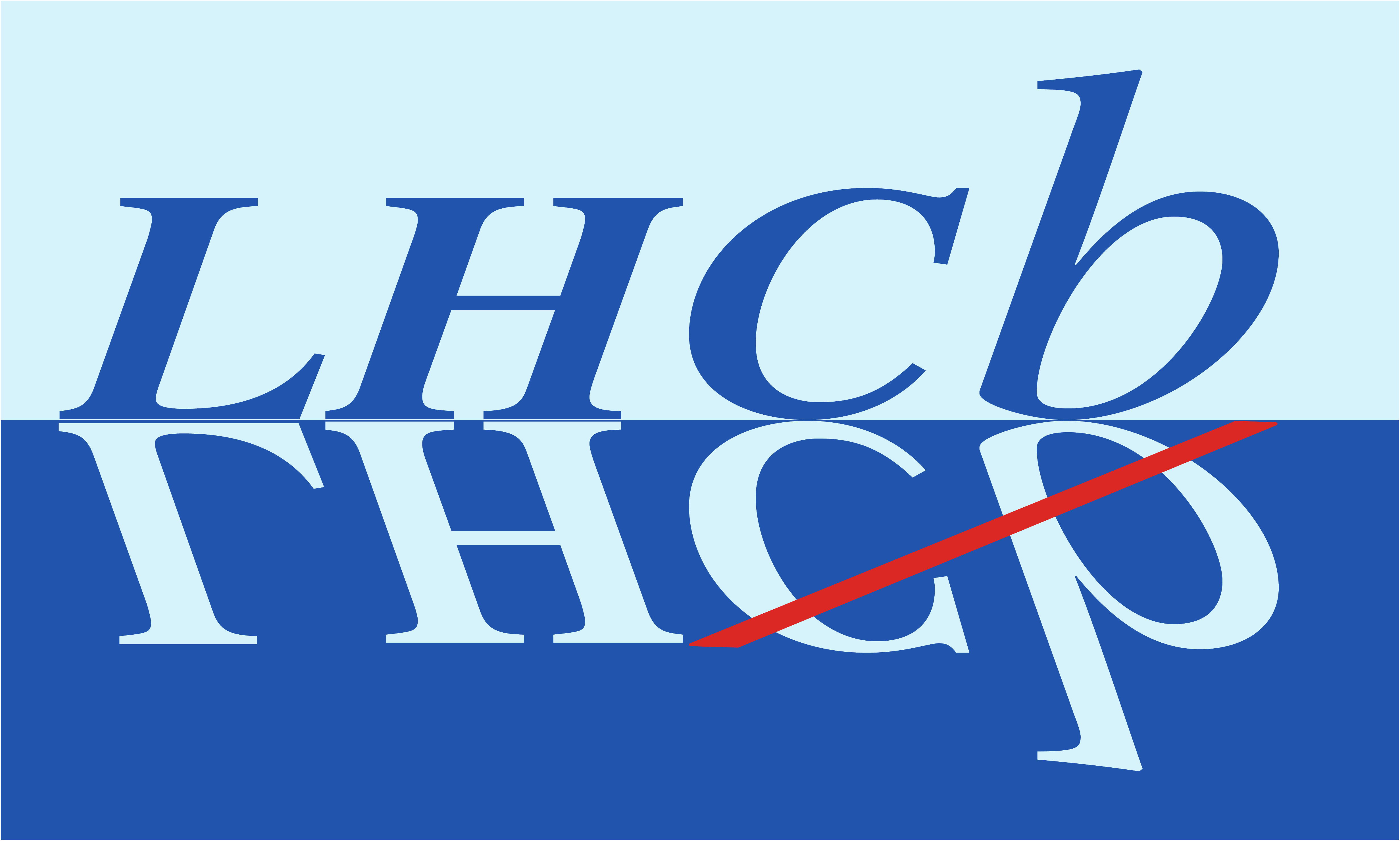}} & &}%
{\vspace*{-1.2cm}\mbox{\!\!\!\includegraphics[width=.12\textwidth]{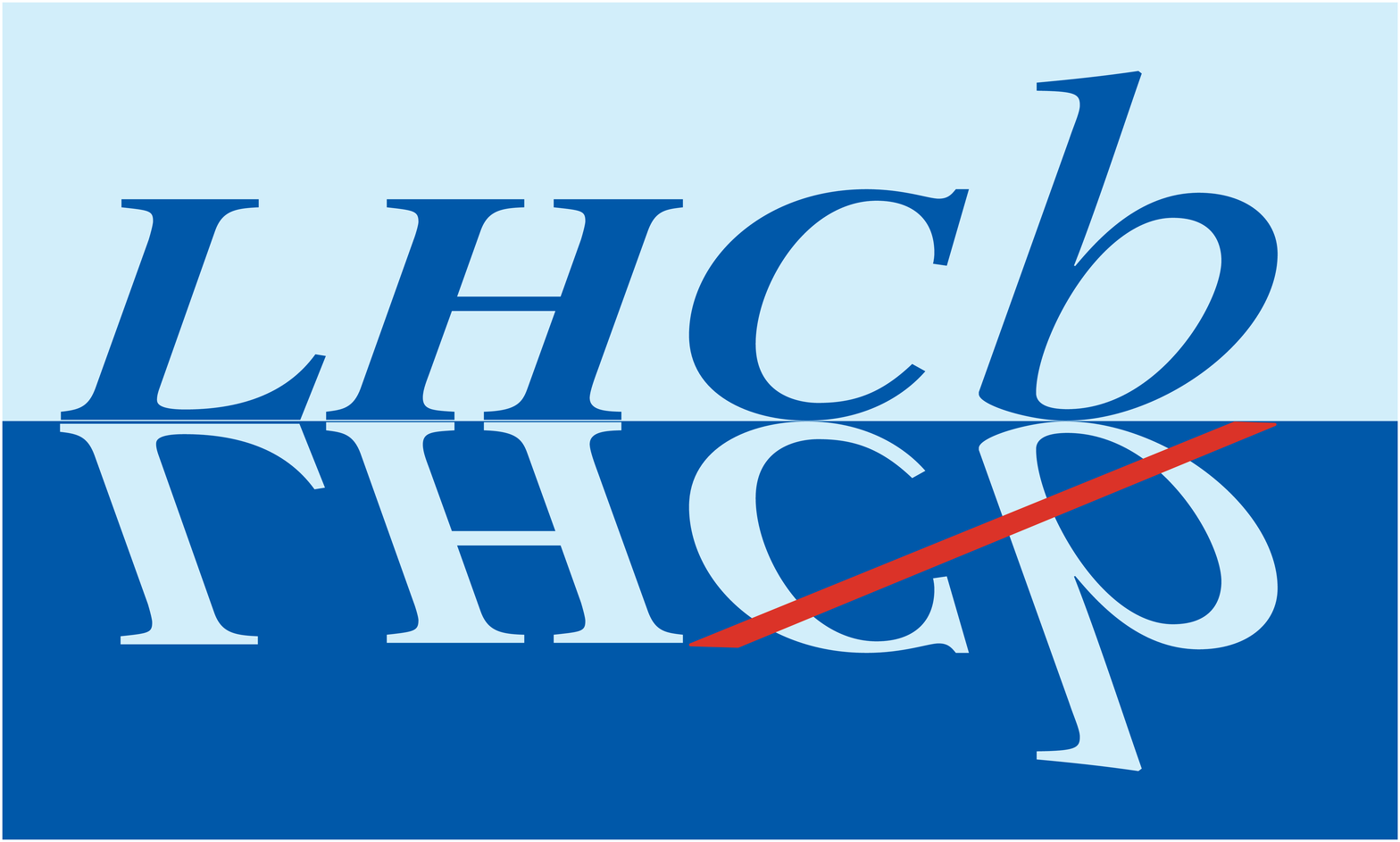}} & &}%
\\
 & & CERN-EP-2021-216\\  
 & & LHCb-PAPER-2021-034 \\  
  & & November 2, 2021 \\ 
\end{tabular*}

\vspace*{3.5cm}

{\normalfont\bfseries\boldmath\huge
\begin{center}
  \papertitle 
\end{center}
}

\vspace*{0.2cm}

\begin{center}
\paperauthors\footnote{Authors are listed at the end of this paper.}
\end{center}


\begin{abstract}
\noindent 
Using proton-proton collision data, 
corresponding to an integrated luminosity of $9\invfb$
collected with the \lhcb detector, 
seven decay modes of the~\Bc~meson into a \jpsi or \psitwos~meson 
and three charged hadrons, kaons or pions,  are studied.
The~decays  
\mbox{\decay{\Bc}{(\psitwosTojpsipipi)\pip}}, 
\BcTopsitwostripi, 
\mbox{\BcTojpsikpipi} and \BcTojpsikkk 
are observed for the first time,
and evidence  for the~\mbox{\BcTopsitwoskkpi}~decay is found,
where $\jpsi$~and $\psitwos$~mesons are 
reconstructed in their dimuon decay modes.
The~ratios of branching fractions 
between the~different  \Bc~decays 
are reported as well as the fractions 
of the~decays proceeding 
via intermediate resonances.
The~results largely support
the~factorisation approach used 
for a~theoretical description 
of the~studied decays.
\end{abstract}

\vspace{\fill}

\begin{center}
  Submitted to JHEP 
\end{center}

\vspace*{0.5cm}

{\footnotesize 
\centerline{\copyright~\papercopyright. \href{\paperlicenceurl}{\paperlicence}.}}
\vspace*{2mm}

\end{titlepage}


\newpage
\setcounter{page}{2}
\mbox{~}
%

\cleardoublepage

%% file: introduction.tex
\section{Introduction}
\label{sec:Introduction}
The~\Bc mesons are unique because they contain two 
different heavy-flavour quarks, charm and beauty. 
The~ground state has a~rich set of weak\nobreakdash-decay modes 
since either of
the~heavy quarks can decay with the~other
behaving as a~spectator quark, 
or both quarks can annihilate via a~virtual \Wp~boson. 
Studies of \Bc decay channels and measurements of their branching fractions 
improve the~understanding of
models describing strong interactions
and test various effective models. 
Experiments at the~Large Hadron Collider\,(LHC) have 
opened a~new era for~\Bc meson investigations. 
The~high \bquark-quark production cross-section at 
the~LHC enables the~\lhcb experiment 
to study the~production, decays and other properties 
of the~\Bc meson~\cite{LHCb-PAPER-2011-044, 
LHCb-PAPER-2012-054, 
LHCb-PAPER-2013-010, 
LHCb-PAPER-2013-021, 
LHCb-PAPER-2013-044, 
LHCb-PAPER-2013-047, 
LHCb-PAPER-2014-009, 
LHCb-PAPER-2014-039, 
LHCb-PAPER-2014-050, 
LHCb-PAPER-2014-060, 
LHCb-PAPER-2015-024, 
LHCb-PAPER-2016-001, 
LHCb-PAPER-2016-022, 
LHCb-PAPER-2016-020, 
LHCb-PAPER-2016-055, 
LHCb-PAPER-2016-058, 
LHCb-PAPER-2017-035, 
LHCb-PAPER-2017-045, 
LHCb-PAPER-2019-033, 
LHCb-PAPER-2020-003, 
LHCb-PAPER-2021-023}.  

Although the~\Bc meson was discovered in 1998 by 
the~\cdf collaboration~\cite{PhysRevLett.81.2432, 
PhysRevD.58.112004}, only two  
\decay{\Bc}{\Ppsi3\Ph^\pm}~decay 
channels,\footnote{Inclusion of charge-conjugate states is implied throughout the paper.} where 
the~symbol \Ppsi 
denotes a~\jpsi or a \psitwos meson 
and $\Ph^{\pm}$ stands for a~charged kaon or pion,
were previously observed by~the~\lhcb~collaboration~\cite{LHCb-PAPER-2011-044, 
LHCb-PAPER-2013-047}. 
The~decays of \Bc mesons into charmonium and light hadrons 
are expected to be well described by the factorisation 
approach~\cite{Bauer:1986bm, WIRBEL198833}.
In~this scheme, the~\decay{\Bc}{\Ppsi3\Ph^{\pm}}~decay 
is characterised by the~form factors of 
the ~\decay{\Bc}{\Ppsi\Wp} transition
and the~universal spectral function 
for the~virtual \Wp boson fragmenting
into light hadrons~\cite{Likhoded:2009ib}.
The~ratios of branching fractions of various \decay{\Bc}{\Ppsi3\Ph^\pm} decays, 
based on this theoretical approach are predicted in  Refs.~\cite{Luchinsky:2013yla, Likhoded:2013iua}.
A~measurement of the~branching fractions of 
the~exclusive \Bc~meson decays into 
the~final states consisting of charmonium and light hadrons 
allows for precise tests of the factorisation approach.

In this paper a~study of the~\Bc meson decaying into 
seven final states, namely
the~Cabibbo\nobreakdash-favoured 
$\BcTojpsitripi$,
$\BcTopsitwostripi$,
$\BcTojpsikkpi$,
$\BcTopsitwoskkpi$,
$\mbox{\decay{\Bc}{(\decay{\psitwos}{\jpsi\pip\pim})}\pip}$~decays, 
and the~Cabibbo\nobreakdash-suppressed 
\mbox{$\BcTojpsikpipi$}
and \mbox{$\BcTojpsikkk$}~decays, 
 is reported.
The~analysis is based 
on proton\nobreakdash-proton\,($\proton\proton$) collision data, 
corresponding to an~integrated luminosity of 9\,\invfb
collected with the \lhcb detector between 2011 and 2018
at~centre-of-mass energies of 7, 8, and 13\,\tev. 
The~\jpsi and \psitwos mesons are reconstructed from their decay into two muons
and the $\psitwosTojpsipipi$ channel is used  for the \BcTopsitwospi decay. 
The~ratios of branching fractions for the decay channels under study are presented.

%% file: detector.tex
\section{Detector and simulation}
\label{sec:Detector}

The \lhcb detector~\cite{Alves:2008zz,LHCb-DP-2014-002} is a single-arm forward
spectrometer covering the~pse\-udora\-pi\-dity range \mbox{$2<\eta <5$},
designed for the study of particles containing $\bquark$~or~$\cquark$~quarks. 
The~detector includes a high-precision tracking system consisting of a 
silicon-strip vertex detector surrounding the \proton\proton interaction
region~\cite{LHCb-DP-2014-001}, a large-area silicon-strip detector located
upstream of a dipole magnet with a bending power of about $4{\mathrm{\,Tm}}$,
and three stations of silicon-strip detectors and straw drift tubes~\cite{LHCb-DP-2013-003,LHCb-DP-2017-001} placed downstream of the magnet. 
The tracking system provides a measurement of the momentum of charged particles
with a relative uncertainty that varies from $0.5\%$ at low momentum to $1.0\%$~at~$200 \gevc$. 
The~momentum scale is calibrated using samples of $\decay{\jpsi}{\mumu}$ 
and $\decay{\Bu}{\jpsi\Kp}$~decays collected concurrently
with the~data sample used for this analysis~\cite{LHCb-PAPER-2012-048,LHCb-PAPER-2013-011}. 
The~relative accuracy of this
procedure is estimated to be $3 \times 10^{-4}$ using samples of other
fully reconstructed $\bquark$~hadrons, $\PUpsilon$~and
$\KS$~mesons.
The~minimum distance between a track and 
a~primary $\proton\proton$\nobreakdash-collision vertex\,(PV), 
the~impact parameter, 
is~measured with a~resolution of $(15+29/\pt)\mum$, where \pt is the component 
of the~momentum transverse to the beam, in\,\gevc. Different types of charged hadrons
are distinguished using information from 
two ring\nobreakdash-imaging Cherenkov 
detectors\,(RICH)~\cite{LHCb-DP-2012-003}. Photons,~electrons and hadrons are identified 
by a~calorimeter system consisting of scintillating\nobreakdash-pad 
and preshower detectors, 
an electromagnetic and 
a~hadronic calorimeter. Muons are~identified by a~system 
composed of alternating layers of iron and multiwire proportional chambers~\cite{LHCb-DP-2012-002}.

The online event selection is performed by a trigger~\cite{LHCb-DP-2012-004}, 
which consists of a hardware stage, based on information from the calorimeter and muon systems,
followed by a~software stage, which performs a full event reconstruction. 
The~hardware trigger selects muon candidates 
with high transverse momentum 
or dimuon candidates with a~high value of 
the~product
of the~individual muon $\pt$. 
In~the~software trigger, two 
oppositely-charged muons are required to form 
a~good\nobreakdash-quality
vertex that is significantly displaced from any~PV,
and to have a dimuon mass 
exceeding~$2.7\gevcc$.

Simulated events are used to describe the~signal  
and to~compute efficiencies needed to determine 
the~branching fraction ratios.
In~the~simulation, \proton\proton collisions are generated 
using \pythia~\cite{Sjostrand:2007gs}  with a~specific \lhcb configuration~\cite{LHCb-PROC-2010-056}. 
Decays of unstable particles are described by 
the~\evtgen 
package~\cite{Lange:2001uf}, 
in which final-state radiation is generated using \photos~\cite{davidson2015photos}. 
The~simulated decays of \mbox{\BcTojpsitripi} and \mbox{\BcTopsitwostripi} 
are produced via an intermediate $\Pa_1(1260)^+$ state, 
followed by $\decay{\aone}{\rhoz\pip}$ decay, 
using the~phenomenological model by
Berezhnoy, Likhoded, and Luchinsky ~\cite{Likhoded:2009ib,
Berezhnoy:2011nx,
Likhoded:2013iua} and referred to as~BLL model 
hereafter.
The~simulated decays of \mbox{\BcTojpsikpipi}, \mbox{\BcTojpsikkpi}, 	
 and \BcTopsitwoskkpi 
include intermediate \Kstarz and \Kstarzb states 
in the~\Kp\pim and \Km\pip systems, respectively.  
In~the~\mbox{\BcTojpsikkk} decay several excited $\PK^{*+}$ states are included, 
with fractions in accordance to the~observations described 
in Refs.~\cite{LHCb-PAPER-2016-018,
LHCb-PAPER-2016-019}. 
All~simulated $\decay{\Bc}{\Ppsi3\Ph^\pm}$ decays are further 
corrected to reproduce 
the~$\pip\pim$, $\Kp\pim$, $\Km\pip$, and~$\Kp\Km$~mass distributions 
observed in data. 
The~interaction of the~generated particles with the~detector,
and its response, are implemented using
the~\geant toolkit~\cite{Allison:2006ve, *Agostinelli:2002hh} 
as described in Ref.~\cite{LHCb-PROC-2011-006}.
To~account for imperfections in the~simulation of
charged\nobreakdash-particle reconstruction, 
the~track\nobreakdash-reconstruction efficiency
determined from simulation 
is corrected using calibration samples~\cite{LHCb-DP-2013-002}.

%% file: selection.tex
\section{Event selection}
\label{sec:Selection}
 
 Candidate \decay{\Bc}{\Ppsi3\Ph^\pm}~decays are reconstructed using 
dimuon decays of the~\jpsi and $\psitwos$~mesons.
The~criteria largely follow 
that described in Refs.~\cite{LHCb-PAPER-2013-047,
 LHCb-PAPER-2014-009}.
The~selection starts from 
 reconstructed charged tracks
of good quality
that are inconsistent 
with being produced in 
a~$\proton\proton$~interaction 
vertex. 
 Muon, pion and kaon candidates are identified 
 by combining information from the~RICH, 
 calorimeter and muon detectors~\cite{LHCb-PROC-2011-008}. 
The~muon candidates are  required to have a~transverse 
momentum larger than 550\mevc.
Pairs of oppositely\nobreakdash-charged 
muons consistent with originating from a~common vertex 
are combined to form \decay{\jpsi}{\mumu} and \decay{\psitwos}{\mumu} candidates. 
The~reconstructed mass of the~\mumu~pair is required to be
in the~range $3.0<m_{\mumu}<3.2~\gevcc$ and 
$3.60<m_{\mumu}<3.73~\gevcc$
for the~\jpsi and \psitwos~candidates, respectively. 
 The~kaon candidates  are  required to have a~transverse  
 momentum larger than 800\mevc for the~\BcTojpsikpipi~candidates 
 and 500\mevc for the other decay modes.
The~transverse momentum of pion candidates is required to be greater than 
500\mevc in the~\BcTojpsikpipi channel and 400\mevc in the~other decay modes.
For~efficient  particle identification,
kaons and pions are required to have a~momentum between 3.2\gevc~and~150\gevc. 
To~reduce combinatorial background, 
only tracks that are inconsistent with originating from any 
reconstructed PV in the~event are considered.

 To form the $\Bc$~candidates, 
 the selected $\Ppsi$~candidates 
 are combined with three charged tracks identified as 
 kaons or pions, 
 requiring a~good-quality reconstructed vertex.
Each~$\Bc$~candidate is associated with
 the~PV that yields the~smallest~$\chisqip$, 
 where \chisqip is defined as the~difference in the~vertex\nobreakdash-fit 
 \chisq of a~given PV 
 reconstructed with and without the~particle under consideration. 
 To~improve the~mass resolution for the~$\Bc$\nobreakdash~candidates,
 a~kinematic fit~\cite{Hulsbergen:2005pu}  is performed. 
 This~fit constrains the~mass of the~$\mumu$~pair 
 to the~known mass of the~$\jpsi$ or \psitwos  meson~\cite{PDG2021} 
 and  constrains the~\Bc~candidate 
 to originate from its associated PV.
 A requirement on
the quality of this fit is applied 
to further suppress combinatorial background.
Such requirement also reduces 
contributions from 
the~\Bc~decays with 
the~intermediate weakly\nobreakdash-decayed hadron, 
such as  
the~\mbox{$\decay{\Bc}{\jpsi(\decay{\Ds}{3\Ph^{\pm}})}$}, 
\mbox{$\decay{\Bc}{\jpsi(\decay{\Dz}{\Ph^+\Ph^-})\Kp}$}, 
or
\mbox{$\decay{\Bc}
{\big( \decay{\B^0_{(\squark)}}{\jpsi\Ph^+\Ph^-} 
\big)\Ph^+}$}~decays~\cite{LHCb-PAPER-2013-010,LHCb-PAPER-2013-044,LHCb-PAPER-2016-055}.

 The~measured decay time of 
 the~$\Bc$~candidate, 
 calculated with respect to the~associated PV, 
 is required to be greater than $175\mum/c$ for 
 the~\BcTojpsitripi, 
 \BcTopsitwostripi, and \BcTojpsikpipi~candidates, 
 and $125\mum/c$ for other decay modes. 
 This~requirement 
 suppresses 
 random combinations of candidates, 
 which include tracks
 originating from the~PV.
 The~mass of selected \Bc candidates 
 is required to be between 6.15~\gevcc~and~6.45~\gevcc.
 
 To~suppress $\Ppsi 3\Ph^{\pm}$~combinations  
 with 
 intermediate \Bs, \Bz, \Dsp, and \Dz~mesons,
 a~veto is applied on the mass of 
 the~corresponding 
 two- or three\nobreakdash-body  
 combinations, namely 
 $\Ppsi\Kp\Km$,
 $\Ppsi\kaon^{\mp}\pion^{\pm}$,  
 $\Kp\Km\pip$, and 
 $\Km\pip$.
 All~candidates having 
 any of~these masses 
 within a~range of approximately
 $\pm3\upsigma_m$ around 
 the~known masses of the~intermediate particles~\cite{PDG2021},
 where $\upsigma_m$
 stands for the~mass resolution, 
 are rejected. 
For~the~\mbox{$\decay{\Bc}{\jpsi\pip\pim\pip}$}~decay
the~contribution from the~\mbox{$\decay{\Bc}{ \left( \decay{\psitwos}{\jpsi\pip\pim}\right) \pip} $}~decays
is removed by rejecting  candidates with 
the~\mbox{$\jpsi\pip\pim$}~mass within
the~range~\mbox{$3.67<m_{\jpsi\pip\pim}<3.70\gevcc$}.
For~the~\BcTopsitwoskkpi decay
 the~mass of the~$\Km\pip$ system is required to be between 
 0.74\gevcc and 1.04\gevcc, consistent with originating from a \Kstarzb~meson.

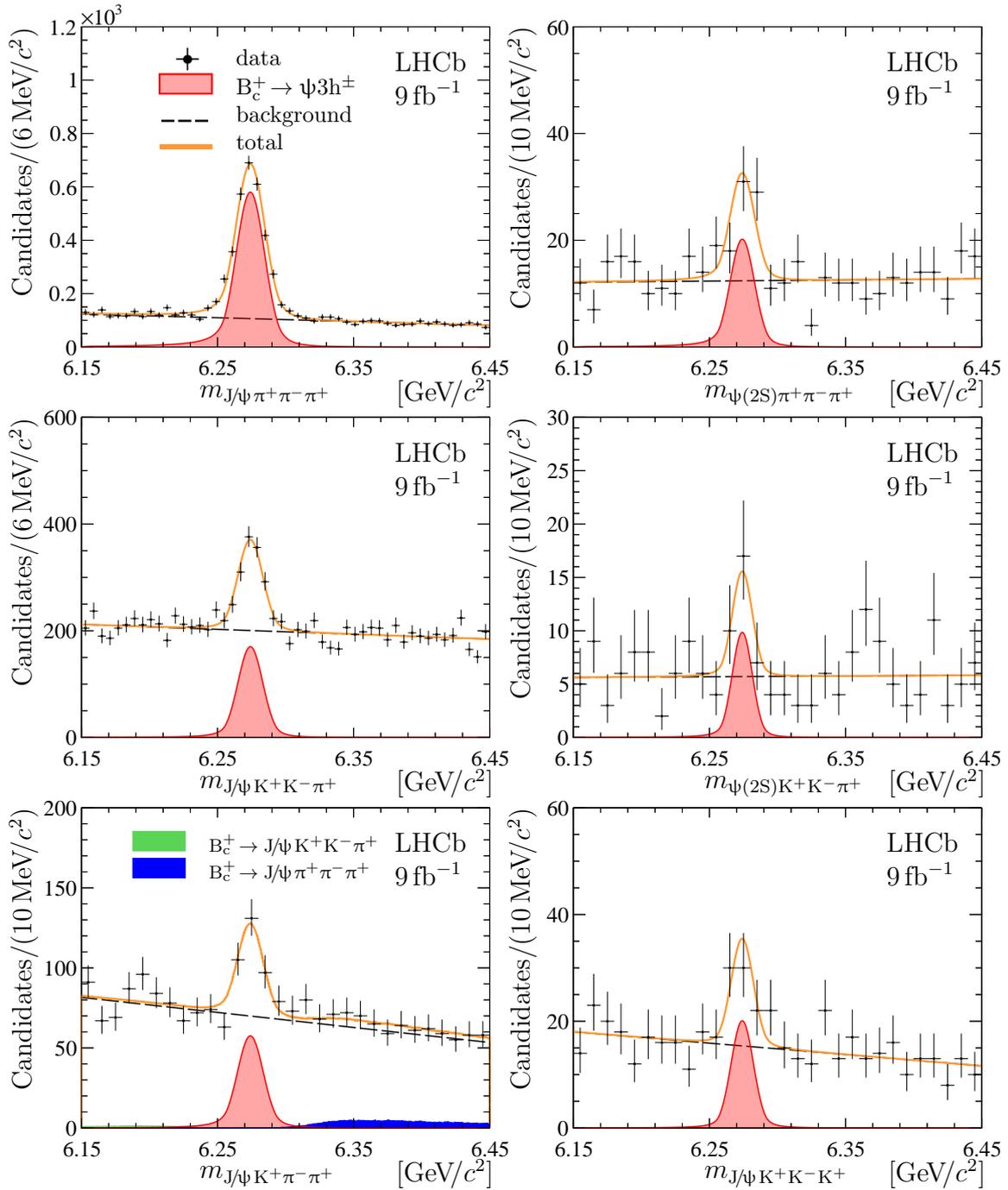
\begin{figure}[t]
	\setlength{\unitlength}{1mm}
	\centering
	\begin{picture}(150,175)
	\definecolor{root8}{rgb}{0.35, 0.83, 0.33}
        \put( 0,120) {\includegraphics*[width=75mm]{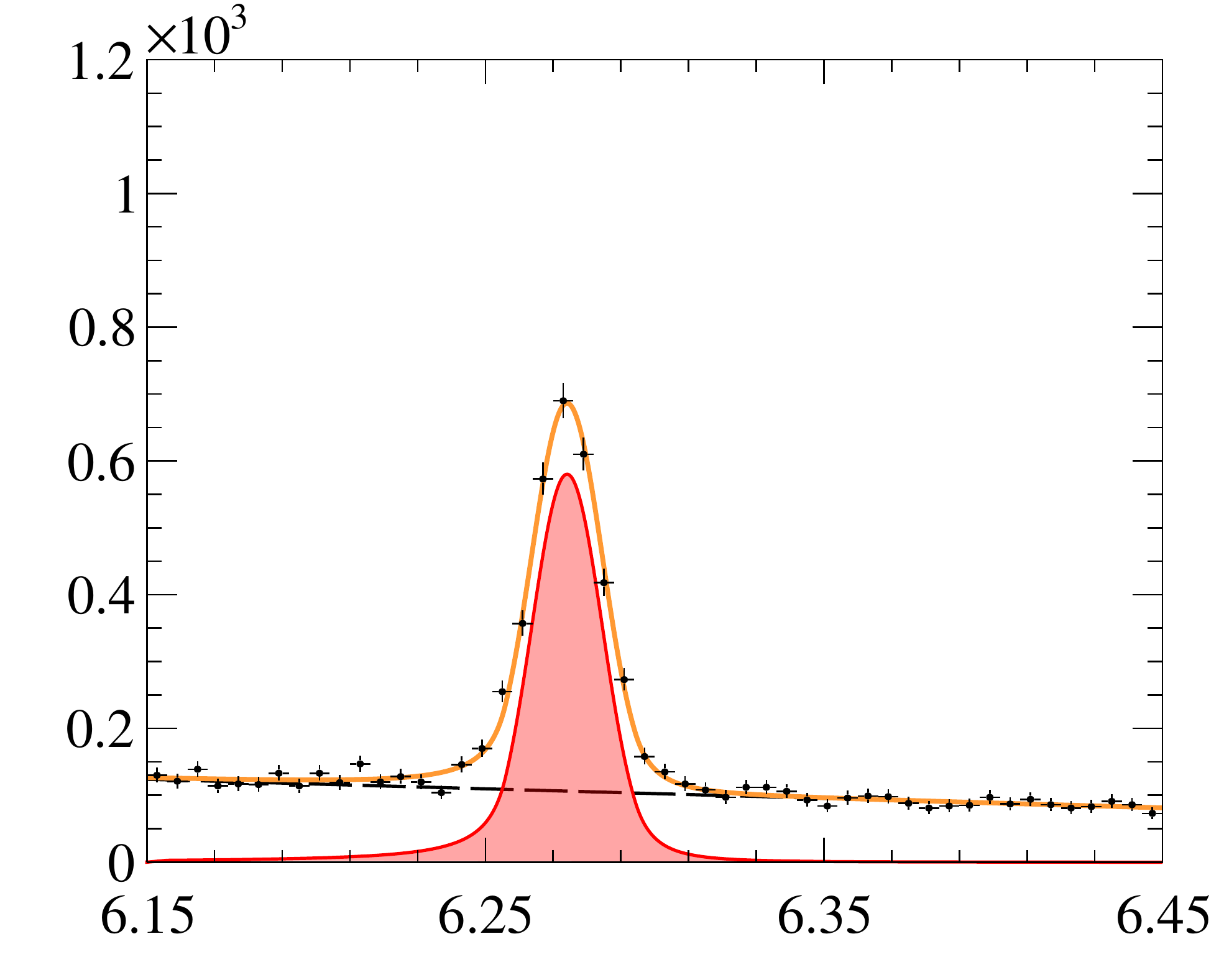}}
        \put(75,120) {\includegraphics*[width=75mm]{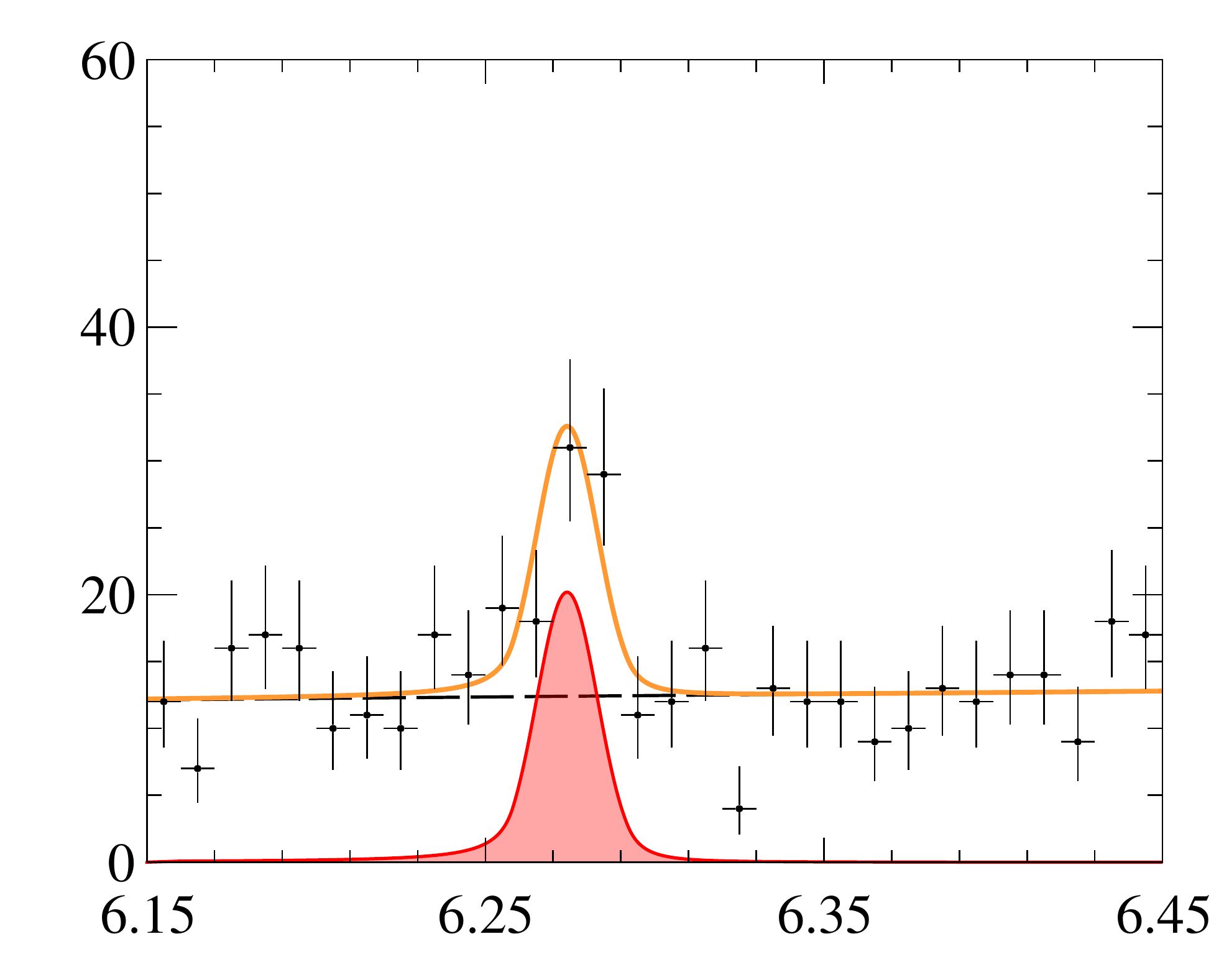}}
        \put( 0, 60) {\includegraphics*[width=75mm]{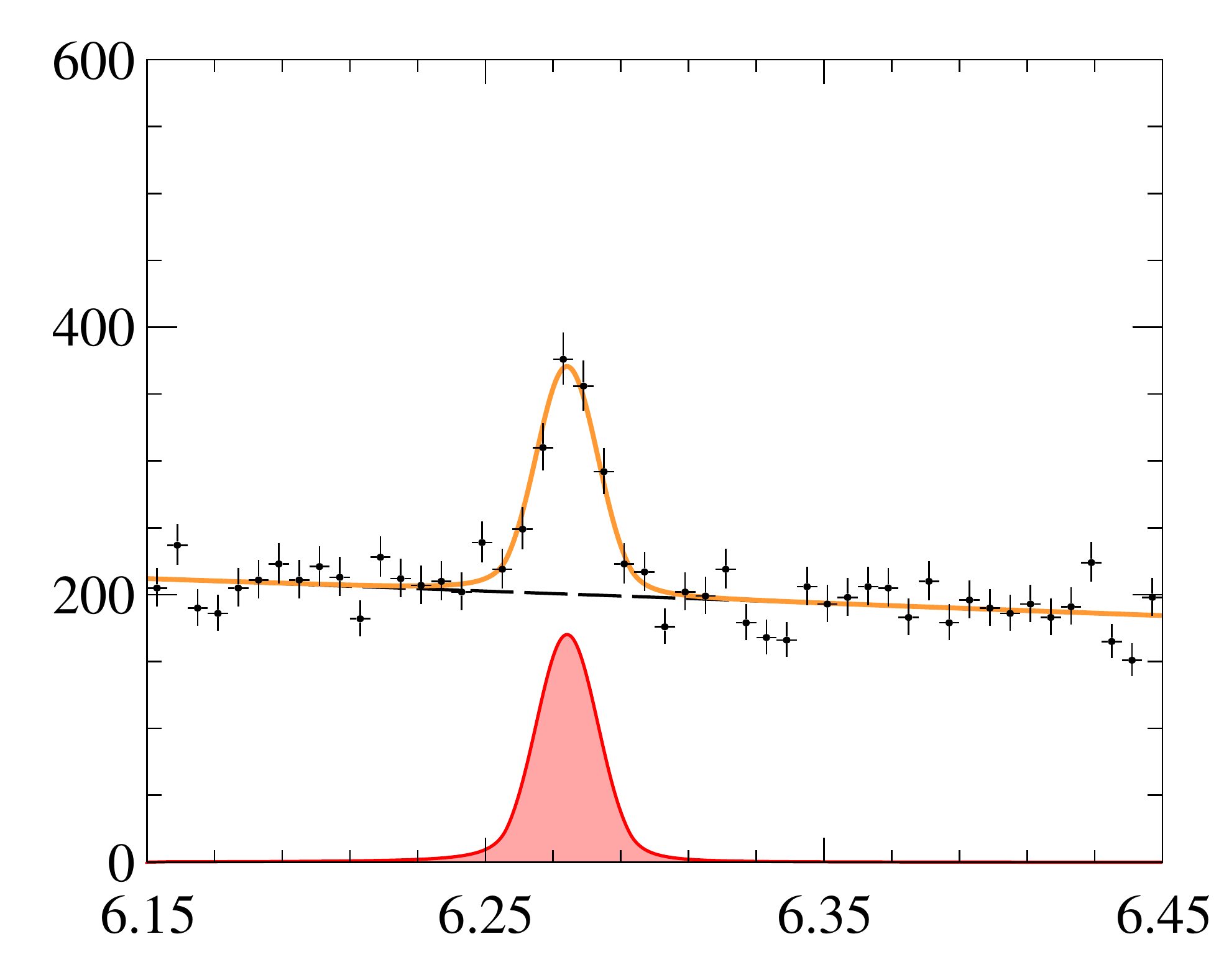}}
	\put(75, 60) {\includegraphics*[width=75mm]{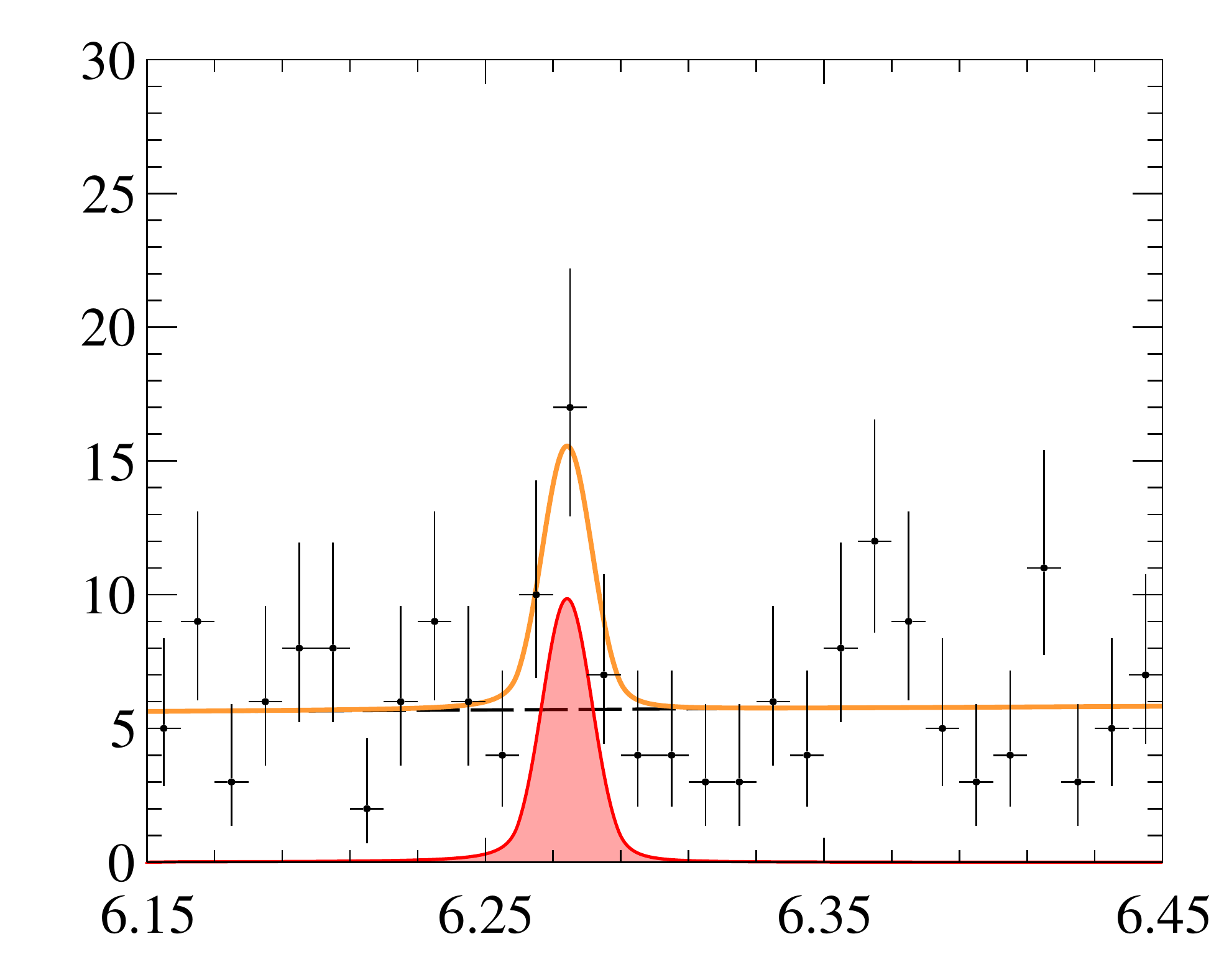}}
	\put( 0,  0) {\includegraphics*[width=75mm]{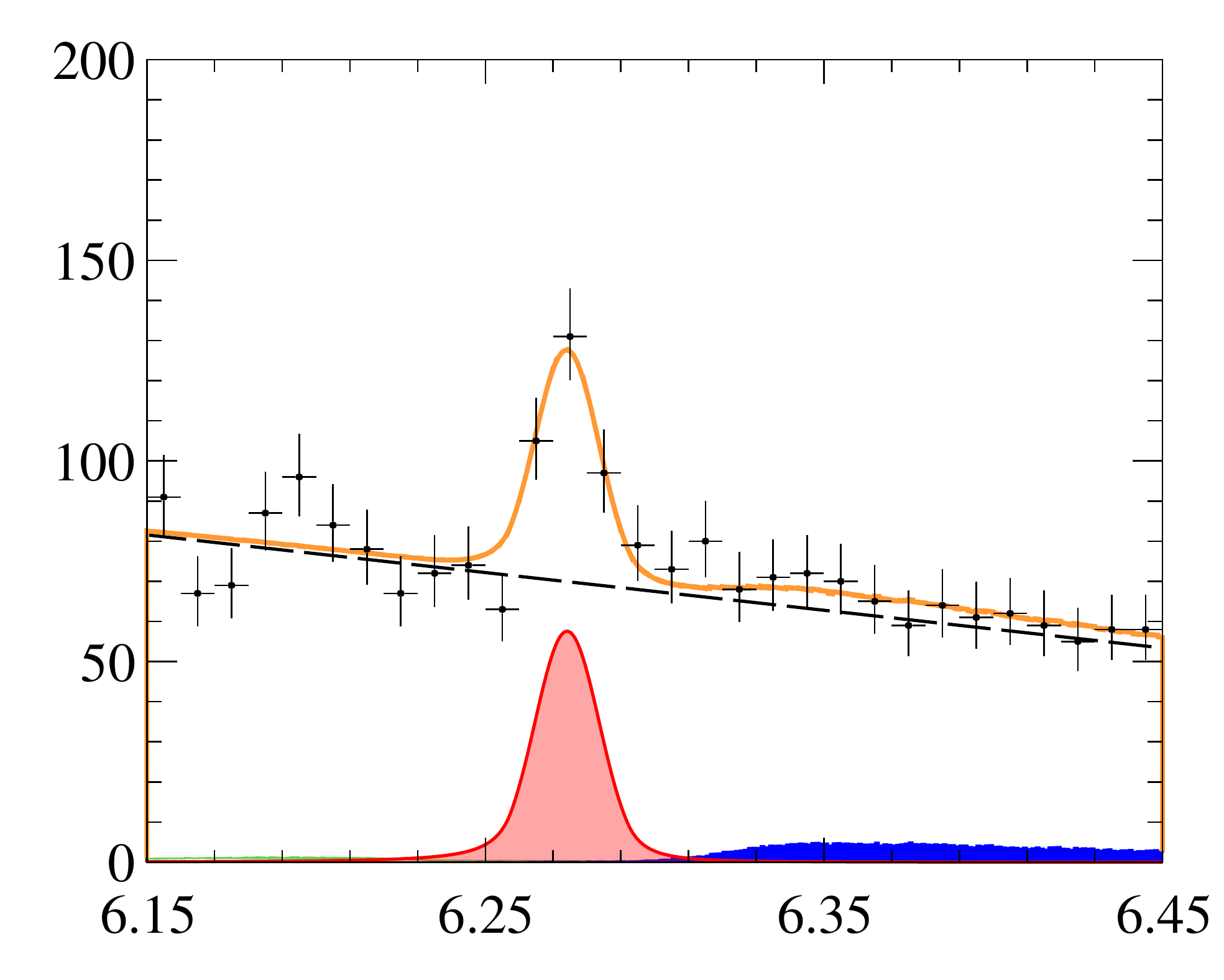}}
	\put(75,  0) {\includegraphics*[width=75mm]{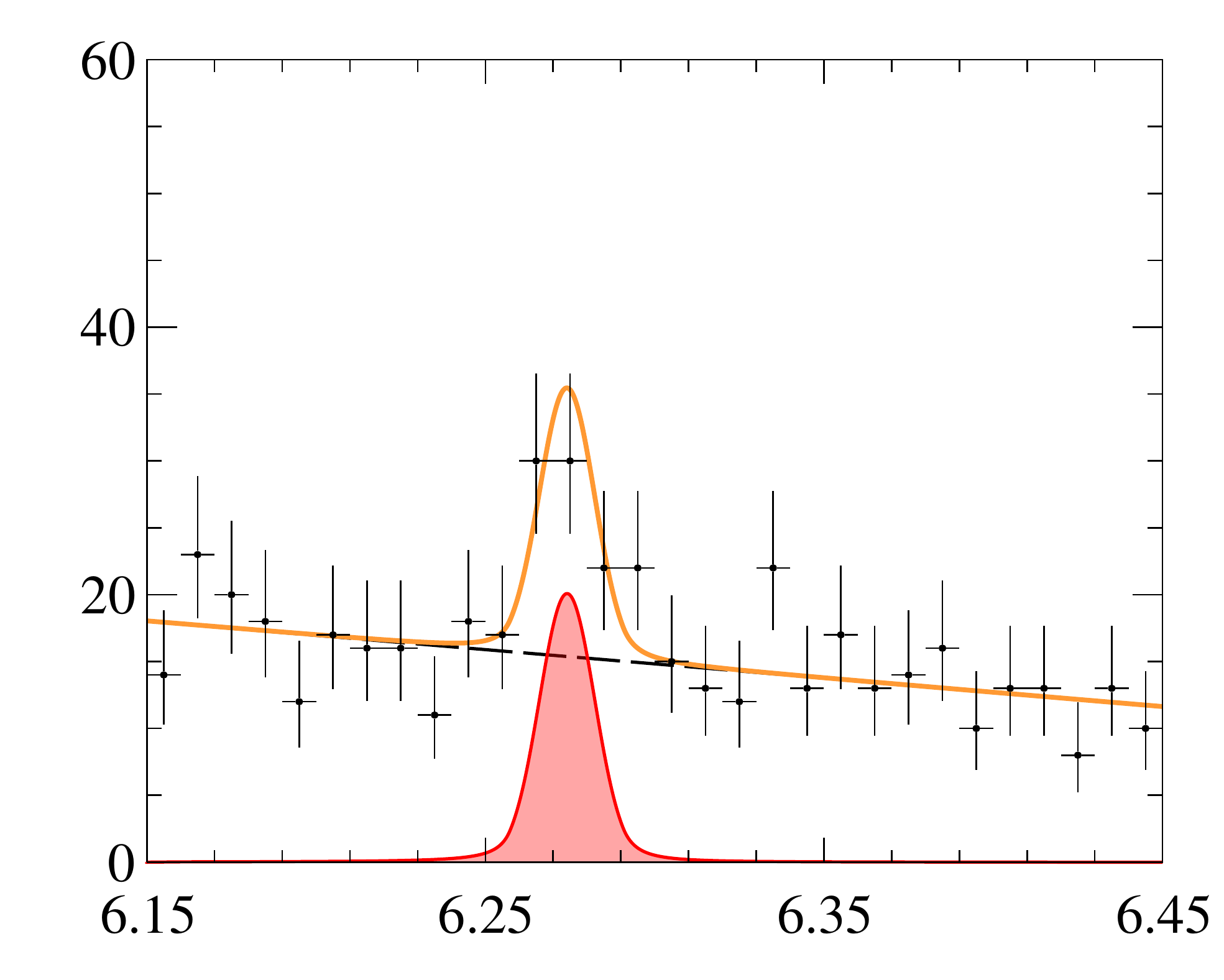}}

	\put( -2,135){\begin{sideways}{Candidates/$(6\mevcc)$}\end{sideways}}
	\put( -2, 75){\begin{sideways}{Candidates/$(6\mevcc)$}\end{sideways}}
	\put( -2, 13){\begin{sideways}{Candidates/$(10\mevcc)$}\end{sideways}}
	
	\put( 74,133){\begin{sideways}{Candidates/$(10\mevcc)$}\end{sideways}}
	\put( 74, 73){\begin{sideways}{Candidates/$(10\mevcc)$}\end{sideways}}
	\put( 74, 13){\begin{sideways}{Candidates/$(10\mevcc)$}\end{sideways}}
	
	\put( 27 ,119.5){$m_{\jpsi\pip\pim\pip}$}
	\put( 27 , 59.5){$m_{\jpsi\Kp\Km\pip}$}
	\put( 27 , -0.5){$m_{\jpsi\Kp\pim\pip}$}
	
	\put(104 ,119.5){$m_{\psitwos\pip\pim\pip}$}
	\put(104 , 59.5){$m_{\psitwos\Kp\Km\pip}$}
	\put(104 , -0.5){$m_{\jpsi\Kp\Km\Kp}$}
	
	\put( 57,119){$\left[\!\gevcc\right]$}
	\put( 57, 59){$\left[\!\gevcc\right]$}
	\put( 57, -1){$\left[\!\gevcc\right]$}
	
	\put( 132,119){$\left[\!\gevcc\right]$}
	\put( 132, 59){$\left[\!\gevcc\right]$}
	\put( 132, -1){$\left[\!\gevcc\right]$}

	\put( 55,167){$\begin{array}{l}\lhcb\\ 9\invfb \end{array}$}
	\put( 55,107){$\begin{array}{l}\lhcb\\ 9\invfb \end{array}$}
    \put( 55, 47){$\begin{array}{l}\lhcb\\ 9\invfb \end{array}$}
	\put(130,167){$\begin{array}{l}\lhcb\\ 9\invfb \end{array}$}
	\put(130,107){$\begin{array}{l}\lhcb\\ 9\invfb \end{array}$}
    \put(130, 47){$\begin{array}{l}\lhcb\\ 9\invfb \end{array}$}

	 \put(19,164){\footnotesize$\begin{array}{cl}
	 \!\bigplus\mkern-18mu\bullet & \mathrm{data} 
	 \\ 
	 \begin{tikzpicture}[x=1mm,y=1mm]\filldraw[fill=red!35!white,draw=red,thick]  (0,0) rectangle (8,3);\end{tikzpicture} & \decay{\Bc}{\Ppsi3\Ph^{\pm}} 
	 \\
	 {\color[RGB]{0,0,0}{\hdashrule[0.0ex][x]{8mm}{1.0pt}{2.0mm 0.3mm}}} & \mathrm{background}
	 \\
	 {\color[RGB]{255,153,51} {\rule{8mm}{2.0pt}}} & \mathrm{total}
	 \end{array}$}
	 
	 \put(15,48){\scriptsize$\begin{array}{cl}
	 \begin{tikzpicture}[x=1mm,y=1mm]\filldraw[thin,root8,color=root8,fill=root8]  (0,0) rectangle (8,3);\end{tikzpicture} & \decay{\Bc}{\jpsi\Kp\Km\pip} 
	 \\
	 \begin{tikzpicture}[x=1mm,y=1mm]\filldraw[thin,blue,fill=blue,pattern color=blue]  (0,0) rectangle (8,3);\end{tikzpicture} & \BcTojpsitripi 
	 \end{array}$}

	\end{picture}
	\caption {\small 
	Mass distributions 
	for selected \mbox{$\decay{\Bc}{\Ppsi 3\Ph^{\pm}}$}~candidates.
    Projections of a~fit, described in the~text, are overlaid.}
	\label{fig:signal_fit_1D}
\end{figure}

\begin{figure}[t]
	\setlength{\unitlength}{1mm}
	\centering
	\begin{picture}(150,60)
	\definecolor{root8}{rgb}{0.35, 0.83, 0.33}
	

	\put( 0,0) {\includegraphics*[width=75mm]{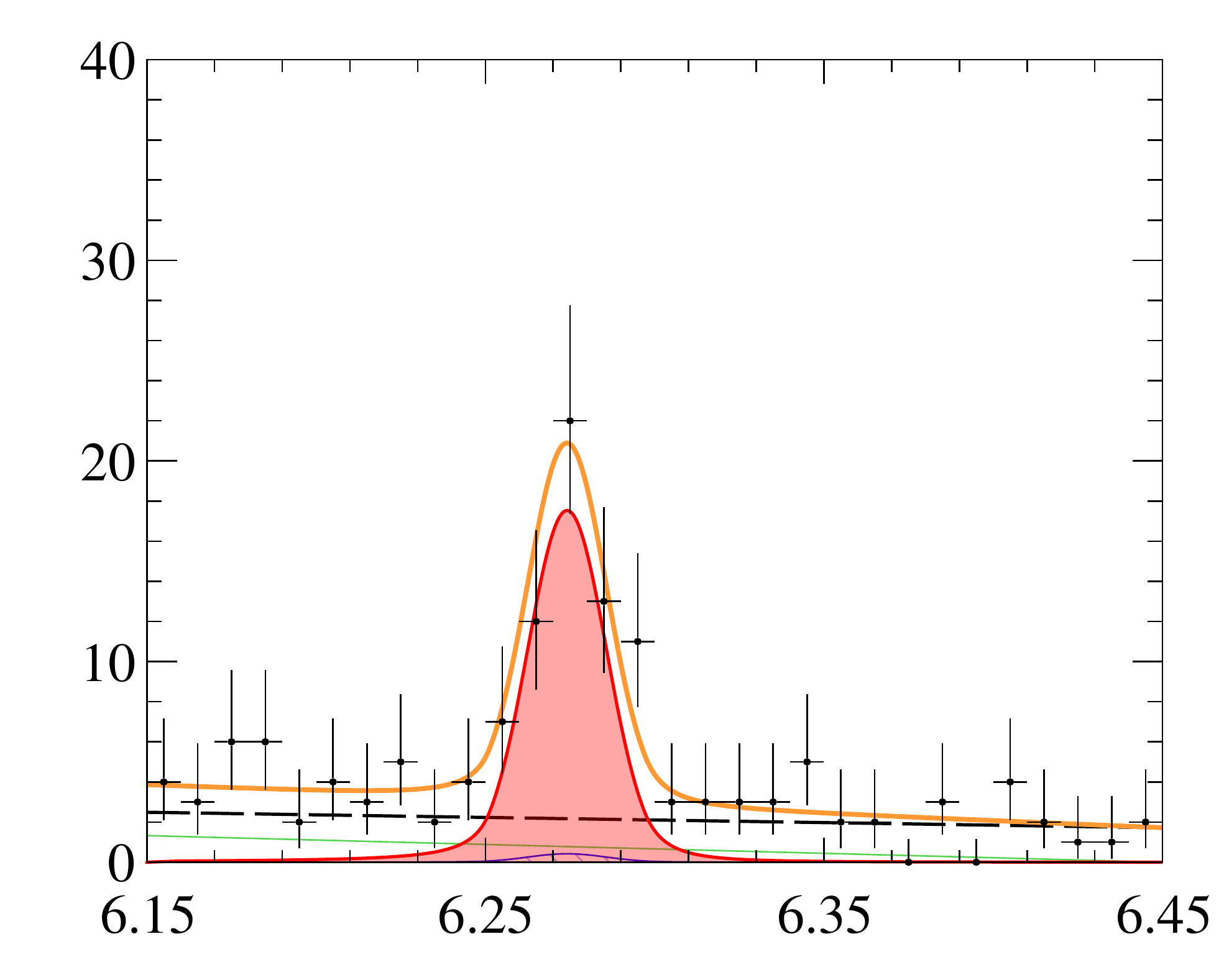}}
	\put(75,0) {\includegraphics*[width=75mm]{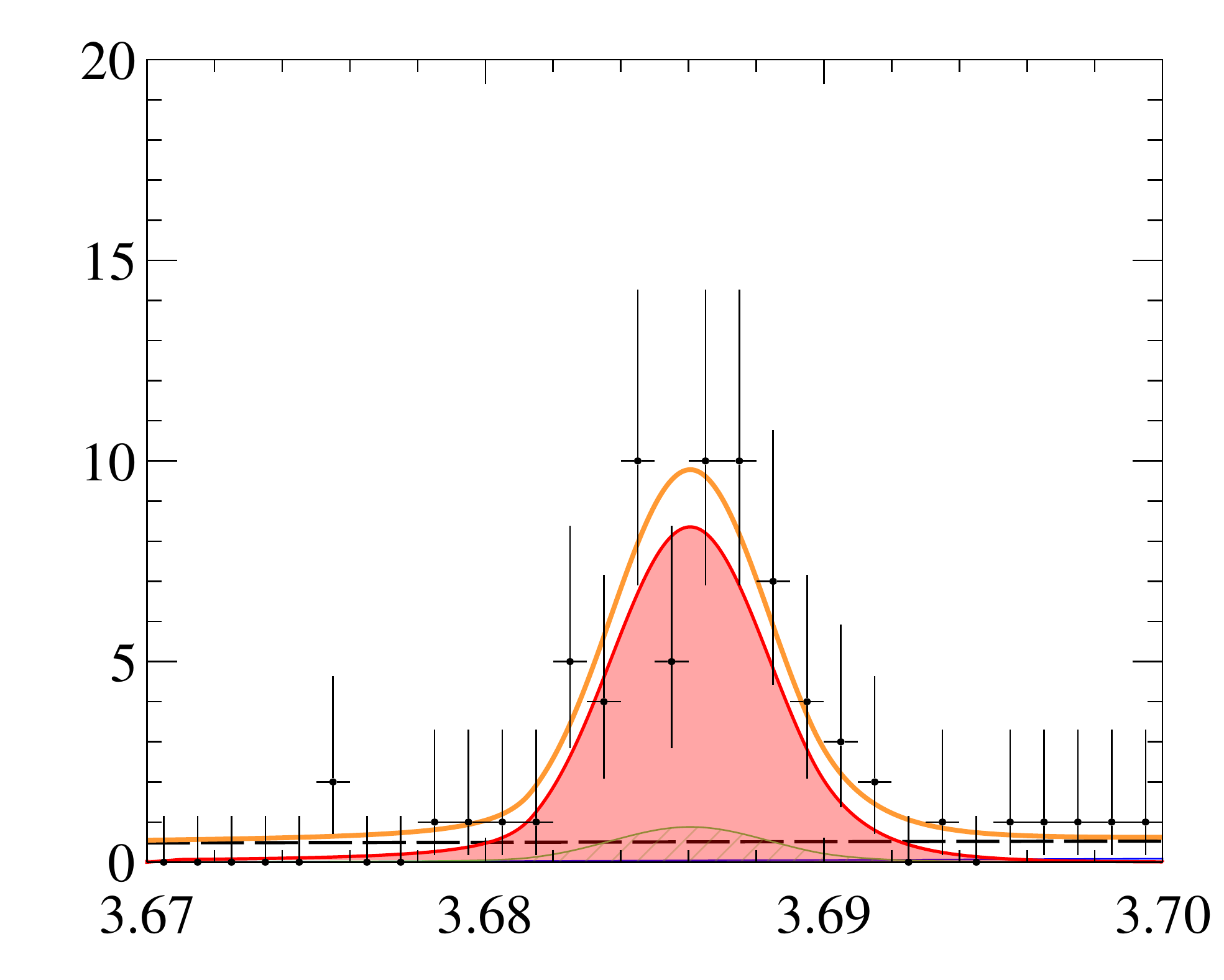}}
	\put(-1,14){\begin{sideways}{Candidates/$(10\mevcc)$}\end{sideways}}
	\put(74,16){\begin{sideways}{Candidates/$(1\mevcc)$}\end{sideways}}
	
	\put(27 ,-1){$m_{\jpsi\pip\pim\pip}$}
	\put(108 ,-1){$m_{\jpsi\pip\pim}$}
	\put( 58,-1.5){$\left[\!\gevcc\right]$}
	\put( 133,-1.5){$\left[\!\gevcc\right]$}

   \put( 12, 47){$\begin{array}{l}\lhcb\\ 9\invfb \end{array}$}
   \put( 87, 47){$\begin{array}{l}\lhcb\\ 9\invfb \end{array}$}

	 \put(35,41){\footnotesize$\begin{array}{cl}
	  \!\bigplus\mkern-18mu\bullet&\hspace{-1.5mm}\mathrm{data} 
	 \\ 
	  \begin{tikzpicture}[x=1mm,y=1mm]
	  \filldraw[fill=red!35!white,draw=red,thick]  (0,0) rectangle  (6,3);\end{tikzpicture} 
	   &\hspace{-1.5mm}\BcTopsitwospi
	   \\
	   \begin{tikzpicture}[x=1mm,y=1mm]
	   \draw[thin,blue,pattern=north west lines, pattern color=blue]  (0,0) rectangle (6,3);\end{tikzpicture} 
       &\hspace{-1.5mm} \decay{\Bc}{\jpsi\Ppi\Ppi\pip} 
       \\ 
	   \begin{tikzpicture}[x=1mm,y=1mm]\draw[thin,root8,pattern=north east lines, pattern color=root8]  (0,0) rectangle (6,3);\end{tikzpicture} 
 	    & \hspace{-1.5mm} \text{comb.}\ \psitwos\pip
 	   \\
 	   \color[RGB]{0,0,0}     {\hdashrule[0.0ex][x]{6mm}{1.0pt}{2.0mm 0.3mm} } 
	   &  \hspace{-1.5mm}\text{background}  
	   \\ 
      \color[RGB]{255,153,51} {\rule{6mm}{2.0pt}}
	   & \hspace{-1.5mm}\text{total} 
	 \end{array}$} 
	 
	\end{picture}
	\caption {\small 
	Distributions of 
	the (left)\,\jpsi\pip\pim\pip and
	(right)\,\jpsi\pip\pim masses 
    for selected 
    \mbox{$\decay{\Bc}{ \left(\decay{\psitwos}{\jpsi\pip\pim}\right)\pip}$}~candidates.
    Projections of a~fit, described in the~text, are overlaid.
    The~right plot corresponds to the~mass range
    $6.24<m_{\jpsi\pip\pim\pip}<6.31\gevcc$.}
	\label{fig:signal_fit_2D}
\end{figure}

Mass distributions for selected 
\mbox{$\decay{\Bc}{\jpsi\pip\pim\pip}$},
\mbox{$\decay{\Bc}{ \psitwos \pip\pim\pip}$},
\mbox{$\decay{\Bc}{\jpsi\Kp\Km\pip}$},
\mbox{$\decay{\Bc}{ \psitwos \Kp\Km\pip}$},
\mbox{$\decay{\Bc}{\jpsi\Kp\pim\pip}$}
and \mbox{$\decay{\Bc}{\jpsi\Kp\Km\Kp}$} candidates 
are shown in Fig.~\ref{fig:signal_fit_1D}.
Figure~\ref{fig:signal_fit_2D}\,(left) shows 
the mass distribution for selected 
\mbox{$\decay{\Bc}{\left(\decay{\psitwos}{\jpsi\pip\pim}\right)\pip}$} candidates,
while the~corresponding mass distribution for
\mbox{$\decay{\psitwos}{\jpsi\pip\pim}$}~candidates 
is shown in Fig.~\ref{fig:signal_fit_2D}\,(right).

%% file: sig_eff.tex
\section{Signal yields}
\label{sec:Sig_eff}

The yields for the~\mbox{\decay{\Bc}{\Ppsi3\Ph^\pm}}~decays 
are determined 
using a~simultaneous
extended unbinned maximum\nobreakdash-likelihood fit
to the six mass distributions of selected 
\mbox{\BcTojpsitripi}, 
\BcTopsitwostripi, 
\BcTojpsikpipi,  
\BcTojpsikkpi, 
\BcTopsitwoskkpi
and 
\BcTojpsikkk~candidates; 
and a~two\nobreakdash-dimensional distribution
of the~\Tojpsitripi  mass, $m_{\jpsi\pip\pim\pip}$, 
versus 
$\jpsi\pip\pim$~mass, $m_{\jpsi\pip\pim}$, for
the~\mbox{\decay{\Bc}{(\decay{\psitwos}{\JpsiPiPi})\pip}}~candidates. 
To~improve the~resolution on 
the~$\jpsi\pip\pim$~mass
for the~\mbox{\decay{\Bc}{(\decay{\psitwos}{\JpsiPiPi})\pip}}~candidates 
and to eliminate a~small correlation between 
$m_{\jpsi\pip\pim\pip}$ and $m_{\JpsiPiPi}$,
following Refs.~\cite{LHCb-PAPER-2020-009,LHCb-PAPER-2020-035}
the~$m_{\JpsiPiPi}$ variable is computed~\cite{Hulsbergen:2005pu} 
by constraining the~mass of the~\Bc~candidate
to its known value~\cite{PDG2021}.
For~each \Bc~mass distribution the~one-dimensional 
fit function consists of two components: 
\begin{enumerate}
	\item signal $\decay{\Bc}{\Ppsi3\Ph^\pm}$ decays
	parameterised by a~modified Gaussian function 
	with power\nobreakdash-law tails on both sides of the~distribution~\cite{LHCb-PAPER-2011-013,Skwarnicki:1986xj}.
	The~tail parameters are fixed to 
    the~values obtained from simulation; 
    \item random $\Ppsi3\Ph^{\pm}$ combinations, 
    modelled by a~first\nobreakdash-order polynomial function.
\end{enumerate}
The two-dimensional fit function for 
the~\mbox{\decay{\Bc}{(\decay{\psitwos}{\JpsiPiPi})\pip}} channel 
is defined as a~sum of four components: 
\begin{enumerate}
	\item signal  \decay{\Bc}{(\decay{\psitwos}{\JpsiPiPi})\pip} decays
	parameterised as a~product of  \Bc~and $\psitwos$~signal functions 
	modelled  by a~modified Gaussian function  with power\nobreakdash-law tails on both sides of the~distribution~\cite{LHCb-PAPER-2011-013,Skwarnicki:1986xj}. 
	The~tail parameters are fixed to 
    the~values obtained from simulation;
	\item 
	contributions 
	from the~decays \mbox{\decay{\Bc}{\left(\jpsi\pip\pim\right)_{\mathrm{NR}}\pip}}
	without proceeding through 
	a~narrow intermediate $\psitwos$~state,
	parameterised as a~product of the~\Bc~signal function and 
	 a~linear function of $m_{\jpsi\pip\pim}$; 
	\item random combinations of  
	$\psitwos$ and \pip~candidates,
     parameterised as a~product of the~$\psitwos$~signal 
     template as obtained from simulation 
     and a~linear function of $m_{\jpsi\pip\pim\pip}$, 
     multiplied 
     by the~two\nobreakdash-body 
     phase\nobreakdash-space function~\cite{Byckling};
	\item random $\jpsi\pip\pim\pip$ combinations, 
	described by a~two\nobreakdash-dimensional 
	positive\nobreakdash-definite  
    first\nobreakdash-order polynomial function.
\end{enumerate}
For the~Cabibbo\nobreakdash-suppressed channel 
\BcTojpsikpipi, components
describing feed\nobreakdash-down contributions 
from the~Cabibbo\nobreakdash-favoured 
\BcTojpsikkpi and \mbox{\BcTojpsitripi} decays, 
where the~kaon is misidentified as a pion or vice versa,
are added into the~fit function.
The~shapes for these 
contributions are taken from simulation,
and their yields are 
constrained 
to the~expected number of misidentified events.

For all \Bc~signal~functions, 
the~peak\nobreakdash-position parameter 
is shared between all decays and allowed to vary in the~fit. 
The~mass-resolution parameters used in the~\Bc and $\psitwos$~signal 
functions are fixed to the~values determined from simulation, 
and corrected by 
scale factors,
$s_{\Bc}$ and $s_{\psitwos}$, 
to account for a~discrepancy  
in the~mass resolution  
between data and simulation~\cite{LHCb-PAPER-2020-008,
LHCb-PAPER-2020-009,
LHCb-PAPER-2020-035}.
These~factors are allowed to vary in the~fit,  
and the~factor $s_{\Bc}$ is shared for all decay modes.
The~factor $s_{\psitwos}$ and 
the~peak\nobreakdash-position parameter 
for the~\psitwos~signal component  
are constrained to the~values
from a~previous \lhcb study~\cite{LHCb-PAPER-2020-009}.
The~mass distributions 
together with projections of the~fit are shown 
in Fig.\,\ref{fig:signal_fit_1D} for
\mbox{\decay{\Bc}{\jpsi\pip\pim\pip}},
\mbox{\decay{\Bc}{\psitwos\pip\pim\pip}},
\mbox{\decay{\Bc}{\jpsi\Kp\Km\pip}},
\mbox{\decay{\Bc}{\psitwos\Kp\Km\pip}},
\mbox{\decay{\Bc}{\jpsi\Kp\pim\pip}}  and
\mbox{\decay{\Bc}{\jpsi\Kp\Km\Kp}}~candidates
and Fig.~\ref{fig:signal_fit_2D} for 
the~\mbox{$\decay{\Bc}{\left( \decay{\psitwos}{\jpsi\pip\pim}\right)\pip}$}~candidates.
The fit parameters of interest with statistical significance of the~observed signals 
are summarised in Table~\ref{tab:sim_fit_res}.
The~resolution correction factors 
are found to be $s_{\Bc} =  1.096 \pm 0.029$ and $s_{\psitwos} = 1.048 \pm 0.004$.

The~statistical significance for previously
unobserved decay modes is estimated 
with a~large number of pseudoexperiments 
produced according to the~background distribution 
observed in data.
These results amount  to 
the~first observation of   
the~decays 
\mbox{$\BcTopsitwostripi$}, 
\mbox{$\BcTojpsikpipi$} and 
\mbox{$\BcTojpsikkk$}~decays, 
and  the~first evidence for 
the~\mbox{\BcTopsitwoskkpi}~decay.
The~decay \mbox{$\decay{\Bc}{\psitwos\pip}$}
is confirmed using 
the~\mbox{$\decay{\psitwos}{\jpsi\pip\pim}$}~mode.

\begin{table}[t]
	\centering
	\caption{\small 
	Parameters of interest from the~simultaneous  
	unbinned extended 
	maximum-likelihood fit.
	The uncertainties are statistical only.
	For previously unobserved decay modes,  
	the~last column shows the~statistical significance 
	estimated using pseudoexperiments 
	in units of standard deviations. 
	}
	\begin{tabular}{lr@{$\,\pm\,$}lc}
	Decay & \multicolumn{2}{c}{Yield}  &  $\mathcal{S}~\left[\upsigma\right]$ 
    \\[1.5mm]
    \hline     \\[-3mm]
  \BcTojpsitripi
  &  2750 & 69  &   
  \\
   \BcTojpsikkpi
   &  686 & 48 &  
   \\
   \BcTojpsikkk
   &  43 & 10  &  \phantom{0}5.2 
   \\
 \BcTojpsikpipi
   &  148 & 22 &  \phantom{0}7.8 
   \\
  \BcTopsitwostripi
   &  49 & 11 &   \phantom{0}5.8 
   \\ 
    \BcTopsitwoskkpi
   &  19 & 6   &  \phantom{0}3.7 
 \\
    \decay{\Bc}{(\decay{\psitwos}{\JpsiPiPi})\pip}
   &  54 & 9  &   11.8  
     \\[1.5mm]
     \hline 
     \\[-3mm]
     Parameter &  \multicolumn{2}{c}{Value} & 
 	\\[1.5mm]
     \hline     \\[-3mm]
     $m_{\Bc}\,\phantom{00}[\mevcc]$   
     &  6274.14 & 0.26 & 
     \\
     $m_{\psitwos}~[\mevcc]$   
     & 3686.05& 0.01 & 
	\end{tabular}
	\label{tab:sim_fit_res}
\end{table}

 \section{Resonance structure}
 
\begin{figure}[t]
  \setlength{\unitlength}{1mm}
  \centering
  \begin{picture}(150,60)
	\put( 2,0) {\includegraphics*[width=75mm]{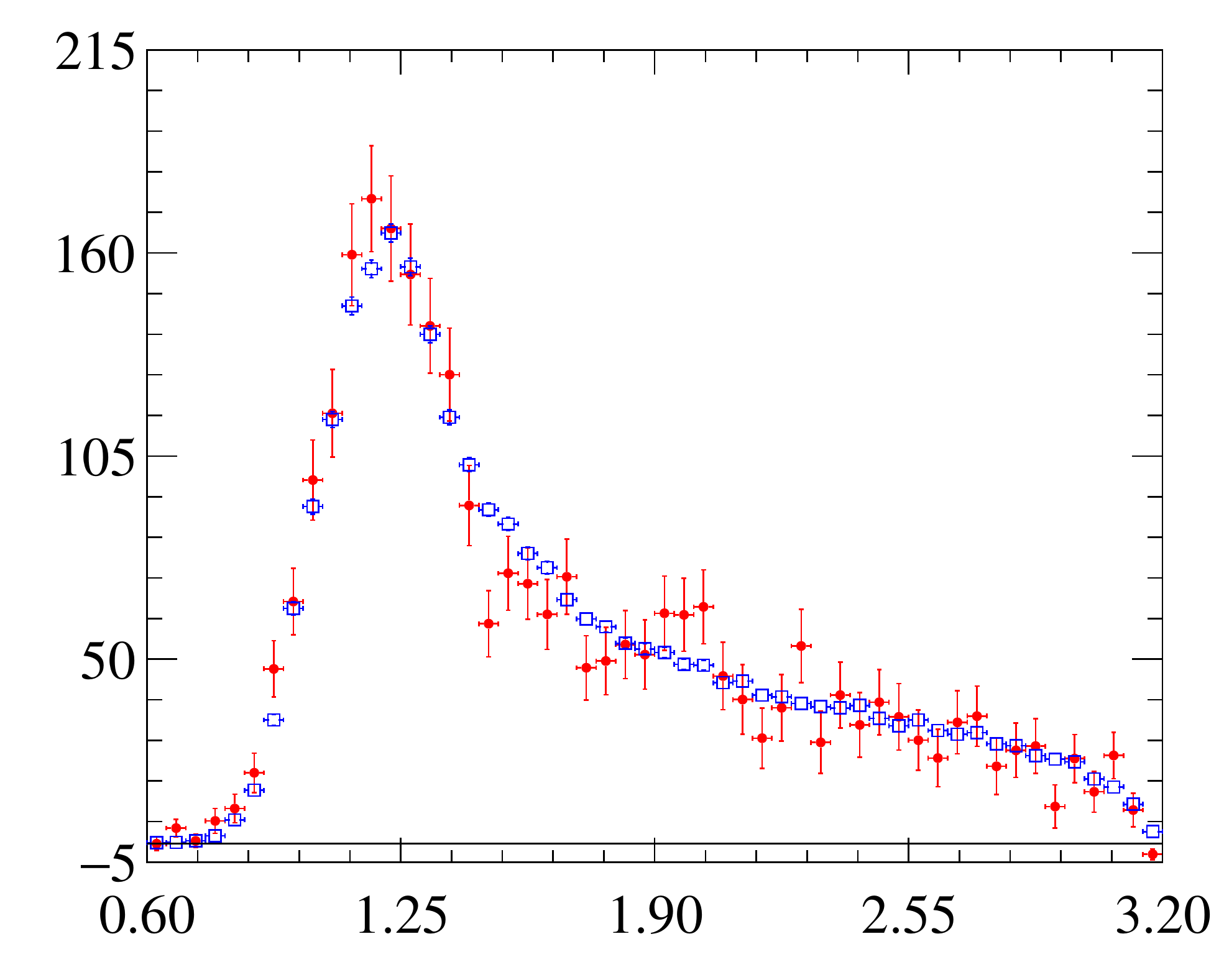}}
	\put(77,0) {\includegraphics*[width=75mm]{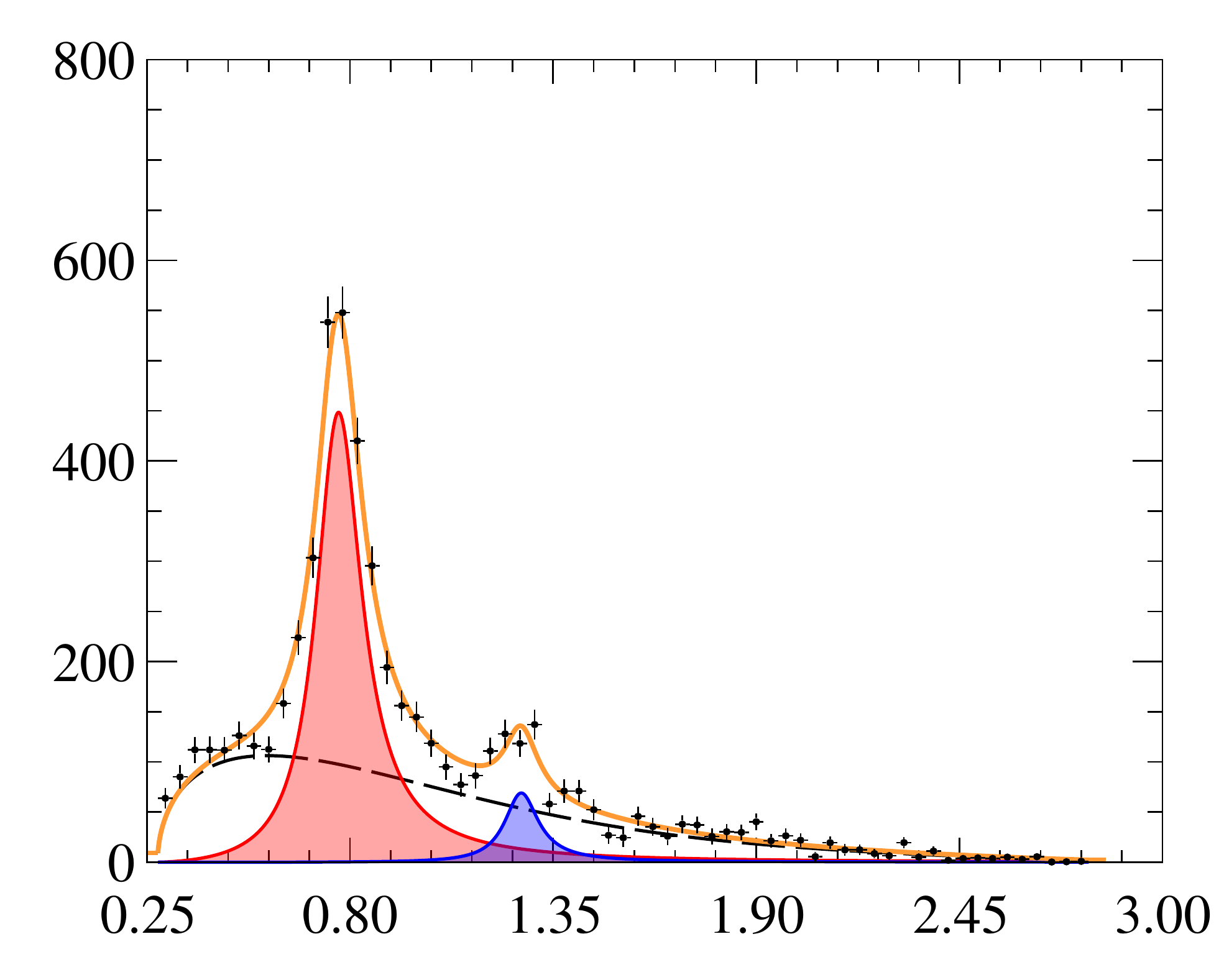}}
	\put(-1,25) {\begin{sideways}Yield/$(50\mevcc)$\end{sideways}}
	\put(75,25) {\begin{sideways}Yield/$(40\mevcc)$\end{sideways}}
	
	\put(35 ,-1){$m_{\pip\pim\pip}$}
	\put(110 ,-1){$m_{\pip\pim}$}
	\put( 59,-1){$\left[\!\gevcc\right]$}
	\put(135,-1){$\left[\!\gevcc\right]$}

	\put( 58,49){$\begin{array}{l} \lhcb \\ 9\invfb \end{array}$} 
	\put(133,49){$\begin{array}{l} \lhcb \\ 9\invfb \end{array}$} 

	\put(35,40){\footnotesize$\begin{array}{cl}
	 \color{red}{\bigplus\mkern-18mu\bullet} & \mathrm{data} 
	 \\
	 \,\color{blue}{\square}   & \mathrm{simulation\,(BLL)}
	\end{array}$}

	\put(102,35){\footnotesize$\begin{array}{cl}
	 \bigplus\mkern-18mu\bullet & \mathrm{data} 
	\\
	\begin{tikzpicture}[x=1mm,y=1mm]\filldraw[fill=red!35!white,draw=red,thick]  (0,0) rectangle (6,3);\end{tikzpicture} 
	& \decay{\Bc}{\jpsi\rhoz\pip}
	\\ 
	\begin{tikzpicture}[x=1mm,y=1mm]    \filldraw[fill=blue!35!white,draw=blue,thick] (0,0) rectangle (6,3);\end{tikzpicture} 
	& \decay{\Bc}{\jpsi\PR \pip}
	\\
	\color[RGB]{0,0,0}     {\hdashrule[0.0ex][x]{6mm}{1.0pt}{2.0mm 0.3mm} }
	& \BcTojpsitripi
	\\
	\color[RGB]{255,153,51} {\rule{6mm}{2.0pt}}
	&  \mathrm{total}
	\end{array}$} 
	
	\end{picture}
	
	\caption {\small 
	Background-subtracted 
	(left)~$\pip\pim\pip$ and 
	(right)~$\pip\pim$ mass distributions
	for selected $\decay{\Bc}{\jpsi\pip\pim\pip}$~candidates.
	In~the~left plot the expectation from 
	simulation for the~\mbox{$\decay{\Bc}{\jpsi\aone}$}~decays 
	with BLL model~\cite{Likhoded:2009ib,Berezhnoy:2011nx,Likhoded:2013iua} is overlaid. 
     }
	\label{fig:three_pions}
\end{figure}

 The $\pip\pim\pip$ and $\pip\pim$~mass distributions 
 from the~$\BcTojpsitripi$~decays  were previously
  studied in Ref.~\cite{LHCb-PAPER-2011-044} and were shown to be 
 compatible with originating from 
 a~$\decay{\Bc}{\jpsi\Pa_1(1260)^+}$, followed by~an~$\decay{\Pa_1(1260)^{+}}
 {\rhoz\pip}$~decay~\cite{Likhoded:2009ib,Berezhnoy:2011nx}.
 The~background\nobreakdash-subtracted 
 $\pip\pim\pip$ and $\pip\pim$ distribution
 for the~$\BcTojpsitripi$~candidates are shown in Fig.~\ref{fig:three_pions}, 
 where the~\sPlot technique is used for background subtraction~\cite{Pivk:2004ty}, 
 using the~$\Tojpsitripi$~mass as a~discriminating variable. 

\begin{table}[t]
	\centering
	\caption{\small
	Results for parameters 
	from the~fit to the~background\protect\nobreakdash-subtracted 
	$\pip\pim$~mass spectrum from the~\mbox{\BcTojpsitripi}~decays.
	The~last row shows the~statistical 
	significance for the~\PR~structure 
	estimated using Wilks' theorem~\cite{Wilks:1938dza}. 
	} 
	\label{tab:f_pi3}
	\vspace{2mm}
	\begin{tabular*}{0.45\textwidth}
	{@{\hspace{3mm}}l@{\extracolsep{\fill}}lc@{\hspace{3mm}}}
	\multicolumn{2}{l}{Parameter}  
	&       Value
   \\[1.5mm]
  \hline 
  \\[-1.5mm]
  $f^{\decay{\Bc}{\jpsi\pip\pim\pip}}_{\rhoz}$   
  & $\left[\%\right]$          &   $\phantom{0}88.1\pm3.0$
  \\
  $f^{\decay{\Bc}{\jpsi\pip\pim\pip}}_{\PR}$       
  & $\left[\%\right]$          &   $\phantom{0}10.4\pm1.4$
  \\
  $m_{\PR}$       & $\left[\!\mevcc\right]$    &    $\phantom{.}1265\pm10\phantom{.}$
  \\
  $\Gamma_{\PR}$  & $\left[\!\mev\right]$      &    $\phantom{0.}110\pm21\phantom{.}$ 
  \\[1.5mm]
  \hline 
  \\[-1.5mm]
  $\mathcal{S}_{\PR}$   &   $\left[\upsigma\right]$  & $\phantom{000000}8\phantom{\pm1.4}$
  \end{tabular*}
  \vspace{3mm}
\end{table}

The~$\pip\pim\pip$~mass distribution agrees well with 
the~BLL model expectations~\mbox{\cite{Likhoded:2009ib,
Berezhnoy:2011nx,Likhoded:2013iua}}.
The~$\pip\pim$~mass distribution exhibits 
a~clear peak from the~$\rhoz$~resonance and 
a~structure near $m_{\pip\pim}\sim1.3\gevcc$, 
that can be due to 
contributions from 
the~decays of 
the~wide $\Pa_1(1260)^+$~resonance  
via 
$\Pf_2(1270)$, 
$\Pf_0(1370)$ or 
$\Prho(1450)$~mesons~\cite{CLEO:1999rzk}, 
jointly referred to as \PR in the~following.
Unbinned maximum\nobreakdash-likelihood fits 
to the~$\pip\pim$~mass distribution are performed 
with functions that contain  three terms:
a~component 
corresponding to the~decay via the~$\rhoz$~resonance; 
a~component corresponding  to the~decays
via S-, P-, or D\nobreakdash-wave 
$\pip\pim$~resonances, 
and a~component corresponding to 
\Bc~meson decays into the~$\Tojpsitripi$~final state
without resonances in the~$\pip\pim$~system. 
The~resonance components 
are parameterised with 
relativistic P- and S\nobreakdash-wave 
Breit\nobreakdash--Wigner functions
with Blatt\nobreakdash--Weisskopf form factors 
with a~meson radius of 
$3.5\gev^{-1}$~\cite{Blatt:1952ije}.\footnote{Blatt--Weisskopf 
form factors with meson radius of $3.5\gev^{-1}$ are used 
for all subsequent fits with 
relativistic Breit\nobreakdash--Wigner functions, 
unless stated otherwise.}
The~non\nobreakdash-resonant component 
is parameterised with a~product of 
the~phase-space function describing
a~two\nobreakdash-body system 
out of the~four\nobreakdash-body final state~\cite{Byckling}
and a~positive second\nobreakdash-order polynomial function,
that accounts
for the~decay dynamics 
via the~intermediate $\Pa_1(1260)^+$~state. 
The~\rhoz~peak position and
the~width are constrained to their~known values~\cite{PDG2021} 
using Gaussian constraints, 
while parameters for the~\PR~structure are free to vary in the~fit.

The~fractions
$f^{\decay{\Bc}{\jpsi\pip\pim\pip}}_{\rhoz}$
and $f^{\decay{\Bc}{\jpsi\pip\pim\pip}}_{\PR}$ 
of the~\Bc~meson decays into 
the~$\Tojpsitripi$~final state
via the~intermediate \rhoz and 
\PR~resonances,  
as well as 
the~Breit\nobreakdash--Wigner
mass and width for the~R~structure,  
$m_{\PR}$ and $\Gamma_{\PR}$, 
are shown in Table~\ref{tab:f_pi3}.
The~results for fractions,
mass and width of the~resonance
are stable with respect to the~choice
of the~orbital momentum used for 
the~Breit--Wigner function:
$f_{\rhoz}$ and $f_{\PR}$
change by 0.002, 
$m_{\PR}$ and 
$\Gamma_{\PR}$ change
by respectively 2\mevcc and 3\mev, 
when the~orbital momentum varies from 
S\nobreakdash-wave to D\nobreakdash-wave. 
The~statistical significance 
for the~structure is estimated using
Wilks' theorem~\cite{Wilks:1938dza}
and is found to be 8~standard deviations.  
The~obtained Breit\nobreakdash--Wigner 
mass and width of the~structure 
are consistent with those 
for the~$\Pf_0(1370)$~state~\cite{PDG2021}. 
The~yield relative to the~yield of decays 
via the~$\rhoz$~resonance, $\left(11.8\pm1.6\right)\%$,  
agrees with that obtained 
by the~CLEO collaboration 
from a~Dalitz analysis 
of the~\mbox{$\decay{\Pa_1(1260)^+}{\pip\piz\piz}$}~decay~\cite{CLEO:1999rzk},
and is much larger than those for 
the~$\Pf_2(1270)$ and $\Prho(1450)$~states.
It~allows interpretation of  
the~\PR~structure as 
the~$\Pf_0(1370)$~resonance, 
however alternative interpretations
such as the~$\Pf_2(1270)$ or $\Prho(1450)$~state 
are also possible.

\begin{figure}[t]
  \setlength{\unitlength}{1mm}
  \centering
  \begin{picture}(150,60)
    \put( 2,0) {\includegraphics*[width=75mm]{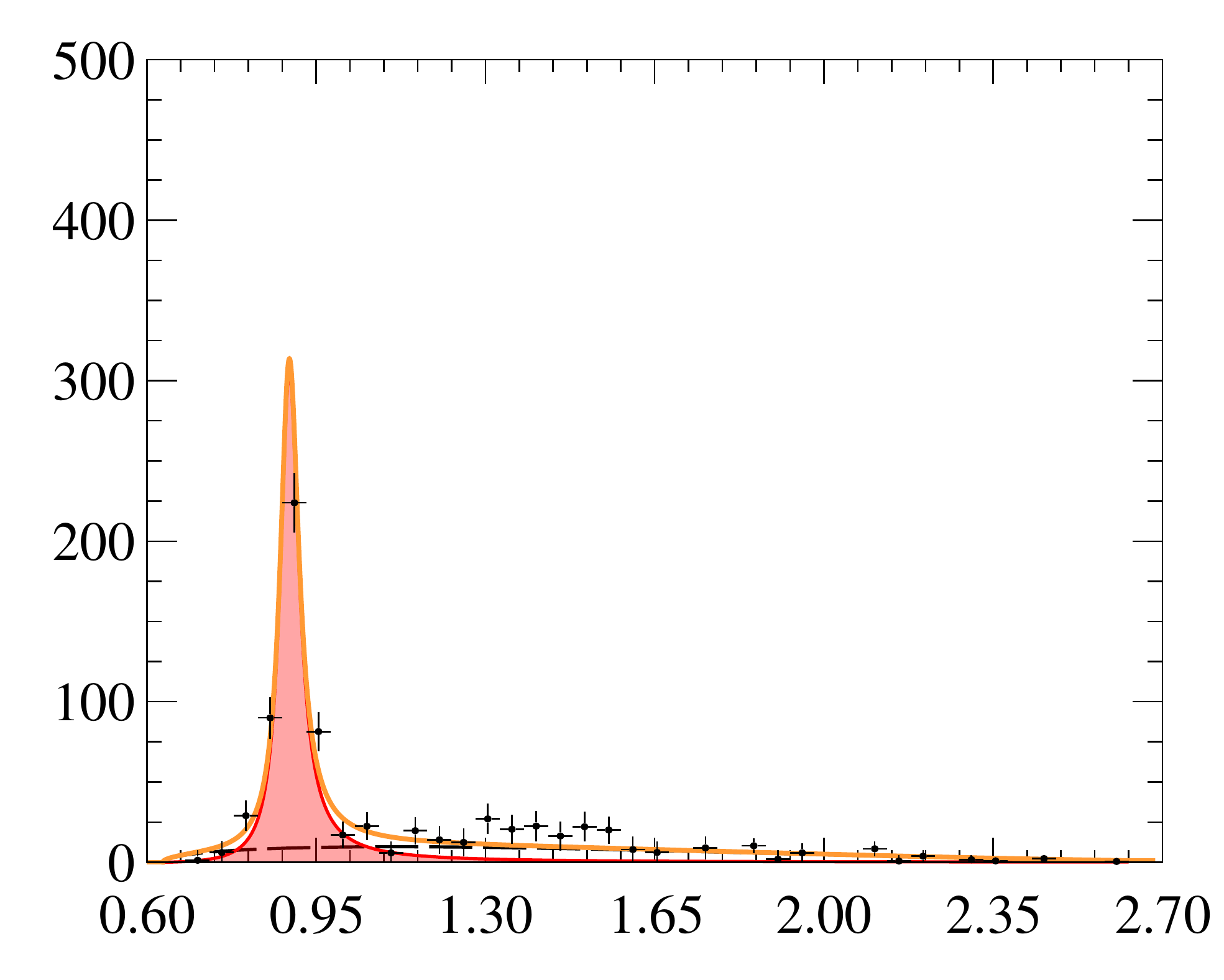}}
	\put(77,0) {\includegraphics*[width=75mm]{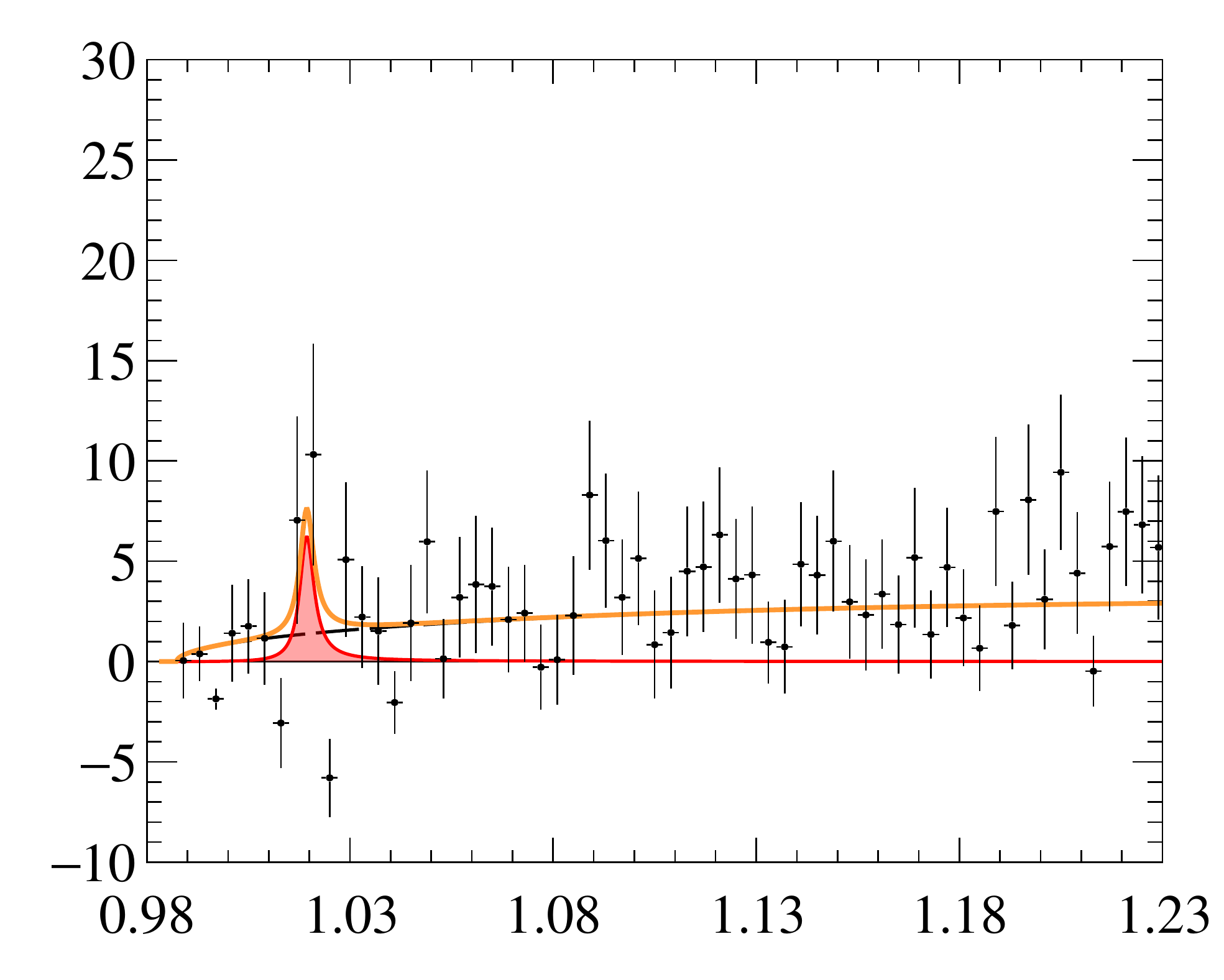}}
	\put(-1,25){\begin{sideways}Yield/$(50\mevcc)$\end{sideways}}
	\put(76,27){\begin{sideways}Yield/$(4\mevcc)$\end{sideways}}
	
	\put(35 ,-1){$m_{\Km\pip}$}
	\put(115 ,-1){$m_{\Kp\Km}$}
	\put( 59,-1){$\left[\!\gevcc\right]$}
	\put(135,-1){$\left[\!\gevcc\right]$}

	\put( 58,49){$\begin{array}{l} \lhcb \\ 9\invfb \end{array}$} 
	\put(133,49){$\begin{array}{l} \lhcb \\ 9\invfb \end{array}$} 


	\put(20,42){\footnotesize$\begin{array}{cl}
	 \bigplus\mkern-18mu\bullet & \mathrm{data} 
	\\
	\begin{tikzpicture}[x=1mm,y=1mm]\filldraw[fill=red!35!white,draw=red,thick]  (0,0) rectangle (6,3);\end{tikzpicture} 
	& \decay{\Bc}{\jpsi\Kp\Kstarzb}
	\\ 
	\color[RGB]{0,0,0}     {\hdashrule[0.0ex][x]{6mm}{1.0pt}{2.0mm 0.3mm} }
	& \decay{\Bc}{\jpsi\Kp\Km\pip}
	\\
	\color[RGB]{255,153,51} {\rule{6mm}{2.0pt}}
	&  \mathrm{total}
	\end{array}$} 

	\put(95,42){\footnotesize$\begin{array}{cl}
	 \bigplus\mkern-18mu\bullet & \mathrm{data} 
	\\
	\begin{tikzpicture}[x=1mm,y=1mm]\filldraw[fill=red!35!white,draw=red,thick]  (0,0) rectangle (6,3);\end{tikzpicture} 
	& \decay{\Bc}{\jpsi\Pphi\pip}
	\\ 
	\color[RGB]{0,0,0}     {\hdashrule[0.0ex][x]{6mm}{1.0pt}{2.0mm 0.3mm} }
	& \decay{\Bc}{\jpsi\Kp\Km\pip}
	\\
	\color[RGB]{255,153,51} {\rule{6mm}{2.0pt}}
	&  \mathrm{total}
	\end{array}$}

	\end{picture}

	\caption {\small 
	Background-subtracted 
	(left)~$\Km\pip$ and 
	(right)~$\Kp\Km$ mass distributions
	for selected $\decay{\Bc}{\jpsi\Kp\Km\pip}$~candidates.
	The~$\Kp\Km$~mass spectrum is fitted
	in the~full accessible \Kp\Km~mass region, \mbox{$m_{\Kp\Km}<2.923\gevcc$}. 
	For better visibility, only 
	a~low\protect\nobreakdash-mass part of the~spectrum is shown.
     }
	\label{fig:two_kaons}
\end{figure}

In Ref.~\cite{LHCb-PAPER-2013-047} it has been demonstrated that 
a~large fraction of the~decays
of the~\Bc~mesons into the~\mbox{$\jpsi\Kp\Km\pip$}~final state  
proceeds via an~intermediate $\Kstarzb$~meson,
while no evidence for 
Okubo\nobreakdash--Zweig\nobreakdash--Iizuka\nobreakdash-suppressed\,(OZI)
decays \mbox{\cite{Okubo:1963fa, Zweig2,Iizuka:1966fk}}
via intermediate $\Pphi$~mesons is found. 
Figure~\ref{fig:two_kaons} shows 
the~background-subtracted 
$\Km\pip$ and $\Kp\Km$~mass distributions 
for the~selected $\decay{\Bc}{\jpsi\Kp\Km\pip}$~candidates.
Unbinned maximum-likelihood fits
to these distributions are performed 
with functions consisting of two terms:
a~component corresponding to 
decays via intermediate 
\mbox{$\decay{\Kstarzb}{\Km\pip}$} or 
\mbox{$\decay{\Pphi}{\Kp\Km}$} decays
and a~component 
without resonances in $\Km\pip$
or $\Kp\Km$~systems.
The~former is parameterised with 
a~relativistic P\nobreakdash-wave  
Breit\nobreakdash--Wigner function,
while
the~latter is  parameterised with 
a~phase-space function 
describing a two-body system 
in a~four\nobreakdash-body final state~\cite{Byckling}.
Masses and widths of the~$\Kstarzb$ and $\Pphi$~resonances 
are allowed to vary in the~fit and 
are Gaussian constrained 
to their known values~\cite{PDG2021}.
Results of the~fits are overlaid 
in Fig.~\ref{fig:two_kaons}.  
The~fractions of the~\Bc~meson decays into 
the~$\jpsi\Kp\Km\pip$~final state
via intermediate $\Kstarzb$ and 
$\Pphi$~resonances are found to be 
\begin{eqnarray*}
   f_{\Kstarzb}^{\decay{\Bc}{\jpsi\Kp\Km\pip}} 
   & = &  \left( 64.5 \pm 4.7 \right)\,\%        \,, \\ 
   f_{\Pphi}^{\decay{\Bc}{\jpsi\Kp\Km\pip}}     
   & = &  \phantom{0}\left(1.6\,{}^{\,+\,0.7}_{\,-\,0.6}\right)\,\% \,,
\end{eqnarray*}
respectively, confirming the previous observations~\cite{LHCb-PAPER-2013-047}.
The~upper limit at 90\,(95)\% confidence level\,(CL) on the~fraction 
$f_{\Pphi}^{\decay{\Bc}{\jpsi\Kp\Km\pip}}$
is set as 
\begin{equation*}
    f_{\Pphi}^{\decay{\Bc}{\jpsi\Kp\Km\pip}}  < 3.9\,(4.5)\%\,.
\end{equation*}

\begin{figure}[t]
  \setlength{\unitlength}{1mm}
  \centering
  \begin{picture}(150,65)
    \put( 2,0) {\includegraphics*[width=75mm]{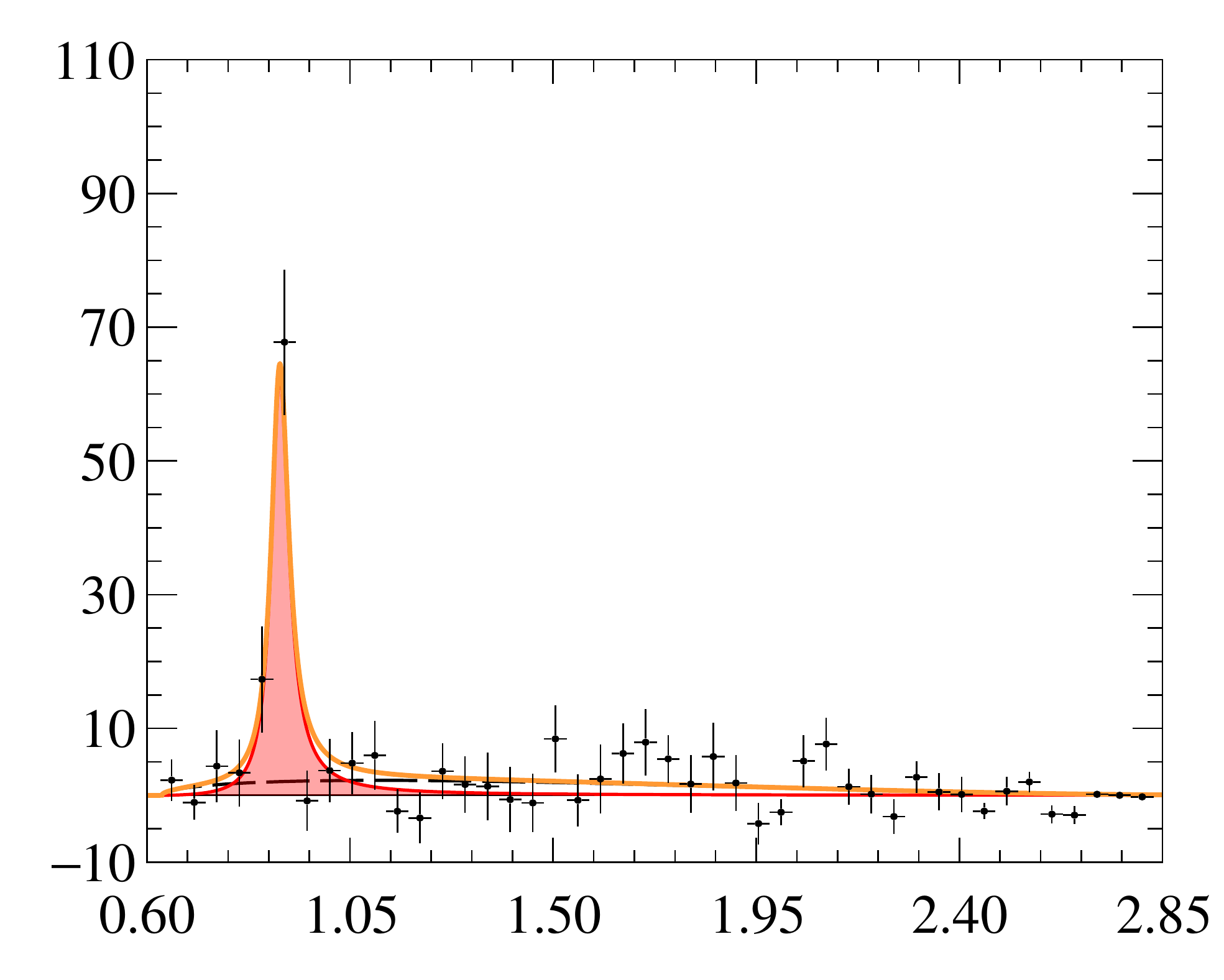}}
	\put(77,0) {\includegraphics*[width=75mm]{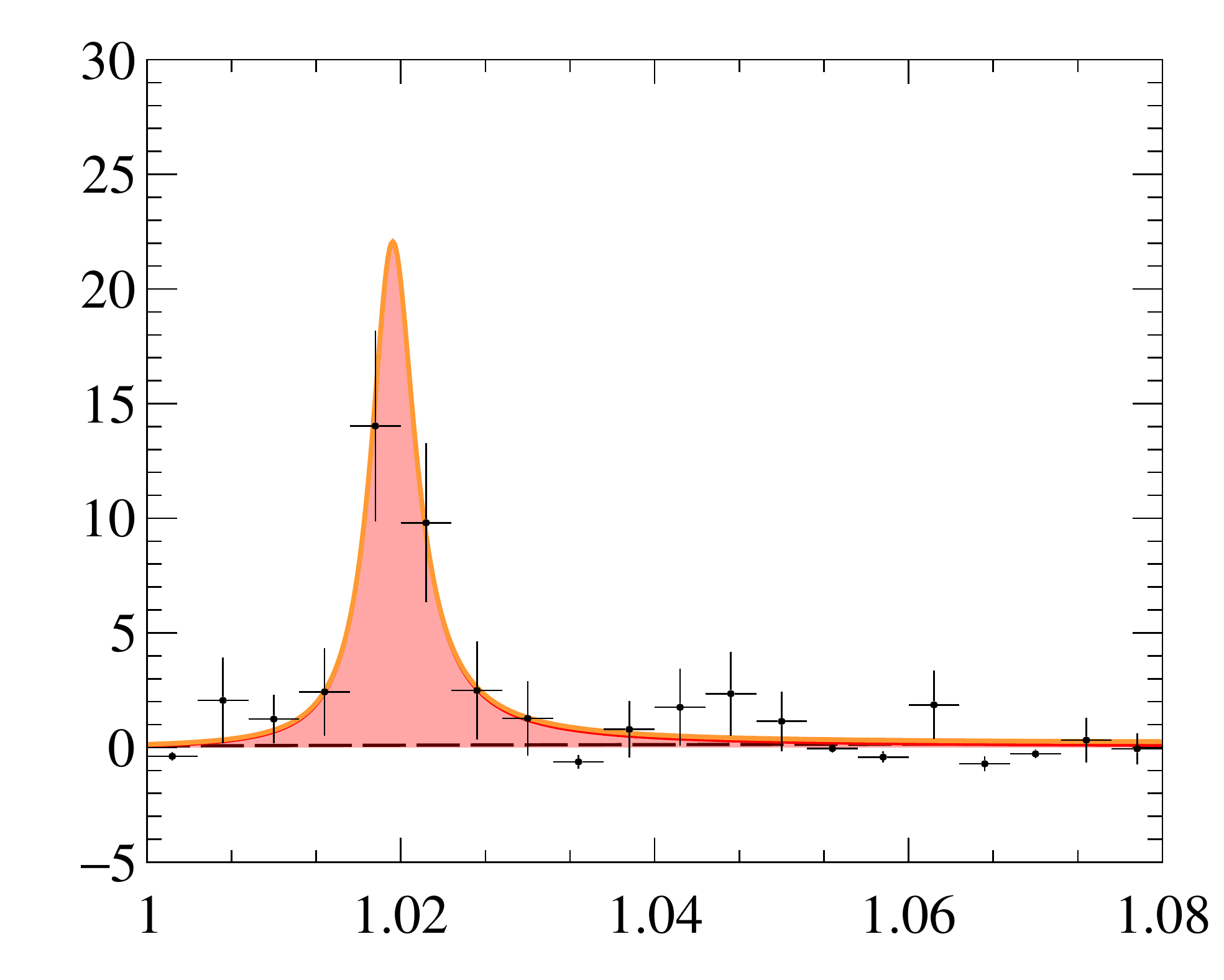}}
	\put(-1,25){\begin{sideways}Yield/$(50\mevcc)$\end{sideways}}
	\put(76,27){\begin{sideways}Yield/$(4\mevcc)$\end{sideways}}
	
	\put(35 ,-1){$m_{\Kp\pim}$}
	\put(115 ,-1){$m_{\Kp\Km}$}
	\put( 59,-1){$\left[\!\gevcc\right]$}
	\put(135,-1){$\left[\!\gevcc\right]$}

	\put( 58,49){$\begin{array}{l} \lhcb \\ 9\invfb \end{array}$} 
	\put(133,49){$\begin{array}{l} \lhcb \\ 9\invfb \end{array}$} 


	\put(25,35){\footnotesize$\begin{array}{cl}
	 \bigplus\mkern-18mu\bullet & \mathrm{data} 
	\\
	\begin{tikzpicture}[x=1mm,y=1mm]\filldraw[fill=red!35!white,draw=red,thick]  (0,0) rectangle (6,3);\end{tikzpicture} 
	& \decay{\Bc}{\jpsi\Kstarz\pip}
	\\ 
	\color[RGB]{0,0,0}     {\hdashrule[0.0ex][x]{6mm}{1.0pt}{2.0mm 0.3mm} }
	& \decay{\Bc}{\jpsi\Kp\pim\pip}
	\\
	\color[RGB]{255,153,51} {\rule{6mm}{2.0pt}}
	&  \mathrm{total}
	\end{array}$} 

	\put(104,35){\footnotesize$\begin{array}{cl}
	 \bigplus\mkern-18mu\bullet & \mathrm{data} 
	\\
	\begin{tikzpicture}[x=1mm,y=1mm]\filldraw[fill=red!35!white,draw=red,thick]  (0,0) rectangle (6,3);\end{tikzpicture} 
	& \decay{\Bc}{\jpsi\Pphi\Kp}
	\\ 
	\color[RGB]{0,0,0}     {\hdashrule[0.0ex][x]{6mm}{1.0pt}{2.0mm 0.3mm} }
	& \BcTojpsikkk
	\\
	\color[RGB]{255,153,51} {\rule{6mm}{2.0pt}}
	&  \mathrm{total}
	\end{array}$} 

	\end{picture}

	\caption {\small 
	Background-subtracted mass distributions of 
	(left)~$\Kp\pim$~pairs
	for the~selected \mbox{\BcTojpsikpipi}~candidates
     and 	
	(right)~$\Kp\Km$~combinations  
	for the~selected~\mbox{\BcTojpsikkk}~candidates.
	The~$\Kp\Km$~mass spectrum is fitted
	in the~full accessible \Kp\Km~mass region, \mbox{$m_{\Kp\Km}<2.2\gevcc$}. 
	For better visibility, only 
	a~low\protect\nobreakdash-mass part of the~spectrum is shown.
     }
	\label{fig:one_three_kaons}
\end{figure}
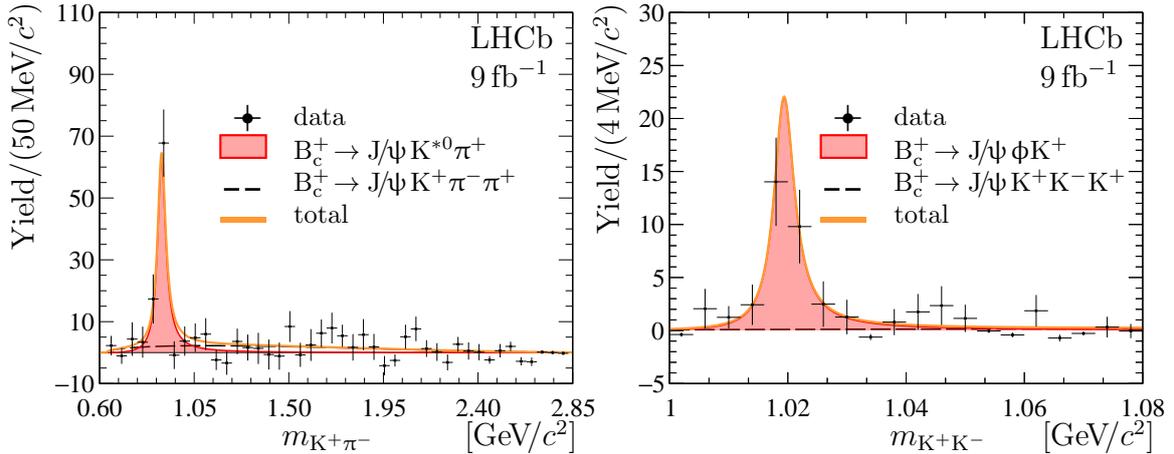

The~background-subtracted $\Kp\pim$
and $\Kp\Km$~mass distributions 
for selected Cabibbo\nobreakdash-suppressed 
\mbox{\BcTojpsikpipi}
and 
\mbox{\BcTojpsikkk}~candidates
are shown in Fig.~\ref{fig:one_three_kaons}.
Fits~to these distributions with  two-component  
functions similar to those described above are performed
and results of the~fits are overlaid in 
Fig.~\ref{fig:one_three_kaons}.
For \Bc~decays into 
 the~\mbox{\Tojpsikpipi}~final state 
 a~large fraction 
 proceeds 
via a~$\Kstarz$~meson,
\begin{equation*}\label{eq:f_Kpipi}
    f_{\Kstarz}^{\BcTojpsikpipi} 
    = \left(61.3\pm5.0\right)\%
\end{equation*}
and for \Bc~decays 
into the~\mbox{\Tojpsikkk}~final state
a~dominant fraction 
proceeds 
via a~$\Pphi$~meson,
\begin{equation*}\label{eq:f_KKK}
    f_{\Pphi}^{\BcTojpsikkk}
    = \left(90\pm19\right)\%\,.
\end{equation*}
All uncertainties for these resonance fractions are statistical only.

 \section{Ratios of branching fractions}

Six ratios of branching fractions are reported in this paper,  
\begin{subequations}
\begingroup
\allowdisplaybreaks
\begin{eqnarray}
\mathcal{R}^{\Tojpsikkk}_{\Tojpsikkpi} 
& \equiv & 
\dfrac{   \BR({\BcTojpsikkk})   } 
      {   \BR({\BcTojpsikkpi})  }     \,, 
\\
\mathcal{R}^{\Tojpsikpipi}_{\Tojpsikkpi} 
& \equiv & 
\dfrac{   \BR({ \BcTojpsikpipi})  } 
      {   \BR({\BcTojpsikkpi})   }     \,, 
\\
\mathcal{R}^{\Topsitwoskkpi}_{\Tojpsikkpi} 
& \equiv & 
\dfrac{   \BR({\BcTopsitwoskkpi}) 
\times    \BR({\decay{\psitwos}{\mumu}}) } 
      {   \BR({\BcTojpsikkpi}) 
\times    \BR({\decay{\jpsi}{\mumu}})  }     \,, 
\\
\mathcal{R}^{\Topsitwostripi}_{\Tojpsitripi} 
& \equiv & 
\dfrac{   \BR({\BcTopsitwostripi}) 
\times    \BR({\decay{\psitwos}{\mumu}}) } 
{         \BR({\BcTojpsitripi}) 
\times    \BR({\decay{\jpsi}{\mumu}}) }     \,, 
\\
\mathcal{R}^{\Topsitwospi}_{\Tojpsitripi} 
& \equiv & 
\dfrac{   \BR({\BcTopsitwospi}) \times 
          \BR({\decay{\psitwos}{\jpsi\pip\pim}}) 
} 
{   \BR({\BcTojpsitripi})  }     \,, 
\\
\mathcal{R}^{\Tojpsikkpi}_{\Tojpsitripi} 
& \equiv & 
\dfrac{   \BR({\BcTojpsikkpi})  } 
{   \BR({\BcTojpsitripi})  }     \,, 
\end{eqnarray}
\endgroup
\end{subequations}
where decays are paired to give 
the~largest cancellation of 
systematic 
uncertainties. 
Each ratio of branching fractions for 
\mbox{$\decay{\Bc}{\PX}$} and 
\mbox{$\decay{\Bc}{\PY}$}~decay modes, $\mathcal{R}^{\mathrm{X}}_{\mathrm{Y}}$,
is calculated as  
\begin{equation}
     \mathcal{R}^{\mathrm{X}}_{\mathrm{Y}} =
     \dfrac{ N_{\mathrm{X}}}
           { N_{\mathrm{Y}}} \times 
      \dfrac{ \upvarepsilon_{\mathrm{Y}}}
           { \upvarepsilon_{\mathrm{X}}}\,, \label{eq:br_rat}
\end{equation}
where $N$ is the~signal yield 
reported in Table~\ref{tab:sim_fit_res} 
and \eps denotes the~efficiency 
of the~corresponding decay.
The~efficiency is defined as the~product of geometric acceptance, 
reconstruction, selection, hadron identification 
and trigger efficiencies. 
All of these~contributions, 
except that 
of the~hadron\nobreakdash-identification efficiency, 
are determined using simulated samples, 
corrected as described above.
The~hadron\nobreakdash-identification efficiency is 
calculated from single-track hadron identification efficiencies 
for kaons and pions, 
determined from large calibration samples of
\mbox{$\decay{\Dstarp}{ \left(\decay{\Dz}{\Km\pip}\right)\pip}$}, 
\mbox{$\decay{\KS}{\pip\pim}$}
and 
\mbox{$\decay{\Ds}{\left(\decay{\Pphi}
{\Kp\Km}\right)\pip}$}~decays~\cite{LHCb-DP-2012-003,
LHCb-DP-2018-001}. 

The~measured ratios of branching fractions are
\begin{subequations}
\begingroup
\allowdisplaybreaks
\begin{eqnarray*}
\mathcal{R}^{\Tojpsikkk}_{\Tojpsikkpi} 
& = &  
\left(7.0\pm1.8\right)\times10^{-2} \,, 
\\
\mathcal{R}^{\Tojpsikpipi}_{\Tojpsikkpi}
& = &  
0.35 \pm 0.06 \,,  
\\
\mathcal{R}^{\Topsitwoskkpi}_{\Tojpsikkpi} 
& = & 
\left ( 3.7 \pm 1.2 \right) \times 10^{-2} \,,
\\
\mathcal{R}^{\Topsitwostripi}_{\Tojpsitripi}  
& = & 
\left( 1.9 \pm 0.4 \right) \times 10^{-2} \,, 
\\
\mathcal{R}^{\Topsitwospi}_{\Tojpsitripi}  
& = &
\left( 3.5 \pm 0.6 \right)\times 10^{-2} \,, 
\\
\mathcal{R}^{\Tojpsikkpi}_{\Tojpsitripi}  
& = &
0.185 \pm 0.013 \,, 
\end{eqnarray*}
\endgroup
\end{subequations}
where uncertainties are statistical only and 
correlation coefficients are listed in Table~\ref{tab:corr_stat}.
Systematic uncertainties are discussed in Sec.~\ref{sec:Systematics}. 

%% file: systematics.tex
\section{Systematic uncertainties}
\label{sec:Systematics}

The~decay channels under study have similar kinematics
and topologies, therefore  
a~large part of systematic uncertainties cancels 
in the~branching fraction ratios~$\mathcal{R}^{\PX}_{\PY}$. 
The~remaining contributions 
to the systematic uncertainty are 
summarised in Table~\ref{tab:systematics} 
and discussed below.

An important source of systematic uncertainty on the
ratios arises from the imperfect knowledge 
of the shapes of signal and background components used in the fits. 
To~estimate this
uncertainty, several alternative models 
are tested.  
For~the~\Bc and \psitwos
signal shapes an~Apollonios function~\cite{Santos:2013gra}
is employed as an alternative
model. The~degree
of the~polynomials used in the~fits is increased by one.
The~systematic uncertainty related to 
 the fit model is estimated 
 by pseudoexperiments
with the baseline fit model and fitted 
with alternative models. Each pseudoexperiment
is approximately 100 times larger than the 
data sample. The maximal deviations 
in the~ratios of the signal yields with 
respect to the baseline model
do not exceed 2.8\% for variation of the~signal model
and 3.7\% for variation of background model, 
and   
are taken
as systematic uncertainties in the~ratios~$\mathcal{R}^{\PX}_{\PY}$.

The~simulated \decay{\Bc}{\Ppsi3\Ph^\pm} 
decays are corrected to reproduce the~two\nobreakdash-dimensional
$\pip\pim$, $\Kp\pim$, $\Km\pip$, and~$\Kp\Km$~mass distributions
observed in data. 
The~uncertainty associated with this~correction procedure and 
related to the~imperfect knowledge of 
the~\Bc decay model is estimated by varying 
the~reference mass~distributions
within their uncertainties.
It~causes small changes in  
the~efficiencies and subsequent changes in  
the~ratios $\mathcal{R}^{\PX}_{\PY}$, that do not exceed 
$0.3\%$. These changes are 
taken as systematic uncertainties related to 
the~\Bc~decay model.

An~additional uncertainty arises from 
differences between  data and  simulation,
in particular differences in the reconstruction efficiency 
of charged\nobreakdash-particle tracks.
The~track\nobreakdash-finding efficiencies obtained from 
simulation  
are corrected \mbox{using} data calibration samples~\cite{LHCb-DP-2013-002}.
The uncertainties related to the correction factors, 
\mbox{together}
with the~uncertainty in the~hadron\nobreakdash-identification efficiency 
due to the~finite size of 
the~calibration samples~\cite{LHCb-DP-2012-003, LHCb-DP-2018-001},
are propagated to the~ratio of total efficiencies using 
pseudoexperiments. The~obtained 
systematic uncertainty for the~$\mathcal{R}^{\PX}_{\PY}$
ratios do not exceed 1.6\%. 

The~systematic uncertainty related 
to the~trigger efficiency is estimated 
by comparing 
the~ratios of trigger
efficiencies in data and simulation
using large samples of  
the~\mbox{$\decay{\Bp}{\jpsi\Kp}$} and 
\mbox{$\decay{\Bp}{\psitwos\Kp}$}~decays~\cite{LHCb-PAPER-2012-010}. 
Another source of uncertainty is a~potential disagreement
between data and simulation in the~estimation of efficiencies, 
due to possible effects not considered above. 
This is studied by varying 
the~selection criteria in ranges that lead 
up to a~$\pm 20\%$ change in the~measured signal yields. 
For~this study, the~high
yield \mbox{\BcTojpsitripi} data sample is used. 
The~resulting
difference between the~efficiencies estimated using data 
and simulation does not exceed $3.0\%$,
which is taken as a~systematic uncertainty 
for the~ratios $\mathcal{R}^{\PX}_{\PY}$.
The~last systematic uncertainty 
considered for the~ratios $\mathcal{R}^{\PX}_{\PY}$ is 
due to the~finite size of the~simulated samples
and it varies between 0.6\% and 2.1\%.

\begin{table}[t]
	\centering
	\caption{\small
	Ranges of relative systematic uncertainties for 
    the~ratios of branching fractions $\mathcal{R}^{\PX}_{\PY}$.
    The~total systematic uncertainty is 
    the~quadratic sum of individual contributions.} 
	\label{tab:systematics}
	\vspace{2mm}
	\begin{tabular*}{0.55\textwidth}
	{@{\hspace{5mm}}l@{\extracolsep{\fill}}c@{\hspace{5mm}}}
	Source  &   Uncertainty~$\left[\%\right]$ 
   \\[1.5mm]
  \hline 
  \\[-1.5mm]
  Fit model                  &   
  \\
  ~~~Signal shape            & $  < 0.1        - 2.8$ 
  \\
  ~~~Background shape        & $\phantom{<}0.2 - 3.7$ 
  \\
  \Bc~decay model            & $  < 0.1        - 0.3$  
  \\ 
  Efficiency corrections     & $ < 0.1         - 1.6$
  \\
  Trigger efficiency         & $ \phantom{<} 1.1$ 
  \\
  Data-simulation difference & $ \phantom{<} 3.0$ 
  \\
  Size of simulated sample   &  $\phantom{<} 0.6 - 2.1 $ 
    \\[1.5mm]
  \hline 
  \\[-1.5mm]
  Total                     &   $\phantom{0} 3.3 - 5.6 $ 
	\end{tabular*}
	\vspace{3mm}
\end{table}

\begin{table}[t]
	\centering
	\caption{\small
	Systematic uncertainties for 
    the~fractions of the~decays
    via resonances.} 
	\label{tab:f_systematics}
	\vspace{2mm}
	\begin{tabular*}{0.45\textwidth}
	{@{\hspace{5mm}}l@{\extracolsep{\fill}}c@{\hspace{5mm}}}
	Fraction  &   Uncertainty~$\left[\%\right]$ 
   \\[1.5mm]
  \hline 
  \\[-1.5mm]
  $f_{\rhoz}^{\BcTojpsitripi}$        &  ${}^{+\,12.0}_{-\,0.3}$  
  \\
    $f_{\PR}^{\BcTojpsitripi}$        &  ${}^{+\,8.0\phantom{0}}_{-\,1.2}$    
  \\
   $f_{\Kstarzb}^{\BcTojpsikkpi}$     &  ${}^{+\,3.9\phantom{0}}_{-\,4.8}$    
  \\
  $f_{\Kstarz}^{\BcTojpsikpipi}$      &  ${}^{+\,7.7\phantom{0}}_{-\,0.3}$   
  \\
 $f_{\Pphi}^{\BcTojpsikkk}$           &  ${}^{+\,5.0\phantom{0}}_{-\,7.0}$   
  \end{tabular*}
  \vspace{3mm}
\end{table}

Systematic uncertainties for the~fractions 
of decays via resonances 
are estimated 
by variation of the~fit models. 
In~particular, the~meson radii are varied 
between $1.5$ and $5\gev^{-1}$,
the~degree of polynomial functions is varied 
from one to three and, 
for \mbox{\BcTojpsitripi}~decays,
the~Gounaris\nobreakdash--Sakurai~function~\cite{Gounaris:1968mw} 
is used for the~$\rhoz$~meson parameterisation. 
For~each alternative model 
the~fit fraction is determined and 
the~maximal difference in the~fractions
with respect to the~default fit model 
is taken as a~corresponding systematic uncertainty.
The~systematic uncertainties for the~fractions 
are summarised in Table~\ref{tab:f_systematics}.
For~the~fraction $f_{\Pphi}^{\decay{\Bc}{\jpsi\Kp\Km\pip}}$, 
the~upper limits are estimated for each alternative model,   
and 
the~largest value is taken. 
The~upper limit at 90\,(95)\%\,CL that accounts 
for the~systematic uncertainty~is 
$f_{\Pphi}^{\decay{\Bc}{\jpsi\Kp\Km\pip}}  < 4.2\,(4.8)\%$. 

%% file: results.tex
\section{Results and summary}
\label{sec:Results}

The~\decay{\Bc}{\Ppsi3\Ph^\pm} decays  are studied
using proton\nobreakdash-proton collision data, 
corresponding to an~integrated luminosity of $9 \invfb$,
collected with the~\lhcb detector.
The~first observation of   
the~decays 
\mbox{$\BcTopsitwostripi$}, 
\mbox{$\BcTojpsikpipi$} and 
\mbox{$\BcTojpsikkk$} is reported. 
The~decay \mbox{$\decay{\Bc}{\psitwos\pip}$}
is confirmed using 
\mbox{$\decay{\psitwos}{\jpsi\pip\pim}$}~mode
and  the~first evidence for 
the~\mbox{\BcTopsitwoskkpi}~decay 
is obtained with a significance 
of $3.7$~standard deviations.

The~three-pion mass distribution 
for the~\mbox{\BcTojpsitripi}~decays 
is found to be consistent
with a~BLL~model for 
the~\mbox{$\decay{\Bc}{\jpsi\Pa_1(1260)^+}$}~decay
based on the~factorisation 
approach~\cite{Likhoded:2009ib,Berezhnoy:2011nx,Likhoded:2013iua},
in agreement with 
previous studies~\cite{LHCb-PAPER-2011-044}. 
The~presence of the~intermediate 
axial $\Pa_1(1260)^+$~meson in this decay 
is further supported  
by a~large fraction of the~\mbox{\BcTojpsitripi}~decays
proceeding via the intermediate $\rhoz$~resonance
\begin{equation*}
     f_{\rhoz}^{\BcTojpsitripi} = \left( 88.1 \pm 3.0 ^{\,+\,12.0}_{\,-\,0.3} \right)\%\,,
\end{equation*}
and the~observation of a~structure in the~\pip\pim~mass spectrum, consistent
with the~\mbox{$\decay{\Pa_1(1260)^+}
{\left( \decay{\Pf_0(1370)}{\pip\pim}\right) \pip}$}~decay.
The~fraction of the~\mbox{\BcTojpsitripi}~decays proceeding 
via this  intermediate state 
is found to be 
\begin{equation*}
     f_{\PR}^{\BcTojpsitripi} = \left( 10.4 \pm 1.4 ^{\,+\,8.0}_{\,-\,1.2} \right)\%\,,
\end{equation*}
which agrees with the~value 
of $\left( 7.40\pm2.71\pm1.26\right)\%$, 
obtained by the CLEO collaboration 
for the~fraction of 
the~\mbox{$\decay{\Pa_1(1260)^+}{\pip\piz\piz}$}~decays
proceeding via intermediate 
$\Pf_0(1370)$~state~\cite{CLEO:1999rzk}.

The~fraction of the~\mbox{$\decay{\Bc}{\jpsi\Kp\Km\pip}$}~decays 
proceeding via intermediate $\Kstarzb$~state is measured to be 
\begin{equation*}
     f_{\Kstarzb}^{\decay{\Bc}{\jpsi\Kp\Km\pip}} = 
     \left( 64.5 \pm 4.7 ^{\,+\,3.9}_{\,-\,4.8} \right)\%\,,
\end{equation*}
while no evidence for OZI-suppressed decays 
\mbox{$\decay{\Bc}{\jpsi\left(\decay{\Pphi}{\Kp\Km}\right)\pip}$}~is found 
and the~upper limit at 90\,(95)\% CL of the~corresponding fraction 
of  
the~\mbox{$\decay{\Bc}{\jpsi\Kp\Km\pip}$}~decays 
proceeding via intermediate $\phiz$~meson 
is set to be 
\begin{equation*}  
f_{\Pphi}^{\decay{\Bc}{\jpsi\Kp\Km\pip}}  < 4.2\,(4.8)\%\,.
\end{equation*}
Both results  are in agreement with the~previous study 
by the~LHCb collaboration~\cite{LHCb-PAPER-2013-047}.  
For~the~\mbox{\BcTojpsikpipi}
and \mbox{\BcTojpsikkk}~decays
large fractions of decays via intermediate \Kstarz and \Pphi~resonances, 
respectively, are found 
\begin{eqnarray*}
   f_{\Kstarz}^{\BcTojpsikpipi} 
    & = & \left(61.3\pm5.0^{\,+\,7.7}_{\,-\,0.3}\right)\%   \,,
    \\
    f_{\Pphi}^{\BcTojpsikkk}
    & = & \left(90\pm19^{\,+\,5}_{\,-\,7} \right)\%   \,.
\end{eqnarray*}

 \begin{table}[t]
\caption{\small 
Comparison of the measured 
ratios $\mathcal{R}^{\PX}_{\PY}$
with their theoretical  predictions 
or derivations from previous measurements.}
\centering
\begin{tabular*}{0.85\textwidth}
	{@{\hspace{2mm}}l@{\extracolsep{\fill}}llll@{\hspace{2mm}}}
\multirow{2}{*}{Ratio} & 
\multirow{2}{*}{Value} &  Prediction, & 
\multirow{2}{*}{Reference}
\\ 
 & & measurement & 
    \\[1.5mm]
    \hline 
    \\[-3mm]
$\mathcal{R}_{\Topsitwostripi}^{\Topsitwoskkpi}$
    & $0.37 \pm 0.15 \pm 0.01 $
    & $0.16$ 
    &  BLL~\cite{Likhoded:2013iua, Luchinsky:2013yla}
    \\
$\mathcal{R}^{\Tojpsikpipi}_{\Tojpsikkpi}$ 
 &  $0.35 \pm  0.06 \pm 0.01$ 
 & $0.37$ & BLL~\cite{Luchinsky:2013yla}
    \\[1.5mm]
 $\mathcal{R}^{\Tojpsikpipi}_{\Tojpsitripi}$
  &  $\left( 6.4 \pm 1.0 \pm 0.2 \right) \times 10^{-2}$ 
    &$7.7\times10^{-2}$   &  BLL~\cite{Luchinsky:2013yla}
    \\[1.5mm]
 $\mathcal{R}^{\Tojpsikkpi  }_{\Tojpsitripi}$ 
  & 
  $0.185 \pm  0.013 \pm  0.006$ 
  & $0.21$ 
  & BLL~\cite{Luchinsky:2013yla, Likhoded:2013iua}
    \\[1.5mm]
    \hline 
    \\[-3mm]
   $\mathcal{R}^{\Topsitwospi}_{\Tojpsikkpi}$
    &$0.19 \pm  0.03 \pm 0.01$
    & $0.18\pm0.04$  
    &  LHCb~\cite{LHCb-PAPER-2015-024, LHCb-PAPER-2013-047}
    \\[1.5mm]
  $\mathcal{R}^{\Topsitwospi}_{\Tojpsitripi}$ 
    & $\left( 3.5 \pm 0.6 \pm 0.2 \right) \times 10^{-2}$ 
    & $\left( 3.9 \pm 0.9 \right) \times 10^{-2}$
    & LHCb~\cite{LHCb-PAPER-2011-044, 
    LHCb-PAPER-2015-024}
    \\[1.5mm]
 $\mathcal{R}^{\Tojpsikkpi  }_{\Tojpsitripi}$ 
  &
  $0.185 \pm  0.013 \pm  0.006$ 
    & $0.22 \pm0.06$ 
    & LHCb~\cite{LHCb-PAPER-2013-047,LHCb-PAPER-2011-044}
\end{tabular*}
\label{tab:bf_ratios_all}
\end{table}

The~six ratios of branching fractions are measured as  
\begingroup
\allowdisplaybreaks
\begin{eqnarray*}
\mathcal{R}^{\Tojpsikkk}_{\Tojpsikkpi} 
& = &  
\left( 7.0 \pm 1.8 \pm 0.2 \right) \times 10^{-2}  \,, 
\\
\mathcal{R}^{\Tojpsikpipi}_{\Tojpsikkpi}
& = &  
0.35 \pm  0.06 \pm 0.01  \,,
\\
\mathcal{R}^{\Topsitwoskkpi}_{\Tojpsikkpi} 
& = & 
\left( 3.7 \pm 1.2  \pm 0.1 \right) \times 10^{-2} \,,  
\\
\mathcal{R}^{\Topsitwostripi}_{\Tojpsitripi}  
& = &  
\left(1.9 \pm 0.4 \pm 0.1\right)\times 10^{-2}\,
\\
\mathcal{R}^{\Topsitwospi}_{\Tojpsitripi}  
& = &   
\left(3.5 \pm 0.6 \pm 0.2 \right)\times 10^{-2} \,, 
\\
\mathcal{R}^{\Tojpsikkpi}_{\Tojpsitripi}  
& = &  
0.185 \pm  0.013 \pm 0.006 \,,  
\end{eqnarray*}
\endgroup
 where the~first uncertainty is statistical and the~second systematic.
 Correlation coefficients for statistical and systematic 
 uncertainties are given in Appendix~\ref{sec:correlations}.
 The~ratios of branching fractions from this measurement 
 are compared in Table~\ref{tab:bf_ratios_all}
with either theoretical predictions~\cite{Likhoded:2013iua, 
Luchinsky:2013yla} or derivations 
from previous 
LHCb measurements~\cite{LHCb-PAPER-2011-044,
LHCb-PAPER-2013-047,
LHCb-PAPER-2015-024}.

 \begin{table}[t]
\caption{\small 
Comparison of the~measured ratios of 
branching fractions for the~Cabibbo\protect\nobreakdash-suppressed 
and Cabibbo\protect\nobreakdash-favoured decays
with the~ratios of 
branching fraction for similar 
decays of \Bc, \Bu, \Bd and 
\Bs~mesons~\cite{LHCb-PAPER-2016-020,
LHCb-PAPER-2011-040,
LHCb-PAPER-2012-046,
LHCb-PAPER-2012-033,
PDG2021}.}
\centering
\begin{tabular*}{0.7\textwidth}
	{@{\hspace{3mm}}l@{\extracolsep{\fill}}cc@{\hspace{3mm}}}
    & Value~$\left[10^{-2}\right]$ & Reference
    \\[2mm]
    \hline 
    \\[-2mm]
    $\mathcal{R}^{\jpsi\Kp\Km\Kp}_{\jpsi\Kp\Km\pip}$
    &  $7.0 \pm 1.8 \pm 0.2$
    &  This paper
    \\[2mm]
    $\mathcal{R}^{\jpsi\Kp\pim\pip}_{\jpsi\pip\pim\pip}$
       &  $6.4 \pm 1.0 \pm 0.2$
    &  This paper
   \\[2mm]
    \hline
    \\[-2mm]
    $\dfrac{\BR({\decay{\Bc}{\jpsi\Kp}})}
           {\BR({\decay{\Bc}{\jpsi\pip}})}$
    &  $7.9 \pm 0.8$ 
    &\cite{LHCb-PAPER-2016-020}
    \\[4mm]
    $\dfrac{\BR({\decay{\Bu}{\Dzb\Kp\pim\pip}})}
           {\BR({\decay{\Bu}{\Dzb\pip\pim\pip}})}$
    &  $9.3\pm 5.1$
    &\cite{PDG2021, LHCb-PAPER-2011-040}
    \\[4mm]
    $\dfrac{\BR({\decay{\Bz}{\Dm\Kp\pim\pip}})}
           {\BR({\decay{\Bz}{\Dm\pip\pim\pip}})}$
    &  $5.8\pm 1.5$
    &\cite{PDG2021, LHCb-PAPER-2011-040}
    \\[4mm]
    $\dfrac{\BR({\decay{\Bz}{\theDstarm\Kp\pim\pip}})}
           {\BR({\decay{\Bz}{\theDstarm\pip\pim\pip}})}$
    &  $6.5\pm 0.6$
    &\cite{PDG2021, LHCb-PAPER-2012-046}
    \\[4mm]
    $\dfrac{\BR({\decay{\Bs}{\Dsm\Kp\pim\pip}})}
          {\BR({\decay{\Bs}{\Dsm\pip\pim\pip}})}$
    &  $5.2\pm 1.3$
    &\cite{PDG2021, LHCb-PAPER-2012-033}
    \end{tabular*}
\label{tab:bf_ratios_cab}
\end{table}

The~ratio of branching fractions 
for the~\Bc~decays via $\psitwos$ and $\jpsi$~mesons, 
$\mathcal{R}^{\Topsitwoskkpi}_{\Tojpsikkpi}$,
agrees with 
the~known ratio of branching fractions  for 
the~\mbox{$\decay{\Bc}{\psitwos\pip}$} and 
\mbox{$\decay{\Bc}{\jpsi\pip}$}~decays
$\mathcal{R}^{\psitwos\pip}_{\jpsi\pip}
=\left(3.54 \pm 0.43\right)\times 10^{-2}$~\cite{LHCb-PAPER-2015-024},
however the~similar ratio 
for the~\mbox{\BcTopsitwostripi}
and \mbox{\BcTojpsitripi}~channels, 
$\mathcal{R}^{\Topsitwostripi}_{\Tojpsitripi}$,
is in tension, at 2.8~standard deviations,
with 
the~measured ratio 
$\mathcal{R}^{\psitwos\pip}_{\jpsi\pip}$~\cite{LHCb-PAPER-2015-024}.
The~ratio $\mathcal{R}^{\Tojpsikkk}_{\Tojpsikkpi}$
of branching fractions for 
the~Cabibbo\nobreakdash-suppressed \mbox{$\decay{\Bc}{\jpsi\Kp\Km\Kp}$}
and Cabibbo\nobreakdash-favoured 
\mbox{$\decay{\Bc}{\jpsi\Kp\Km\pip}$}~decays 
agrees  
within uncertainties with 
the~similar ratio
$\mathcal{R}^{\jpsi\Kp\pim\pip}_{\jpsi\pip\pim\pip}$ and 
the~known ratio of branching fractions  for 
the~Cabibbo\nobreakdash-suppressed  \mbox{$\decay{\Bc}{\jpsi\Kp}$} and  
Cabibbo\nobreakdash-favoured 
\mbox{$\decay{\Bc}{\jpsi\pip}$}~decays~\cite{LHCb-PAPER-2016-020}. 
The~ratios 
$\mathcal{R}^{\Tojpsikkk}_{\Tojpsikkpi}$ and 
$\mathcal{R}^{\jpsi\Kp\pim\pip}_{\jpsi\pip\pim\pip}$ 
also agree with the~ratios of branching 
fraction for the~multibody decays
of \Bu, \Bd and \Bs~mesons, see Table~\ref{tab:bf_ratios_cab}.
This~pattern supports the~factorisation 
hypothesis for the~\mbox{$\decay{\Bc}{\Ppsi3\Ph^{\pm}}$}~decays. 

%% file: acknowledgements.tex
\section*{Acknowledgements}
%
%
\noindent 
We~thank A.\,K.~Likhoded and A.\,V.~Luchinsky for providing
us with the~code for modelling
the~$\decay{\Bc}{\Ppsi3\Ph^\pm}$~decays.
We~express our gratitude to our colleagues in the CERN
accelerator departments for the~excellent performance of the~LHC. 
We~thank the technical and administrative staff at the~LHCb
institutes.
We~acknowledge support from CERN and from the~national agencies:
CAPES, CNPq, FAPERJ and FINEP\,(Brazil); 
MOST and NSFC\,(China); 
CNRS/IN2P3\,(France); 
BMBF, DFG and MPG\,(Germany); 
INFN\,(Italy); 
NWO\,(Netherlands); 
MNiSW and NCN\,(Poland); 
MEN/IFA\,(Romania); 
MSHE\,(Russia); 
MICINN\,(Spain); 
SNSF and SER\,(Switzerland); 
NASU\,(Ukraine); 
STFC\,(United Kingdom); 
DOE NP and NSF\,(USA).
We~acknowledge the~computing resources that are provided by CERN, 
IN2P3\,(France), 
KIT and DESY\,(Germany), 
INFN\,(Italy), 
SURF\,(Netherlands),
PIC\,(Spain), 
GridPP\,(United Kingdom), 
RRCKI and Yandex LLC\,(Russia), 
CSCS\,(Switzerland), 
IFIN\nobreakdash-HH\,(Romania), 
CBPF\,(Brazil),
PL\nobreakdash-GRID\,(Poland) and 
NERSC\,(USA).
We~are indebted to the communities behind 
the~multiple open\nobreakdash-source
software packages on which we depend.
Individual groups or members have received support from
ARC and ARDC\,(Australia);
AvH Foundation\,(Germany);
EPLANET, Marie Sk\l{}odowska\nobreakdash-Curie Actions and ERC\,(European Union);
A*MIDEX, ANR, IPhU and Labex P2IO, and R\'{e}gion Auvergne\nobreakdash-Rh\^{o}ne-Alpes\,(France);
Key Research Program of Frontier Sciences of CAS, CAS PIFI, CAS CCEPP, 
Fundamental Research Funds for the~Central Universities, 
and Sci. \& Tech. Program of Guangzhou\,(China);
RFBR, RSF and Yandex LLC\,(Russia);
GVA, XuntaGal and GENCAT\,(Spain);
the~Leverhulme Trust, the~Royal Society
 and UKRI\,(United Kingdom).

%% file: covariances.tex
 \renewcommand{\thetable}{A.\arabic{table}}
 \renewcommand{\thefigure}{A.\arabic{figure}}
 \renewcommand{\theequation}{A.\arabic{equation}}
 \setcounter{figure}{0}
 \setcounter{table}{0}
 \setcounter{equation}{0}

\section{Correlation matrices}\label{sec:correlations}

Correlation coefficients for the~measured ratios $\mathcal{R}^{\PX}_{\PY}$
are shown in Tables~\ref{tab:corr_stat} 
and~\ref{tab:corr_syst} for statistical and systematic
uncertainties, respectively.

\begin{table}[htb]
	\centering
	\caption{\small
	Off-diagonal correlation 
	coefficients\,($\left[\%\right]$)
	for statistical uncertainties for 
	the~measured ratios $\mathcal{R}^{\PX}_{\PY}$.
   } 
	\label{tab:corr_stat}
	\vspace{2mm}
	\begin{tabular*}{0.75\textwidth}
	{@{\hspace{5mm}}l@{\extracolsep{\fill}}ccccc@{\hspace{5mm}}}
	&  
    \begin{sideways}$\mathcal{R}^{\Tojpsikpipi}_{\Tojpsikkpi}$\end{sideways} 
    & 
    \begin{sideways}$\mathcal{R}^{\Topsitwoskkpi}_{\Tojpsikkpi}$\end{sideways} 
    &
    \begin{sideways}$\mathcal{R}^{\Topsitwostripi}_{\Tojpsitripi}$ \end{sideways} 
    &
    \begin{sideways}$\mathcal{R}^{\Topsitwospi}_{\Tojpsitripi}$ \end{sideways} 
    &
    \begin{sideways}$\mathcal{R}^{\Tojpsikkpi}_{\Tojpsitripi}$ \end{sideways} 
   \\[1.5mm]
  \hline 
  \\[-1.5mm]
  $\mathcal{R}^{\Tojpsikkk}_{\Tojpsikkpi}$
  &   $-13$ & $+0$ & $-8$  & $+2$   & $-24$ 
  \\
  $\mathcal{R}^{\Tojpsikpipi}_{\Tojpsikkpi}$ 
  &          & $+4$ & $-7$  & $+1$   & $-33$     
  \\ 
  $\mathcal{R}^{\Topsitwoskkpi}_{\Tojpsikkpi}$
   &          &        & $-1$  & $-3$   & $-15$ 
 \\ 
  $\mathcal{R}^{\Topsitwostripi}_{\Tojpsitripi}$ 
   &          &        &         & $-4$   & $+12$ 
 \\
  $\mathcal{R}^{\Topsitwospi}_{\Tojpsitripi}$ 
   &          &        &         &          &  $+0\phantom{0}$
  \end{tabular*}
  \vspace{3mm}
\end{table}

\begin{table}[htb]
	\centering
	\caption{\small
	Off-diagonal correlation 
	coefficients\,($\left[\%\right]$)
	for systematic uncertainties for 
	the~measured ratios~$\mathcal{R}^{\PX}_{\PY}$. 
   } 
	\label{tab:corr_syst}
	\vspace{2mm}
	\begin{tabular*}{0.75\textwidth}
	{@{\hspace{5mm}}l@{\extracolsep{\fill}}ccccc@{\hspace{5mm}}}
	&  
    \begin{sideways}$\mathcal{R}^{\Tojpsikpipi}_{\Tojpsikkpi}$\end{sideways} 
    & 
    \begin{sideways}$\mathcal{R}^{\Topsitwoskkpi}_{\Tojpsikkpi}$\end{sideways} 
    &
    \begin{sideways}$\mathcal{R}^{\Topsitwostripi}_{\Tojpsitripi}$ \end{sideways} 
    &
    \begin{sideways}$\mathcal{R}^{\Topsitwospi}_{\Tojpsitripi}$ \end{sideways} 
    &
    \begin{sideways}$\mathcal{R}^{\Tojpsikkpi}_{\Tojpsitripi}$ \end{sideways} 
   \\[1.5mm]
  \hline 
  \\[-1.5mm]
  $\mathcal{R}^{\Tojpsikkk}_{\Tojpsikkpi}$
  &   $+39$ & $+18$ & $+2$  & $-10$   & $-52$ 
  \\
  $\mathcal{R}^{\Tojpsikpipi}_{\Tojpsikkpi}$ 
  &          & $+27$  & $-7$  & $-4\phantom{0}$   & $-60$     
  \\ 
  $\mathcal{R}^{\Topsitwoskkpi}_{\Tojpsikkpi}$
  &          &          & $+9$  & $-20$   & $-59$ 
   \\ 
  $\mathcal{R}^{\Topsitwostripi}_{\Tojpsitripi}$ 
  &          &         &         & $+27$   & $+30$ 
 \\
  $\mathcal{R}^{\Topsitwospi}_{\Tojpsitripi}$ 
  &          &         &         &          &  $+37$
  \end{tabular*}
  \vspace{3mm}
\end{table}

%% file: Authorship_LHCb-PAPER-2021-034.tex
\centerline
{\large\bf LHCb collaboration}
\begin
{flushleft}
\small
R.~Aaij$^{32}$,
A.S.W.~Abdelmotteleb$^{56}$,
C.~Abell{\'a}n~Beteta$^{50}$,
F.J.~Abudinen~Gallego$^{56}$,
T.~Ackernley$^{60}$,
B.~Adeva$^{46}$,
M.~Adinolfi$^{54}$,
H.~Afsharnia$^{9}$,
C.~Agapopoulou$^{13}$,
C.A.~Aidala$^{87}$
% ...

{\footnotesize \it

% ...
$^{9}$Universit{\'e} Clermont Auvergne, CNRS/IN2P3, LPC, Clermont-Ferrand, France\\
% ...
$^{13}$LPNHE, Sorbonne Universit{\'e}, Paris Diderot Sorbonne Paris Cit{\'e}, CNRS/IN2P3, Paris, France\\
% ...
$^{32}$Nikhef National Institute for Subatomic Physics, Amsterdam, Netherlands\\
% ...
$^{46}$Instituto Galego de F{\'\i}sica de Altas Enerx{\'\i}as (IGFAE), Universidade de Santiago de Compostela, Santiago de Compostela, Spain\\
% ...
$^{50}$Physik-Institut, Universit{\"a}t Z{\"u}rich, Z{\"u}rich, Switzerland\\
% ...
$^{54}$H.H. Wills Physics Laboratory, University of Bristol, Bristol, United Kingdom\\
% ...
$^{56}$Department of Physics, University of Warwick, Coventry, United Kingdom\\
% ...
$^{60}$Oliver Lodge Laboratory, University of Liverpool, Liverpool, United Kingdom\\
% ...
$^{68}$Syracuse University, Syracuse, NY, United States\\
% ...
$^{87}$University of Michigan, Ann Arbor, United States, associated to $^{68}$\\
\bigskip
% ...
}
\end{flushleft}

%% file: main.bbl
\ifx\mcitethebibliography\mciteundefinedmacro
\PackageError{LHCb.bst}{mciteplus.sty has not been loaded}
{This bibstyle requires the use of the mciteplus package.}\fi
\providecommand{\href}[2]{#2}
\begin{mcitethebibliography}{10}
\mciteSetBstSublistMode{n}
\mciteSetBstMaxWidthForm{subitem}{\alph{mcitesubitemcount})}
\mciteSetBstSublistLabelBeginEnd{\mcitemaxwidthsubitemform\space}
{\relax}{\relax}

\bibitem{LHCb-PAPER-2011-044}
LHCb collaboration, R.~Aaij {\em et~al.},
  \ifthenelse{\boolean{articletitles}}{\emph{{First observation of the decay
  \mbox{\decay{\Bc}{\jpsi\pip\pim\pip}}}},
  }{}\href{https://doi.org/10.1103/PhysRevLett.108.251802}{Phys.\ Rev.\ Lett.\
  \textbf{108} (2012) 251802},
  \href{http://arxiv.org/abs/1204.0079}{{\normalfont\ttfamily
  arXiv:1204.0079}}\relax
\mciteBstWouldAddEndPuncttrue
\mciteSetBstMidEndSepPunct{\mcitedefaultmidpunct}
{\mcitedefaultendpunct}{\mcitedefaultseppunct}\relax
\EndOfBibitem
\bibitem{LHCb-PAPER-2012-054}
LHCb collaboration, R.~Aaij {\em et~al.},
  \ifthenelse{\boolean{articletitles}}{\emph{{Observation of the decay
  \mbox{\decay{\Bc}{\psitwos\pip}}}},
  }{}\href{https://doi.org/10.1103/PhysRevD.87.071103}{Phys.\ Rev.\
  \textbf{D87} (2013) 071103(R)},
  \href{http://arxiv.org/abs/1303.1737}{{\normalfont\ttfamily
  arXiv:1303.1737}}\relax
\mciteBstWouldAddEndPuncttrue
\mciteSetBstMidEndSepPunct{\mcitedefaultmidpunct}
{\mcitedefaultendpunct}{\mcitedefaultseppunct}\relax
\EndOfBibitem
\bibitem{LHCb-PAPER-2013-010}
LHCb collaboration, R.~Aaij {\em et~al.},
  \ifthenelse{\boolean{articletitles}}{\emph{{Observation of
  \mbox{\decay{\Bc}{\jpsi\Dsp}} and \mbox{\decay{\Bc}{\jpsi\Dssp}} decays}},
  }{}\href{https://doi.org/10.1103/PhysRevD.87.112012}{Phys.\ Rev.\
  \textbf{D87} (2013) 112012}, Erratum
  \href{https://doi.org/10.1103/PhysRevD.89.019901}{ibid.\   \textbf{D89}
  (2014) 019901(E)},
  \href{http://arxiv.org/abs/1304.4530}{{\normalfont\ttfamily
  arXiv:1304.4530}}\relax
\mciteBstWouldAddEndPuncttrue
\mciteSetBstMidEndSepPunct{\mcitedefaultmidpunct}
{\mcitedefaultendpunct}{\mcitedefaultseppunct}\relax
\EndOfBibitem
\bibitem{LHCb-PAPER-2013-021}
LHCb collaboration, R.~Aaij {\em et~al.},
  \ifthenelse{\boolean{articletitles}}{\emph{{First observation of the decay
  \mbox{\decay{\Bc}{\jpsi\Kp}}}},
  }{}\href{https://doi.org/10.1007/JHEP09(2013)075}{JHEP \textbf{09} (2013)
  075}, \href{http://arxiv.org/abs/1306.6723}{{\normalfont\ttfamily
  arXiv:1306.6723}}\relax
\mciteBstWouldAddEndPuncttrue
\mciteSetBstMidEndSepPunct{\mcitedefaultmidpunct}
{\mcitedefaultendpunct}{\mcitedefaultseppunct}\relax
\EndOfBibitem
\bibitem{LHCb-PAPER-2013-044}
LHCb collaboration, R.~Aaij {\em et~al.},
  \ifthenelse{\boolean{articletitles}}{\emph{{Observation of the decay
  \mbox{\decay{\Bc}{\Bs\pip}}}},
  }{}\href{https://doi.org/10.1103/PhysRevLett.111.181801}{Phys.\ Rev.\ Lett.\
  \textbf{111} (2013) 181801},
  \href{http://arxiv.org/abs/1308.4544}{{\normalfont\ttfamily
  arXiv:1308.4544}}\relax
\mciteBstWouldAddEndPuncttrue
\mciteSetBstMidEndSepPunct{\mcitedefaultmidpunct}
{\mcitedefaultendpunct}{\mcitedefaultseppunct}\relax
\EndOfBibitem
\bibitem{LHCb-PAPER-2013-047}
LHCb collaboration, R.~Aaij {\em et~al.},
  \ifthenelse{\boolean{articletitles}}{\emph{{Observation of the decay
  \mbox{\decay{\Bc}{\jpsi\Kp\Km\pip}}}},
  }{}\href{https://doi.org/10.1007/JHEP11(2013)094}{JHEP \textbf{11} (2013)
  094}, \href{http://arxiv.org/abs/1309.0587}{{\normalfont\ttfamily
  arXiv:1309.0587}}\relax
\mciteBstWouldAddEndPuncttrue
\mciteSetBstMidEndSepPunct{\mcitedefaultmidpunct}
{\mcitedefaultendpunct}{\mcitedefaultseppunct}\relax
\EndOfBibitem
\bibitem{LHCb-PAPER-2014-009}
LHCb collaboration, R.~Aaij {\em et~al.},
  \ifthenelse{\boolean{articletitles}}{\emph{{Evidence for the decay
  \mbox{\decay{\Bc}{\jpsi 3\pip 2\pim}}}},
  }{}\href{https://doi.org/10.1007/JHEP05(2014)148}{JHEP \textbf{05} (2014)
  148}, \href{http://arxiv.org/abs/1404.0287}{{\normalfont\ttfamily
  arXiv:1404.0287}}\relax
\mciteBstWouldAddEndPuncttrue
\mciteSetBstMidEndSepPunct{\mcitedefaultmidpunct}
{\mcitedefaultendpunct}{\mcitedefaultseppunct}\relax
\EndOfBibitem
\bibitem{LHCb-PAPER-2014-039}
LHCb collaboration, R.~Aaij {\em et~al.},
  \ifthenelse{\boolean{articletitles}}{\emph{{First observation of a baryonic
  \Bcp decay}}, }{}\href{https://doi.org/10.1103/PhysRevLett.113.152003}{Phys.\
  Rev.\ Lett.\  \textbf{113} (2014) 152003},
  \href{http://arxiv.org/abs/1408.0971}{{\normalfont\ttfamily
  arXiv:1408.0971}}\relax
\mciteBstWouldAddEndPuncttrue
\mciteSetBstMidEndSepPunct{\mcitedefaultmidpunct}
{\mcitedefaultendpunct}{\mcitedefaultseppunct}\relax
\EndOfBibitem
\bibitem{LHCb-PAPER-2014-050}
LHCb collaboration, R.~Aaij {\em et~al.},
  \ifthenelse{\boolean{articletitles}}{\emph{{Measurement of \Bcp production in
  proton-proton collisions at \mbox{$\sqs=$8\tev}}},
  }{}\href{https://doi.org/10.1103/PhysRevLett.114.132001}{Phys.\ Rev.\ Lett.\
  \textbf{114} (2015) 132001},
  \href{http://arxiv.org/abs/1411.2943}{{\normalfont\ttfamily
  arXiv:1411.2943}}\relax
\mciteBstWouldAddEndPuncttrue
\mciteSetBstMidEndSepPunct{\mcitedefaultmidpunct}
{\mcitedefaultendpunct}{\mcitedefaultseppunct}\relax
\EndOfBibitem
\bibitem{LHCb-PAPER-2014-060}
LHCb collaboration, R.~Aaij {\em et~al.},
  \ifthenelse{\boolean{articletitles}}{\emph{{Measurement of the lifetime of
  the \Bcp meson using the \mbox{\decay{\Bc}{\jpsi\pip}} decay mode}},
  }{}\href{https://doi.org/10.1016/j.physletb.2015.01.010}{Phys.\ Lett.\
  \textbf{B742} (2015) 29},
  \href{http://arxiv.org/abs/1411.6899}{{\normalfont\ttfamily
  arXiv:1411.6899}}\relax
\mciteBstWouldAddEndPuncttrue
\mciteSetBstMidEndSepPunct{\mcitedefaultmidpunct}
{\mcitedefaultendpunct}{\mcitedefaultseppunct}\relax
\EndOfBibitem
\bibitem{LHCb-PAPER-2015-024}
LHCb collaboration, R.~Aaij {\em et~al.},
  \ifthenelse{\boolean{articletitles}}{\emph{{Measurement of the branching
  fraction ratio
  \mbox{$\BF(\decay{\Bcp}{\psitwos\pip})/\BF(\decay{\Bcp}{\jpsi\pip})$}}},
  }{}\href{https://doi.org/10.1103/PhysRevD.92.072007}{Phys.\ Rev.\
  \textbf{D92} (2015) 072007},
  \href{http://arxiv.org/abs/1507.03516}{{\normalfont\ttfamily
  arXiv:1507.03516}}\relax
\mciteBstWouldAddEndPuncttrue
\mciteSetBstMidEndSepPunct{\mcitedefaultmidpunct}
{\mcitedefaultendpunct}{\mcitedefaultseppunct}\relax
\EndOfBibitem
\bibitem{LHCb-PAPER-2016-001}
LHCb collaboration, R.~Aaij {\em et~al.},
  \ifthenelse{\boolean{articletitles}}{\emph{{Search for \Bcp decays to the
  $\proton\antiproton\pip$ final state}},
  }{}\href{https://doi.org/10.1016/j.physletb.2016.05.074}{Phys.\ Lett.\
  \textbf{B759} (2016) 313},
  \href{http://arxiv.org/abs/1603.07037}{{\normalfont\ttfamily
  arXiv:1603.07037}}\relax
\mciteBstWouldAddEndPuncttrue
\mciteSetBstMidEndSepPunct{\mcitedefaultmidpunct}
{\mcitedefaultendpunct}{\mcitedefaultseppunct}\relax
\EndOfBibitem
\bibitem{LHCb-PAPER-2016-022}
LHCb collaboration, R.~Aaij {\em et~al.},
  \ifthenelse{\boolean{articletitles}}{\emph{{Study of \Bcp decays to the
  $\Kp\Km\pip$ final state and evidence for the decay
  \mbox{\decay{\Bc}{\chiczero\pip}}}},
  }{}\href{https://doi.org/10.1103/PhysRevD.94.091102}{Phys.\ Rev.\
  \textbf{D94} (2016) 091102(R)},
  \href{http://arxiv.org/abs/1607.06134}{{\normalfont\ttfamily
  arXiv:1607.06134}}\relax
\mciteBstWouldAddEndPuncttrue
\mciteSetBstMidEndSepPunct{\mcitedefaultmidpunct}
{\mcitedefaultendpunct}{\mcitedefaultseppunct}\relax
\EndOfBibitem
\bibitem{LHCb-PAPER-2016-020}
LHCb collaboration, R.~Aaij {\em et~al.},
  \ifthenelse{\boolean{articletitles}}{\emph{{Measurement of the ratio of
  branching fractions
  \mbox{$\BF(\decay{\Bcp}{\jpsi\Kp})/\BF(\decay{\Bcp}{\jpsi\pip})$}}},
  }{}\href{https://doi.org/10.1007/JHEP09(2016)153}{JHEP \textbf{09} (2016)
  153}, \href{http://arxiv.org/abs/1607.06823}{{\normalfont\ttfamily
  arXiv:1607.06823}}\relax
\mciteBstWouldAddEndPuncttrue
\mciteSetBstMidEndSepPunct{\mcitedefaultmidpunct}
{\mcitedefaultendpunct}{\mcitedefaultseppunct}\relax
\EndOfBibitem
\bibitem{LHCb-PAPER-2016-055}
LHCb collaboration, R.~Aaij {\em et~al.},
  \ifthenelse{\boolean{articletitles}}{\emph{{Observation of
  \mbox{\decay{\Bc}{\jpsi \PD^{(\ast)} \PK^{(\ast)}}} decays}},
  }{}\href{https://doi.org/10.1103/PhysRevD.95.032005}{Phys.\ Rev.\
  \textbf{D95} (2017) 032005},
  \href{http://arxiv.org/abs/1612.07421}{{\normalfont\ttfamily
  arXiv:1612.07421}}\relax
\mciteBstWouldAddEndPuncttrue
\mciteSetBstMidEndSepPunct{\mcitedefaultmidpunct}
{\mcitedefaultendpunct}{\mcitedefaultseppunct}\relax
\EndOfBibitem
\bibitem{LHCb-PAPER-2016-058}
LHCb collaboration, R.~Aaij {\em et~al.},
  \ifthenelse{\boolean{articletitles}}{\emph{{Observation of
  \mbox{\decay{\Bc}{\Dz\Kp}} decays}},
  }{}\href{https://doi.org/10.1103/PhysRevLett.118.111803}{Phys.\ Rev.\ Lett.\
  \textbf{118} (2017) 111803},
  \href{http://arxiv.org/abs/1701.01856}{{\normalfont\ttfamily
  arXiv:1701.01856}}\relax
\mciteBstWouldAddEndPuncttrue
\mciteSetBstMidEndSepPunct{\mcitedefaultmidpunct}
{\mcitedefaultendpunct}{\mcitedefaultseppunct}\relax
\EndOfBibitem
\bibitem{LHCb-PAPER-2017-035}
LHCb collaboration, R.~Aaij {\em et~al.},
  \ifthenelse{\boolean{articletitles}}{\emph{{Measurement of the ratio of
  branching fractions
  \mbox{$\mathcal{B}(\decay{\Bc}{\jpsi\taup\Pnu_{\Ptau}})/\mathcal{B}(\decay{\Bc}{\jpsi\mup\Pnu_{\Pmu}})$}}},
  }{}\href{https://doi.org/10.1103/PhysRevLett.120.121801}{Phys.\ Rev.\ Lett.\
  \textbf{120} (2018) 121801},
  \href{http://arxiv.org/abs/1711.05623}{{\normalfont\ttfamily
  arXiv:1711.05623}}\relax
\mciteBstWouldAddEndPuncttrue
\mciteSetBstMidEndSepPunct{\mcitedefaultmidpunct}
{\mcitedefaultendpunct}{\mcitedefaultseppunct}\relax
\EndOfBibitem
\bibitem{LHCb-PAPER-2017-045}
LHCb collaboration, R.~Aaij {\em et~al.},
  \ifthenelse{\boolean{articletitles}}{\emph{{Search for \Bc decays to two
  charm mesons}},
  }{}\href{https://doi.org/10.1016/j.nuclphysb.2018.03.015}{Nucl.\ Phys.\
  \textbf{B930} (2018) 563},
  \href{http://arxiv.org/abs/1712.04702}{{\normalfont\ttfamily
  arXiv:1712.04702}}\relax
\mciteBstWouldAddEndPuncttrue
\mciteSetBstMidEndSepPunct{\mcitedefaultmidpunct}
{\mcitedefaultendpunct}{\mcitedefaultseppunct}\relax
\EndOfBibitem
\bibitem{LHCb-PAPER-2019-033}
LHCb collaboration, R.~Aaij {\em et~al.},
  \ifthenelse{\boolean{articletitles}}{\emph{{Measurement of the \Bcm
  production fraction and asymmetry in 7 and 13\tev $\proton\proton$
  collisions}}, }{}\href{https://doi.org/10.1103/PhysRevD.100.112006}{Phys.\
  Rev.\  \textbf{D100} (2019) 112006},
  \href{http://arxiv.org/abs/1910.13404}{{\normalfont\ttfamily
  arXiv:1910.13404}}\relax
\mciteBstWouldAddEndPuncttrue
\mciteSetBstMidEndSepPunct{\mcitedefaultmidpunct}
{\mcitedefaultendpunct}{\mcitedefaultseppunct}\relax
\EndOfBibitem
\bibitem{LHCb-PAPER-2020-003}
LHCb collaboration, R.~Aaij {\em et~al.},
  \ifthenelse{\boolean{articletitles}}{\emph{{Precision measurement of the \Bc
  meson mass}}, }{}\href{https://doi.org/10.1007/JHEP07(2020)123}{JHEP
  \textbf{07} (2020) 123},
  \href{http://arxiv.org/abs/2004.08163}{{\normalfont\ttfamily
  arXiv:2004.08163}}\relax
\mciteBstWouldAddEndPuncttrue
\mciteSetBstMidEndSepPunct{\mcitedefaultmidpunct}
{\mcitedefaultendpunct}{\mcitedefaultseppunct}\relax
\EndOfBibitem
\bibitem{LHCb-PAPER-2021-023}
LHCb collaboration, R.~Aaij {\em et~al.},
  \ifthenelse{\boolean{articletitles}}{\emph{{Updated search for $\Bc$ decays
  to two charm mesons}},
  }{}\href{http://arxiv.org/abs/2109.00488}{{\normalfont\ttfamily
  arXiv:2109.00488}}, {submitted to JHEP}\relax
\mciteBstWouldAddEndPuncttrue
\mciteSetBstMidEndSepPunct{\mcitedefaultmidpunct}
{\mcitedefaultendpunct}{\mcitedefaultseppunct}\relax
\EndOfBibitem
\bibitem{PhysRevLett.81.2432}
\cdf collaboration, F.~Abe {\em et~al.},
  \ifthenelse{\boolean{articletitles}}{\emph{Observation of the ${\PB}_{\Pc}$
  meson in $\proton\antiproton$ collisions at $\sqrt{s}=1.8\,\tev$},
  }{}\href{https://doi.org/10.1103/PhysRevLett.81.2432}{Phys.\ Rev.\ Lett.\
  \textbf{81} (1998) 2432},
  \href{http://arxiv.org/abs/hep-ex/9805034}{{\normalfont\ttfamily
  arXiv:hep-ex/9805034}}\relax
\mciteBstWouldAddEndPuncttrue
\mciteSetBstMidEndSepPunct{\mcitedefaultmidpunct}
{\mcitedefaultendpunct}{\mcitedefaultseppunct}\relax
\EndOfBibitem
\bibitem{PhysRevD.58.112004}
CDF Collaboration, F.~Abe {\em et~al.},
  \ifthenelse{\boolean{articletitles}}{\emph{Observation of ${\PB}_{\Pc}$
  mesons in $\proton\antiproton$ collisions at $\sqrt{s}=1.8\,\tev$},
  }{}\href{https://doi.org/10.1103/PhysRevD.58.112004}{Phys.\ Rev.\
  \textbf{D58} (1998) 112004},
  \href{http://arxiv.org/abs/hep-ex/9804014}{{\normalfont\ttfamily
  arXiv:hep-ex/9804014}}\relax
\mciteBstWouldAddEndPuncttrue
\mciteSetBstMidEndSepPunct{\mcitedefaultmidpunct}
{\mcitedefaultendpunct}{\mcitedefaultseppunct}\relax
\EndOfBibitem
\bibitem{Bauer:1986bm}
M.~Bauer, B.~Stech, and M.~Wirbel,
  \ifthenelse{\boolean{articletitles}}{\emph{{Exclusive non-leptonic decays of
  \PD-, $\PD_{\Ps}$-, and \PB-mesons}},
  }{}\href{https://doi.org/10.1007/BF01561122}{Z.\ Phys.\  \textbf{C34} (1987)
  103}\relax
\mciteBstWouldAddEndPuncttrue
\mciteSetBstMidEndSepPunct{\mcitedefaultmidpunct}
{\mcitedefaultendpunct}{\mcitedefaultseppunct}\relax
\EndOfBibitem
\bibitem{WIRBEL198833}
M.~Wirbel, \ifthenelse{\boolean{articletitles}}{\emph{{Description of weak
  decays of \PD and \PB mesons}},
  }{}\href{https://doi.org/https://doi.org/10.1016/0146-6410(88)90031-2}{Prog.\
  Part.\ Nucl.\ Phys.\  \textbf{21} (1988) 33}\relax
\mciteBstWouldAddEndPuncttrue
\mciteSetBstMidEndSepPunct{\mcitedefaultmidpunct}
{\mcitedefaultendpunct}{\mcitedefaultseppunct}\relax
\EndOfBibitem
\bibitem{Likhoded:2009ib}
A.~K. Likhoded and A.~V. Luchinsky,
  \ifthenelse{\boolean{articletitles}}{\emph{{Light hadron production in
  \decay{\PB_{\Pc}}{\jpsi + \PX} decays}},
  }{}\href{https://doi.org/10.1103/PhysRevD.81.014015}{Phys.\ Rev.\
  \textbf{D81} (2010) 014015},
  \href{http://arxiv.org/abs/0910.3089}{{\normalfont\ttfamily
  arXiv:0910.3089}}\relax
\mciteBstWouldAddEndPuncttrue
\mciteSetBstMidEndSepPunct{\mcitedefaultmidpunct}
{\mcitedefaultendpunct}{\mcitedefaultseppunct}\relax
\EndOfBibitem
\bibitem{Luchinsky:2013yla}
A.~V. Luchinsky, \ifthenelse{\boolean{articletitles}}{\emph{{Production of \PK
  mesons in exclusive $\PB_{\Pc}$ decays}},
  }{}\href{http://arxiv.org/abs/1307.0953}{{\normalfont\ttfamily
  arXiv:1307.0953}}\relax
\mciteBstWouldAddEndPuncttrue
\mciteSetBstMidEndSepPunct{\mcitedefaultmidpunct}
{\mcitedefaultendpunct}{\mcitedefaultseppunct}\relax
\EndOfBibitem
\bibitem{Likhoded:2013iua}
A.~K. Likhoded and A.~V. Luchinsky,
  \ifthenelse{\boolean{articletitles}}{\emph{{Production of a pion system in
  exclusive $\decay{\PB_{\Pc}}{\PV(\PP) + \Pn\pion}$ decays}},
  }{}\href{https://doi.org/10.1134/S1063778813050062}{Phys.\ Atom.\ Nucl.\
  \textbf{76} (2013) 787}\relax
\mciteBstWouldAddEndPuncttrue
\mciteSetBstMidEndSepPunct{\mcitedefaultmidpunct}
{\mcitedefaultendpunct}{\mcitedefaultseppunct}\relax
\EndOfBibitem
\bibitem{Alves:2008zz}
\lhcb collaboration, A.~A. Alves~Jr.\ {\em et~al.},
  \ifthenelse{\boolean{articletitles}}{\emph{{The \lhcb detector at the LHC}},
  }{}\href{https://doi.org/10.1088/1748-0221/3/08/S08005}{JINST \textbf{3}
  (2008) S08005}\relax
\mciteBstWouldAddEndPuncttrue
\mciteSetBstMidEndSepPunct{\mcitedefaultmidpunct}
{\mcitedefaultendpunct}{\mcitedefaultseppunct}\relax
\EndOfBibitem
\bibitem{LHCb-DP-2014-002}
LHCb collaboration, R.~Aaij {\em et~al.},
  \ifthenelse{\boolean{articletitles}}{\emph{{\lhcb detector performance}},
  }{}\href{https://doi.org/10.1142/S0217751X15300227}{Int.\ J.\ Mod.\ Phys.\
  \textbf{A30} (2015) 1530022},
  \href{http://arxiv.org/abs/1412.6352}{{\normalfont\ttfamily
  arXiv:1412.6352}}\relax
\mciteBstWouldAddEndPuncttrue
\mciteSetBstMidEndSepPunct{\mcitedefaultmidpunct}
{\mcitedefaultendpunct}{\mcitedefaultseppunct}\relax
\EndOfBibitem
\bibitem{LHCb-DP-2014-001}
R.~Aaij {\em et~al.}, \ifthenelse{\boolean{articletitles}}{\emph{{Performance
  of the LHCb Vertex Locator}},
  }{}\href{https://doi.org/10.1088/1748-0221/9/09/P09007}{JINST \textbf{9}
  (2014) P09007}, \href{http://arxiv.org/abs/1405.7808}{{\normalfont\ttfamily
  arXiv:1405.7808}}\relax
\mciteBstWouldAddEndPuncttrue
\mciteSetBstMidEndSepPunct{\mcitedefaultmidpunct}
{\mcitedefaultendpunct}{\mcitedefaultseppunct}\relax
\EndOfBibitem
\bibitem{LHCb-DP-2013-003}
R.~Arink {\em et~al.}, \ifthenelse{\boolean{articletitles}}{\emph{{Performance
  of the LHCb Outer Tracker}},
  }{}\href{https://doi.org/10.1088/1748-0221/9/01/P01002}{JINST \textbf{9}
  (2014) P01002}, \href{http://arxiv.org/abs/1311.3893}{{\normalfont\ttfamily
  arXiv:1311.3893}}\relax
\mciteBstWouldAddEndPuncttrue
\mciteSetBstMidEndSepPunct{\mcitedefaultmidpunct}
{\mcitedefaultendpunct}{\mcitedefaultseppunct}\relax
\EndOfBibitem
\bibitem{LHCb-DP-2017-001}
P.~d'Argent {\em et~al.}, \ifthenelse{\boolean{articletitles}}{\emph{{Improved
  performance of the LHCb Outer Tracker in LHC Run~2}},
  }{}\href{https://doi.org/10.1088/1748-0221/12/11/P11016}{JINST \textbf{12}
  (2017) P11016}, \href{http://arxiv.org/abs/1708.00819}{{\normalfont\ttfamily
  arXiv:1708.00819}}\relax
\mciteBstWouldAddEndPuncttrue
\mciteSetBstMidEndSepPunct{\mcitedefaultmidpunct}
{\mcitedefaultendpunct}{\mcitedefaultseppunct}\relax
\EndOfBibitem
\bibitem{LHCb-PAPER-2012-048}
LHCb collaboration, R.~Aaij {\em et~al.},
  \ifthenelse{\boolean{articletitles}}{\emph{{Measurement of the \Lb, \Xibm,
  and \Omegab baryon masses}},
  }{}\href{https://doi.org/10.1103/PhysRevLett.110.182001}{Phys.\ Rev.\ Lett.\
  \textbf{110} (2013) 182001},
  \href{http://arxiv.org/abs/1302.1072}{{\normalfont\ttfamily
  arXiv:1302.1072}}\relax
\mciteBstWouldAddEndPuncttrue
\mciteSetBstMidEndSepPunct{\mcitedefaultmidpunct}
{\mcitedefaultendpunct}{\mcitedefaultseppunct}\relax
\EndOfBibitem
\bibitem{LHCb-PAPER-2013-011}
LHCb collaboration, R.~Aaij {\em et~al.},
  \ifthenelse{\boolean{articletitles}}{\emph{{Precision measurement of \D meson
  mass differences}}, }{}\href{https://doi.org/10.1007/JHEP06(2013)065}{JHEP
  \textbf{06} (2013) 065},
  \href{http://arxiv.org/abs/1304.6865}{{\normalfont\ttfamily
  arXiv:1304.6865}}\relax
\mciteBstWouldAddEndPuncttrue
\mciteSetBstMidEndSepPunct{\mcitedefaultmidpunct}
{\mcitedefaultendpunct}{\mcitedefaultseppunct}\relax
\EndOfBibitem
\bibitem{LHCb-DP-2012-003}
M.~Adinolfi {\em et~al.},
  \ifthenelse{\boolean{articletitles}}{\emph{{Performance of the \lhcb \rich
  detector at the LHC}},
  }{}\href{https://doi.org/10.1140/epjc/s10052-013-2431-9}{Eur.\ Phys.\ J.\
  \textbf{C73} (2013) 2431},
  \href{http://arxiv.org/abs/1211.6759}{{\normalfont\ttfamily
  arXiv:1211.6759}}\relax
\mciteBstWouldAddEndPuncttrue
\mciteSetBstMidEndSepPunct{\mcitedefaultmidpunct}
{\mcitedefaultendpunct}{\mcitedefaultseppunct}\relax
\EndOfBibitem
\bibitem{LHCb-DP-2012-002}
A.~A. Alves~Jr.\ {\em et~al.},
  \ifthenelse{\boolean{articletitles}}{\emph{{Performance of the LHCb muon
  system}}, }{}\href{https://doi.org/10.1088/1748-0221/8/02/P02022}{JINST
  \textbf{8} (2013) P02022},
  \href{http://arxiv.org/abs/1211.1346}{{\normalfont\ttfamily
  arXiv:1211.1346}}\relax
\mciteBstWouldAddEndPuncttrue
\mciteSetBstMidEndSepPunct{\mcitedefaultmidpunct}
{\mcitedefaultendpunct}{\mcitedefaultseppunct}\relax
\EndOfBibitem
\bibitem{LHCb-DP-2012-004}
R.~Aaij {\em et~al.}, \ifthenelse{\boolean{articletitles}}{\emph{{The \lhcb
  trigger and its performance in 2011}},
  }{}\href{https://doi.org/10.1088/1748-0221/8/04/P04022}{JINST \textbf{8}
  (2013) P04022}, \href{http://arxiv.org/abs/1211.3055}{{\normalfont\ttfamily
  arXiv:1211.3055}}\relax
\mciteBstWouldAddEndPuncttrue
\mciteSetBstMidEndSepPunct{\mcitedefaultmidpunct}
{\mcitedefaultendpunct}{\mcitedefaultseppunct}\relax
\EndOfBibitem
\bibitem{Sjostrand:2007gs}
T.~Sj\"{o}strand, S.~Mrenna, and P.~Skands,
  \ifthenelse{\boolean{articletitles}}{\emph{{A brief introduction to
  $\pythia\,8.1$}},
  }{}\href{https://doi.org/10.1016/j.cpc.2008.01.036}{Comput.\ Phys.\ Commun.\
  \textbf{178} (2008) 852},
  \href{http://arxiv.org/abs/0710.3820}{{\normalfont\ttfamily
  arXiv:0710.3820}}\relax
\mciteBstWouldAddEndPuncttrue
\mciteSetBstMidEndSepPunct{\mcitedefaultmidpunct}
{\mcitedefaultendpunct}{\mcitedefaultseppunct}\relax
\EndOfBibitem
\bibitem{LHCb-PROC-2010-056}
I.~Belyaev {\em et~al.}, \ifthenelse{\boolean{articletitles}}{\emph{{Handling
  of the generation of primary events in \gauss, the \lhcb simulation
  framework}}, }{}\href{https://doi.org/10.1088/1742-6596/331/3/032047}{J.\
  Phys.\ Conf.\ Ser.\  \textbf{331} (2011) 032047}\relax
\mciteBstWouldAddEndPuncttrue
\mciteSetBstMidEndSepPunct{\mcitedefaultmidpunct}
{\mcitedefaultendpunct}{\mcitedefaultseppunct}\relax
\EndOfBibitem
\bibitem{Lange:2001uf}
D.~J. Lange, \ifthenelse{\boolean{articletitles}}{\emph{{The \evtgen particle
  decay simulation package}},
  }{}\href{https://doi.org/10.1016/S0168-9002(01)00089-4}{Nucl.\ Instrum.\
  Meth.\  \textbf{A462} (2001) 152}\relax
\mciteBstWouldAddEndPuncttrue
\mciteSetBstMidEndSepPunct{\mcitedefaultmidpunct}
{\mcitedefaultendpunct}{\mcitedefaultseppunct}\relax
\EndOfBibitem
\bibitem{davidson2015photos}
N.~Davidson, T.~Przedzinski, and Z.~Was,
  \ifthenelse{\boolean{articletitles}}{\emph{{PHOTOS interface in C++:
  Technical and physics documentation}},
  }{}\href{https://doi.org/https://doi.org/10.1016/j.cpc.2015.09.013}{Comp.\
  Phys.\ Comm.\  \textbf{199} (2016) 86},
  \href{http://arxiv.org/abs/1011.0937}{{\normalfont\ttfamily
  arXiv:1011.0937}}\relax
\mciteBstWouldAddEndPuncttrue
\mciteSetBstMidEndSepPunct{\mcitedefaultmidpunct}
{\mcitedefaultendpunct}{\mcitedefaultseppunct}\relax
\EndOfBibitem
\bibitem{Berezhnoy:2011nx}
A.~V. Berezhnoy, A.~K. Likhoded, and A.~V. Luchinsky,
  \ifthenelse{\boolean{articletitles}}{\emph{{ {\tt{BC\_NPI}} module for
  the~analysis of $\decay{\PB_{\Pc}}{\jpsi+\mathrm{n}\pion}$ and
  $\decay{\PB_{\Pc}}{\Bs+\mathrm{n}\pion}$~decays within the
  {\sc{EvtGen}}~package}},
  }{}\href{http://arxiv.org/abs/1104.0808}{{\normalfont\ttfamily
  arXiv:1104.0808}}\relax
\mciteBstWouldAddEndPuncttrue
\mciteSetBstMidEndSepPunct{\mcitedefaultmidpunct}
{\mcitedefaultendpunct}{\mcitedefaultseppunct}\relax
\EndOfBibitem
\bibitem{LHCb-PAPER-2016-018}
LHCb collaboration, R.~Aaij {\em et~al.},
  \ifthenelse{\boolean{articletitles}}{\emph{{Observation of $\jpsi\Pphi$
  structures consistent with exotic states from amplitude analysis of
  \mbox{\decay{\Bp}{\jpsi\Pphi\Kp}} decays}},
  }{}\href{https://doi.org/10.1103/PhysRevLett.118.022003}{Phys.\ Rev.\ Lett.\
  \textbf{118} (2017) 022003},
  \href{http://arxiv.org/abs/1606.07895}{{\normalfont\ttfamily
  arXiv:1606.07895}}\relax
\mciteBstWouldAddEndPuncttrue
\mciteSetBstMidEndSepPunct{\mcitedefaultmidpunct}
{\mcitedefaultendpunct}{\mcitedefaultseppunct}\relax
\EndOfBibitem
\bibitem{LHCb-PAPER-2016-019}
LHCb collaboration, R.~Aaij {\em et~al.},
  \ifthenelse{\boolean{articletitles}}{\emph{{Amplitude analysis of
  \mbox{\decay{\Bp}{\jpsi\phiz\Kp}} decays}},
  }{}\href{https://doi.org/10.1103/PhysRevD.95.012002}{Phys.\ Rev.\
  \textbf{D95} (2017) 012002},
  \href{http://arxiv.org/abs/1606.07898}{{\normalfont\ttfamily
  arXiv:1606.07898}}\relax
\mciteBstWouldAddEndPuncttrue
\mciteSetBstMidEndSepPunct{\mcitedefaultmidpunct}
{\mcitedefaultendpunct}{\mcitedefaultseppunct}\relax
\EndOfBibitem
\bibitem{Allison:2006ve}
Geant4 collaboration, J.~Allison {\em et~al.},
  \ifthenelse{\boolean{articletitles}}{\emph{{\geant developments and
  applications}}, }{}\href{https://doi.org/10.1109/TNS.2006.869826}{IEEE
  Trans.\ Nucl.\ Sci.\  \textbf{53} (2006) 270}\relax
\mciteBstWouldAddEndPuncttrue
\mciteSetBstMidEndSepPunct{\mcitedefaultmidpunct}
{\mcitedefaultendpunct}{\mcitedefaultseppunct}\relax
\EndOfBibitem
\bibitem{Agostinelli:2002hh}
Geant4 collaboration, S.~Agostinelli {\em et~al.},
  \ifthenelse{\boolean{articletitles}}{\emph{{\geant~--~a simulation toolkit}},
  }{}\href{https://doi.org/10.1016/S0168-9002(03)01368-8}{Nucl.\ Instrum.\
  Meth.\  \textbf{A506} (2003) 250}\relax
\mciteBstWouldAddEndPuncttrue
\mciteSetBstMidEndSepPunct{\mcitedefaultmidpunct}
{\mcitedefaultendpunct}{\mcitedefaultseppunct}\relax
\EndOfBibitem
\bibitem{LHCb-PROC-2011-006}
M.~Clemencic {\em et~al.}, \ifthenelse{\boolean{articletitles}}{\emph{{The
  \lhcb simulation application, \gauss: design, evolution and experience}},
  }{}\href{https://doi.org/10.1088/1742-6596/331/3/032023}{J.\ Phys.\ Conf.\
  Ser.\  \textbf{331} (2011) 032023}\relax
\mciteBstWouldAddEndPuncttrue
\mciteSetBstMidEndSepPunct{\mcitedefaultmidpunct}
{\mcitedefaultendpunct}{\mcitedefaultseppunct}\relax
\EndOfBibitem
\bibitem{LHCb-DP-2013-002}
LHCb collaboration, R.~Aaij {\em et~al.},
  \ifthenelse{\boolean{articletitles}}{\emph{{Measurement of the track
  reconstruction efficiency at LHCb}},
  }{}\href{https://doi.org/10.1088/1748-0221/10/02/P02007}{JINST \textbf{10}
  (2015) P02007}, \href{http://arxiv.org/abs/1408.1251}{{\normalfont\ttfamily
  arXiv:1408.1251}}\relax
\mciteBstWouldAddEndPuncttrue
\mciteSetBstMidEndSepPunct{\mcitedefaultmidpunct}
{\mcitedefaultendpunct}{\mcitedefaultseppunct}\relax
\EndOfBibitem
\bibitem{LHCb-PROC-2011-008}
A.~Powell {\em et~al.}, \ifthenelse{\boolean{articletitles}}{\emph{{Particle
  identification at \lhcb}}, }{}PoS \textbf{ICHEP2010} (2010) 020,
  \href{https://cdsweb.cern.ch/record/1322666?ln=en}{LHCb-PROC-2011-008}\relax
\mciteBstWouldAddEndPuncttrue
\mciteSetBstMidEndSepPunct{\mcitedefaultmidpunct}
{\mcitedefaultendpunct}{\mcitedefaultseppunct}\relax
\EndOfBibitem
\bibitem{Hulsbergen:2005pu}
W.~D. Hulsbergen, \ifthenelse{\boolean{articletitles}}{\emph{{Decay chain
  fitting with a Kalman filter}},
  }{}\href{https://doi.org/10.1016/j.nima.2005.06.078}{Nucl.\ Instrum.\ Meth.\
  \textbf{A552} (2005) 566},
  \href{http://arxiv.org/abs/physics/0503191}{{\normalfont\ttfamily
  arXiv:physics/0503191}}\relax
\mciteBstWouldAddEndPuncttrue
\mciteSetBstMidEndSepPunct{\mcitedefaultmidpunct}
{\mcitedefaultendpunct}{\mcitedefaultseppunct}\relax
\EndOfBibitem
\bibitem{PDG2021}
Particle Data Group, P.~A. Zyla {\em et~al.},
  \ifthenelse{\boolean{articletitles}}{\emph{{\href{http://pdg.lbl.gov/}{Review
  of particle physics}}}, }{}\href{https://doi.org/10.1093/ptep/ptaa104}{Prog.\
  Theor.\ Exp.\ Phys.\  \textbf{2020} (2020) 083C01}, and
  {\href{http://pdglive.lbl.gov/}{2021 update}}\relax
\mciteBstWouldAddEndPuncttrue
\mciteSetBstMidEndSepPunct{\mcitedefaultmidpunct}
{\mcitedefaultendpunct}{\mcitedefaultseppunct}\relax
\EndOfBibitem
\bibitem{LHCb-PAPER-2020-009}
LHCb collaboration, R.~Aaij {\em et~al.},
  \ifthenelse{\boolean{articletitles}}{\emph{{Study of the $\psitwod$ and
  $\chiconex$ states in $\decay{\Bp}{(\jpsi \pip\pim)\Kp}$ decays}},
  }{}\href{https://doi.org/10.1007/JHEP08(2020)123}{JHEP \textbf{08} (2020)
  123}, \href{http://arxiv.org/abs/2005.13422}{{\normalfont\ttfamily
  arXiv:2005.13422}}\relax
\mciteBstWouldAddEndPuncttrue
\mciteSetBstMidEndSepPunct{\mcitedefaultmidpunct}
{\mcitedefaultendpunct}{\mcitedefaultseppunct}\relax
\EndOfBibitem
\bibitem{LHCb-PAPER-2020-035}
LHCb collaboration, R.~Aaij {\em et~al.},
  \ifthenelse{\boolean{articletitles}}{\emph{{Study of $\Bs\to \jpsi \pip \pim
  \Kp \Km$ decays}}, }{}\href{https://doi.org/10.1007/JHEP02(2021)024}{JHEP
  \textbf{02} (2021) 024},
  \href{http://arxiv.org/abs/2011.01867}{{\normalfont\ttfamily
  arXiv:2011.01867}}\relax
\mciteBstWouldAddEndPuncttrue
\mciteSetBstMidEndSepPunct{\mcitedefaultmidpunct}
{\mcitedefaultendpunct}{\mcitedefaultseppunct}\relax
\EndOfBibitem
\bibitem{LHCb-PAPER-2011-013}
LHCb collaboration, R.~Aaij {\em et~al.},
  \ifthenelse{\boolean{articletitles}}{\emph{{Observation of \jpsi-pair
  production in \proton\proton collisions at \mbox{$\sqs=$7\tev}}},
  }{}\href{https://doi.org/10.1016/j.physletb.2011.12.015}{Phys.\ Lett.\
  \textbf{B707} (2012) 52},
  \href{http://arxiv.org/abs/1109.0963}{{\normalfont\ttfamily
  arXiv:1109.0963}}\relax
\mciteBstWouldAddEndPuncttrue
\mciteSetBstMidEndSepPunct{\mcitedefaultmidpunct}
{\mcitedefaultendpunct}{\mcitedefaultseppunct}\relax
\EndOfBibitem
\bibitem{Skwarnicki:1986xj}
T.~Skwarnicki, {\em {A~study of the~radiative cascade transitions between
  the~$\PUpsilon^{\prime}$ and $\PUpsilon$~resonances}}, PhD thesis, Institute
  of Nuclear Physics, Krakow, 1986,
  {\href{http://inspirehep.net/record/230779/}{DESY-F31-86-02}}\relax
\mciteBstWouldAddEndPuncttrue
\mciteSetBstMidEndSepPunct{\mcitedefaultmidpunct}
{\mcitedefaultendpunct}{\mcitedefaultseppunct}\relax
\EndOfBibitem
\bibitem{Byckling}
E.~Byckling and K.~Kajantie, {\em Particle kinematics}, John Wiley \& Sons
  Inc., New York, 1973\relax
\mciteBstWouldAddEndPuncttrue
\mciteSetBstMidEndSepPunct{\mcitedefaultmidpunct}
{\mcitedefaultendpunct}{\mcitedefaultseppunct}\relax
\EndOfBibitem
\bibitem{LHCb-PAPER-2020-008}
LHCb collaboration, R.~Aaij {\em et~al.},
  \ifthenelse{\boolean{articletitles}}{\emph{{Study of the lineshape of the
  $\chiconex$ state}},
  }{}\href{https://doi.org/10.1103/PhysRevD.102.092005}{Phys.\ Rev.\
  \textbf{D102} (2020) 092005},
  \href{http://arxiv.org/abs/2005.13419}{{\normalfont\ttfamily
  arXiv:2005.13419}}\relax
\mciteBstWouldAddEndPuncttrue
\mciteSetBstMidEndSepPunct{\mcitedefaultmidpunct}
{\mcitedefaultendpunct}{\mcitedefaultseppunct}\relax
\EndOfBibitem
\bibitem{Pivk:2004ty}
M.~Pivk and F.~R. Le~Diberder,
  \ifthenelse{\boolean{articletitles}}{\emph{{sPlot: a statistical tool to
  unfold data distributions}},
  }{}\href{https://doi.org/10.1016/j.nima.2005.08.106}{Nucl.\ Instrum.\ Meth.\
  \textbf{A555} (2005) 356},
  \href{http://arxiv.org/abs/physics/0402083}{{\normalfont\ttfamily
  arXiv:physics/0402083}}\relax
\mciteBstWouldAddEndPuncttrue
\mciteSetBstMidEndSepPunct{\mcitedefaultmidpunct}
{\mcitedefaultendpunct}{\mcitedefaultseppunct}\relax
\EndOfBibitem
\bibitem{Wilks:1938dza}
S.~S. Wilks, \ifthenelse{\boolean{articletitles}}{\emph{{The large-sample
  distribution of the likelihood ratio for testing composite hypotheses}},
  }{}\href{https://doi.org/10.1214/aoms/1177732360}{Ann.\ Math.\ Stat.\
  \textbf{9} (1938) 60}\relax
\mciteBstWouldAddEndPuncttrue
\mciteSetBstMidEndSepPunct{\mcitedefaultmidpunct}
{\mcitedefaultendpunct}{\mcitedefaultseppunct}\relax
\EndOfBibitem
\bibitem{CLEO:1999rzk}
CLEO collaboration, D.~M. Asner {\em et~al.},
  \ifthenelse{\boolean{articletitles}}{\emph{{Hadronic structure in the~decay
  \mbox{$\decay{\taum}{\neut \pim\piz\piz}$} and the~sign of the~tau neutrino
  helicity}}, }{}\href{https://doi.org/10.1103/PhysRevD.61.012002}{Phys.\ Rev.\
   \textbf{D61} (1999) 012002},
  \href{http://arxiv.org/abs/hep-ex/9902022}{{\normalfont\ttfamily
  arXiv:hep-ex/9902022}}\relax
\mciteBstWouldAddEndPuncttrue
\mciteSetBstMidEndSepPunct{\mcitedefaultmidpunct}
{\mcitedefaultendpunct}{\mcitedefaultseppunct}\relax
\EndOfBibitem
\bibitem{Blatt:1952ije}
J.~M. Blatt and V.~F. Weisskopf, {\em {Theoretical nuclear physics}},
  \href{https://doi.org/10.1007/978-1-4612-9959-2}{ Springer, New York,
  1952}\relax
\mciteBstWouldAddEndPuncttrue
\mciteSetBstMidEndSepPunct{\mcitedefaultmidpunct}
{\mcitedefaultendpunct}{\mcitedefaultseppunct}\relax
\EndOfBibitem
\bibitem{Okubo:1963fa}
S.~Okubo, \ifthenelse{\boolean{articletitles}}{\emph{{$\Pphi$\nobreakdash-meson
  and unitary symmetry model}},
  }{}\href{https://doi.org/10.1016/S0375-9601(63)92548-9}{Phys.\ Lett.\
  \textbf{5} (1963) 165}\relax
\mciteBstWouldAddEndPuncttrue
\mciteSetBstMidEndSepPunct{\mcitedefaultmidpunct}
{\mcitedefaultendpunct}{\mcitedefaultseppunct}\relax
\EndOfBibitem
\bibitem{Zweig2}
G.~Zweig, \ifthenelse{\boolean{articletitles}}{\emph{{An SU$_3$ model for
  strong interaction symmetry and its breaking; Version 2}}}{}
  \href{http://cds.cern.ch/record/570209}{CERN-TH-412}, CERN, Geneva,
  1964\relax
\mciteBstWouldAddEndPuncttrue
\mciteSetBstMidEndSepPunct{\mcitedefaultmidpunct}
{\mcitedefaultendpunct}{\mcitedefaultseppunct}\relax
\EndOfBibitem
\bibitem{Iizuka:1966fk}
J.~Iizuka, \ifthenelse{\boolean{articletitles}}{\emph{{A systematics and
  phenomenology of meson family}},
  }{}\href{https://doi.org/10.1143/PTPS.37.21}{Suppl.\ Prog.\ Theor.\ Phys.\
  \textbf{37} (1966) 21}\relax
\mciteBstWouldAddEndPuncttrue
\mciteSetBstMidEndSepPunct{\mcitedefaultmidpunct}
{\mcitedefaultendpunct}{\mcitedefaultseppunct}\relax
\EndOfBibitem
\bibitem{LHCb-DP-2018-001}
R.~Aaij {\em et~al.}, \ifthenelse{\boolean{articletitles}}{\emph{{Selection and
  processing of calibration samples to measure the particle identification
  performance of the LHCb experiment in Run 2}},
  }{}\href{https://doi.org/10.1140/epjti/s40485-019-0050-z}{EPJ Tech.\
  Instrum.\  \textbf{6} (2019) 1},
  \href{http://arxiv.org/abs/1803.00824}{{\normalfont\ttfamily
  arXiv:1803.00824}}\relax
\mciteBstWouldAddEndPuncttrue
\mciteSetBstMidEndSepPunct{\mcitedefaultmidpunct}
{\mcitedefaultendpunct}{\mcitedefaultseppunct}\relax
\EndOfBibitem
\bibitem{Santos:2013gra}
D.~Mart{\'\i}nez~Santos and F.~Dupertuis,
  \ifthenelse{\boolean{articletitles}}{\emph{{Mass distributions marginalized
  over per-event errors}},
  }{}\href{https://doi.org/10.1016/j.nima.2014.06.081}{Nucl.\ Instrum.\ Meth.\
  \textbf{A764} (2014) 150},
  \href{http://arxiv.org/abs/1312.5000}{{\normalfont\ttfamily
  arXiv:1312.5000}}\relax
\mciteBstWouldAddEndPuncttrue
\mciteSetBstMidEndSepPunct{\mcitedefaultmidpunct}
{\mcitedefaultendpunct}{\mcitedefaultseppunct}\relax
\EndOfBibitem
\bibitem{LHCb-PAPER-2012-010}
LHCb collaboration, R.~Aaij {\em et~al.},
  \ifthenelse{\boolean{articletitles}}{\emph{{Measurement of relative branching
  fractions of \B decays to \psitwos and \jpsi mesons}},
  }{}\href{https://doi.org/10.1140/epjc/s10052-012-2118-7}{Eur.\ Phys.\ J.\
  \textbf{C72} (2012) 2118},
  \href{http://arxiv.org/abs/1205.0918}{{\normalfont\ttfamily
  arXiv:1205.0918}}\relax
\mciteBstWouldAddEndPuncttrue
\mciteSetBstMidEndSepPunct{\mcitedefaultmidpunct}
{\mcitedefaultendpunct}{\mcitedefaultseppunct}\relax
\EndOfBibitem
\bibitem{Gounaris:1968mw}
G.~J. Gounaris and J.~J. Sakurai,
  \ifthenelse{\boolean{articletitles}}{\emph{{Finite-width corrections to the
  vector-meson-dominance prediction for \mbox{$\decay{\rhoz}{\epem}$} }},
  }{}\href{https://doi.org/10.1103/PhysRevLett.21.244}{Phys.\ Rev.\ Lett.\
  \textbf{21} (1968) 244}\relax
\mciteBstWouldAddEndPuncttrue
\mciteSetBstMidEndSepPunct{\mcitedefaultmidpunct}
{\mcitedefaultendpunct}{\mcitedefaultseppunct}\relax
\EndOfBibitem
\bibitem{LHCb-PAPER-2011-040}
LHCb collaboration, R.~Aaij {\em et~al.},
  \ifthenelse{\boolean{articletitles}}{\emph{{First observation of the decays
  \mbox{\decay{\Bzb}{\Dp\Km\pip\pim}} and \mbox{\decay{\Bm}{\Dz\Km\pip\pim}}}},
  }{}\href{https://doi.org/10.1103/PhysRevLett.108.161801}{Phys.\ Rev.\ Lett.\
  \textbf{108} (2012) 161801},
  \href{http://arxiv.org/abs/1201.4402}{{\normalfont\ttfamily
  arXiv:1201.4402}}\relax
\mciteBstWouldAddEndPuncttrue
\mciteSetBstMidEndSepPunct{\mcitedefaultmidpunct}
{\mcitedefaultendpunct}{\mcitedefaultseppunct}\relax
\EndOfBibitem
\bibitem{LHCb-PAPER-2012-046}
LHCb collaboration, R.~Aaij {\em et~al.},
  \ifthenelse{\boolean{articletitles}}{\emph{{Study of
  \mbox{\decay{\Bz}{\Dstarm\pip\pim\pip}} and
  \mbox{\decay{\Bz}{\Dstarm\Kp\pim\pip}} decays}},
  }{}\href{https://doi.org/10.1103/PhysRevD.87.092001}{Phys.\ Rev.\
  \textbf{D87} (2013) 092001},
  \href{http://arxiv.org/abs/1303.6861}{{\normalfont\ttfamily
  arXiv:1303.6861}}\relax
\mciteBstWouldAddEndPuncttrue
\mciteSetBstMidEndSepPunct{\mcitedefaultmidpunct}
{\mcitedefaultendpunct}{\mcitedefaultseppunct}\relax
\EndOfBibitem
\bibitem{LHCb-PAPER-2012-033}
LHCb collaboration, R.~Aaij {\em et~al.},
  \ifthenelse{\boolean{articletitles}}{\emph{{First observation of the decays
  \mbox{\decay{\Bdsb}{\Dsp\Km\pip\pim}} and
  \mbox{\decay{\Bsb}{\PD_{\Ps1}(2536)^+\pim}}}},
  }{}\href{https://doi.org/10.1103/PhysRevD.86.112005}{Phys.\ Rev.\
  \textbf{D86} (2012) 112005},
  \href{http://arxiv.org/abs/1211.1541}{{\normalfont\ttfamily
  arXiv:1211.1541}}\relax
\mciteBstWouldAddEndPuncttrue
\mciteSetBstMidEndSepPunct{\mcitedefaultmidpunct}
{\mcitedefaultendpunct}{\mcitedefaultseppunct}\relax
\EndOfBibitem
\end{mcitethebibliography}
